\documentclass[stslayout, preprint]{imsart}
\usepackage{graphicx}
\usepackage{amssymb, amsmath}
\usepackage{epstopdf}
\usepackage{natbib}
\usepackage{subfigure}
\usepackage[colorlinks,citecolor=blue,urlcolor=blue]{hyperref}
\usepackage{algorithm}
\usepackage[noend]{algpseudocode}
\usepackage[table]{xcolor}
\usepackage{multirow}

\usepackage{tikz}
\usetikzlibrary{intersections, backgrounds, calc}

\definecolor{light}{RGB}{220, 188, 188}
\definecolor{mid}{RGB}{185, 124, 124}
\definecolor{dark}{RGB}{143, 39, 39}
\definecolor{highlight}{RGB}{0, 255, 0}
\definecolor{gray80}{gray}{0.8}
\definecolor{gray90}{gray}{0.9}
\definecolor{gray95}{gray}{0.95}
\definecolor{comment}{gray}{0.50}

\usepackage[thmmarks,amsmath]{ntheorem}

\theoremnumbering{arabic}
\theoremstyle{plain}
\RequirePackage{latexsym}

\theoremstyle{plain}
\newtheorem{definition}{Definition}

\DeclareFontFamily{U}{mathx}{\hyphenchar\font45}
\DeclareFontShape{U}{mathx}{m}{n}{
      <5> <6> <7> <8> <9> <10>
      <10.95> <12> <14.4> <17.28> <20.74> <24.88>
      mathx10
      }{}
\DeclareSymbolFont{mathx}{U}{mathx}{m}{n}
\DeclareFontSubstitution{U}{mathx}{m}{n}
\DeclareMathAccent{\widecheck}{0}{mathx}{"71}

\newcommand{\dd}{ \mathrm{d} }

\newcommand{\LS}{\ensuremath { H^{-1} \! \left( E \right) } }
\newcommand{\sLS}{\ensuremath { H^{-1} \left( E \right) } }
\newcommand{\MLS}{\ensuremath { \widetilde{H}^{-1} \! \left( E \right) } }

\newcommand\T{\mathcal{T}}

\graphicspath{{figures/}}

\begin{document}

\begin{frontmatter}

\title{Identifying the Optimal 
        Integration Time in
        \\ Hamiltonian Monte Carlo}
\runtitle{Optimal Integration Time}

\begin{aug}
  \author{Michael Betancourt%
  \ead[label=e1]{betanalpha@gmail.com}}

  \runauthor{Betancourt}

  \address{Department of Statistics, University of Warwick, 
  Coventry CV4 7AL, UK \\ \printead{e1}.}

\end{aug}

\begin{abstract}
By leveraging the natural geometry of a smooth probabilistic 
system, Hamiltonian Monte Carlo yields computationally efficient
Markov Chain Monte Carlo estimation.  At least provided that the 
algorithm is sufficiently well-tuned.  In this paper I show how the 
geometric foundations of Hamiltonian Monte Carlo implicitly identify 
the optimal choice of these parameters, especially the integration time.
I then consider the practical consequences of these principles in
both existing algorithms and a new implementation called
\textit{Exhaustive Hamiltonian Monte Carlo} before demonstrating
the utility of these ideas in some illustrative examples.
\end{abstract}

\end{frontmatter}

One of the most ubiquitous computational challenges in
statistics is the estimation of expectations of a function
with respect to a given target distribution, $\pi$.  For
example, we might need to compute expectations with
respect to a sampling distribution in a frequentist analysis
or expectations with respect to a posterior distribution
in a Bayesian analysis.  

Fueled by the proliferation of accessible computing resources
and its applicability to many different target distributions,
Markov chain Monte Carlo~\citep{RobertEtAl:1999, BrooksEtAl:2011}
has become one of the most popular strategies for estimating
these expectations.  Here a Markov chain generated by a
Markov kernel explores the target distribution, progressively
building up better and better expectation estimates.  Ensuring
that this strategy can be scaled up to the high-dimensional
and elaborate target distributions of applied interest, however, 
requires Markov kernels capable of efficiently exploring even 
the most complex distributions.

When the target distribution is smooth, Hamiltonian Monte 
Carlo~\citep{DuaneEtAl:1987, Neal:2011, BetancourtEtAl:2014} 
can be employed.  Here the target probabilistic system is mapped
into a Hamiltonian system whose canonical measure-preserving flow 
generates a powerful Markov transition.  The ultimate performance of 
the resulting Markov chain, however, depends crucially on for how
long we integrate along that flow: if we integrate for only a short
time then the chain devolves into diffusive exploration, but long
integration times offer only diminishing returns and potentially wasteful
computation.

In this paper I exploit the geometry inherent to Hamiltonian Monte Carlo 
to isolate the relationship between integration time and effective 
exploration.  After discussing how to implement various schemes
for choosing the integration time in practice, I use this relationship to 
construct a natural choice of integration times known as an \textit{exhaustion} 
and finally demonstrate the performance of the resulting \textit{exhaustive 
Hamiltonian Monte Carlo} algorithm with some illustrative examples.

\section{Hamiltonian Monte Carlo in Theory}

The key to optimizing implementations of Hamiltonian Monte Carlo lies 
in its geometric foundations.  In this section I survey the theoretical 
construction of the algorithm and then demonstrate how the latent 
microcanonical geometry naturally motivates optimality criteria.

\subsection{Constructing a Generic Hamiltonian Kernel}

In this paper I will consider the smooth probabilistic system 
$\left(Q, \mathcal{B} \! \left( Q \right), \pi \right)$, where the sample space, 
$Q$, is a positively-oriented and smooth $N$-dimensional manifold, 
$\mathcal{B} \! \left( Q \right)$ is the canonical Borel $\sigma$-algebra, 
and $\pi$ is a smooth probability distribution.  Our ultimate goal is to 
compute expectations of functions $f : Q \rightarrow \mathbb{R}$ with 
respect to $\pi$, which we'll approximate using Markov chain Monte 
Carlo estimators.  The resulting computational challenge is to develop 
a Markov kernel that efficiently explores the target distribution, $\pi$.

\begin{figure*}
\centering
\subfigure[]{
\begin{tikzpicture}[scale=0.18, thick]
 
\draw[color=white] (-8, 0) -- (8, 0);
 
\draw[] (0, 0) ellipse (5 and 1);
\node at (0, -13) { $Q = \mathbb{S}^{1} $ };
\node[] at (0,0) {\includegraphics[width=1.95cm]{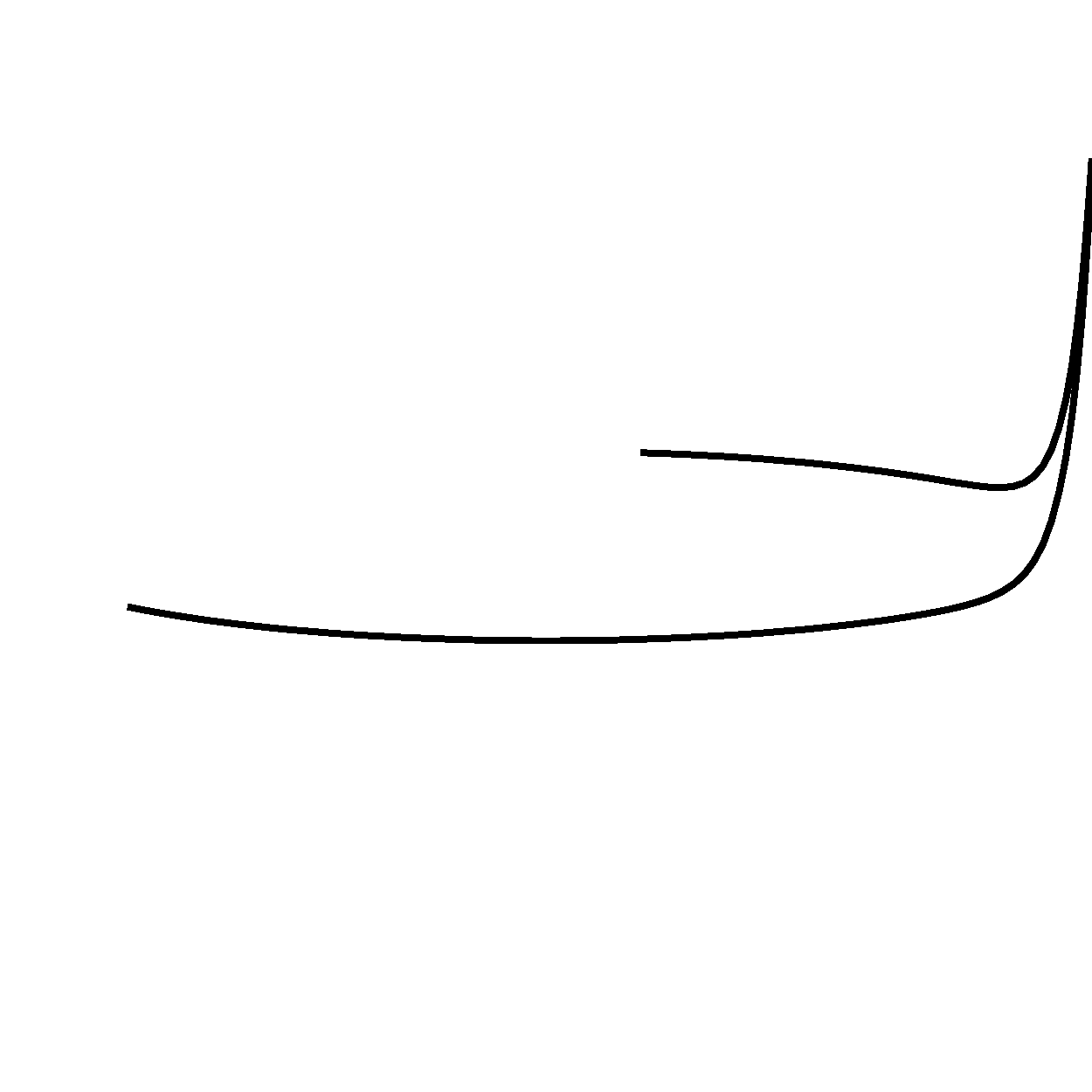}};
 
\end{tikzpicture}
}
\subfigure[]{
\begin{tikzpicture}[scale=0.18, thick]
  
\draw[color=white] (-8, 0) -- (8, 0);
  
\foreach \x in {0,20,...,180} {
   \draw[color=gray95, style=dashed] 
     ({5 * cos(\x + 10)}, {-10 + sin(\x)}) -- ({5 * cos(\x + 10)}, {-7 + sin(\x)});
   \draw[color=gray95] 
     ({5 * cos(\x + 10)}, {-7 + sin(\x)}) -- ({5 * cos(\x + 10)}, {7 + sin(\x)});
   \draw[color=gray95, style=dashed] 
     ({5 * cos(\x + 10)}, {7 + sin(\x)}) -- ({5 * cos(\x + 10)}, {10 + sin(\x)});
}

\draw[color=gray95] (-5, 0) arc (180:0:5 and 1);  
 
\node[] at (0,0) {\includegraphics[width=2.25cm]{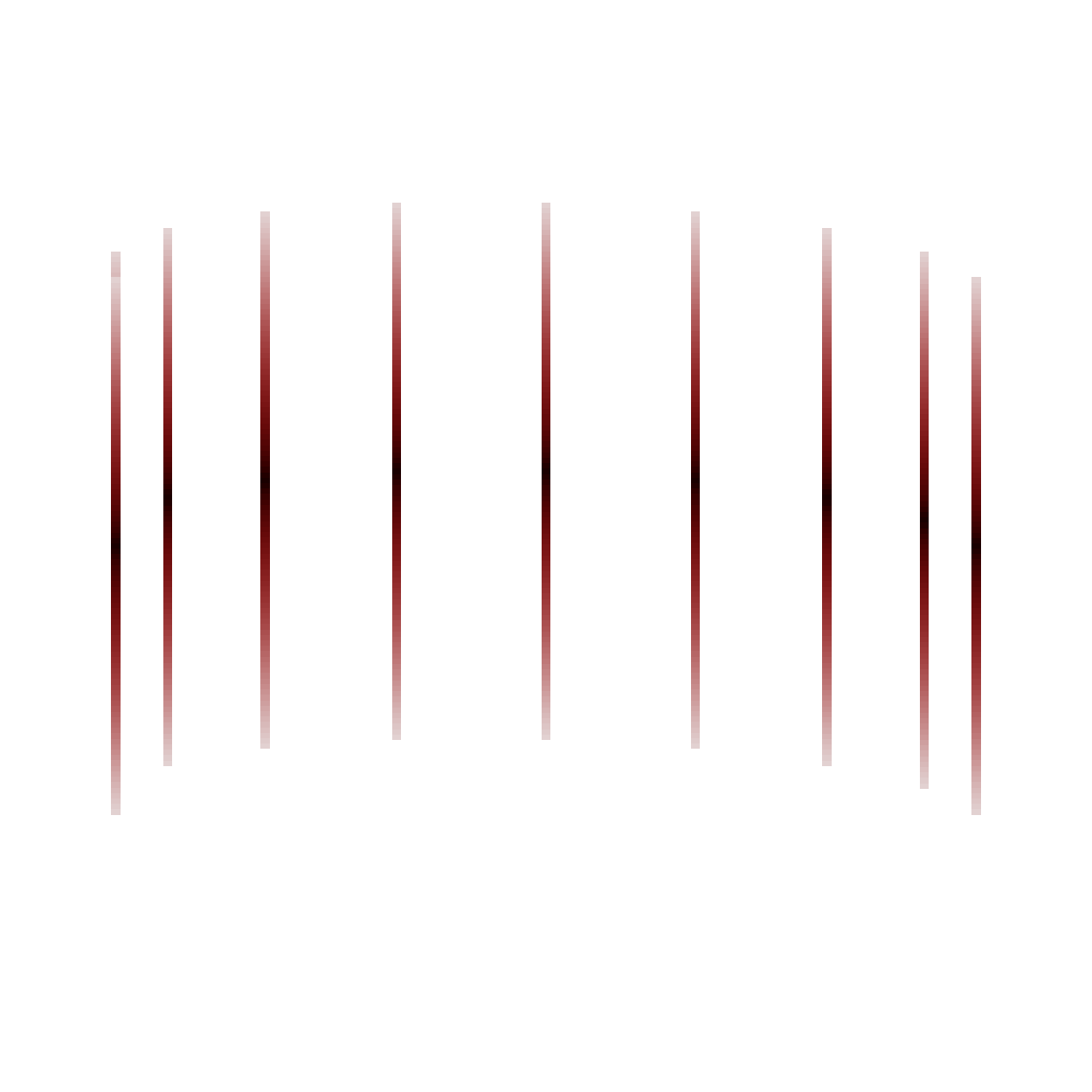}};
 
\draw[color=gray90] (-5, 0) arc (180:360:5 and 1);

\foreach \x in {0,20,...,180} {
   \draw[color=gray90, style=dashed] 
     ({5 * cos(\x)}, {-10 - sin(\x)}) -- ({5 * cos(\x)}, {-7 - sin(\x)});
   \draw[color=gray90] 
     ({5 * cos(\x)}, {-7 - sin(\x)}) -- ({5 * cos(\x)}, {7 - sin(\x)}); 
   \draw[color=gray90, style=dashed] 
     ({5 * cos(\x)}, {7 - sin(\x)}) -- ({5 * cos(\x)}, {10 - sin(\x)});
}
 
\node at (0, -13) { $T^{*}Q \approx \mathbb{S}^{1} \times \mathbb{R}$ };
 
\node[] at (0,0) {\includegraphics[width=2.25cm]{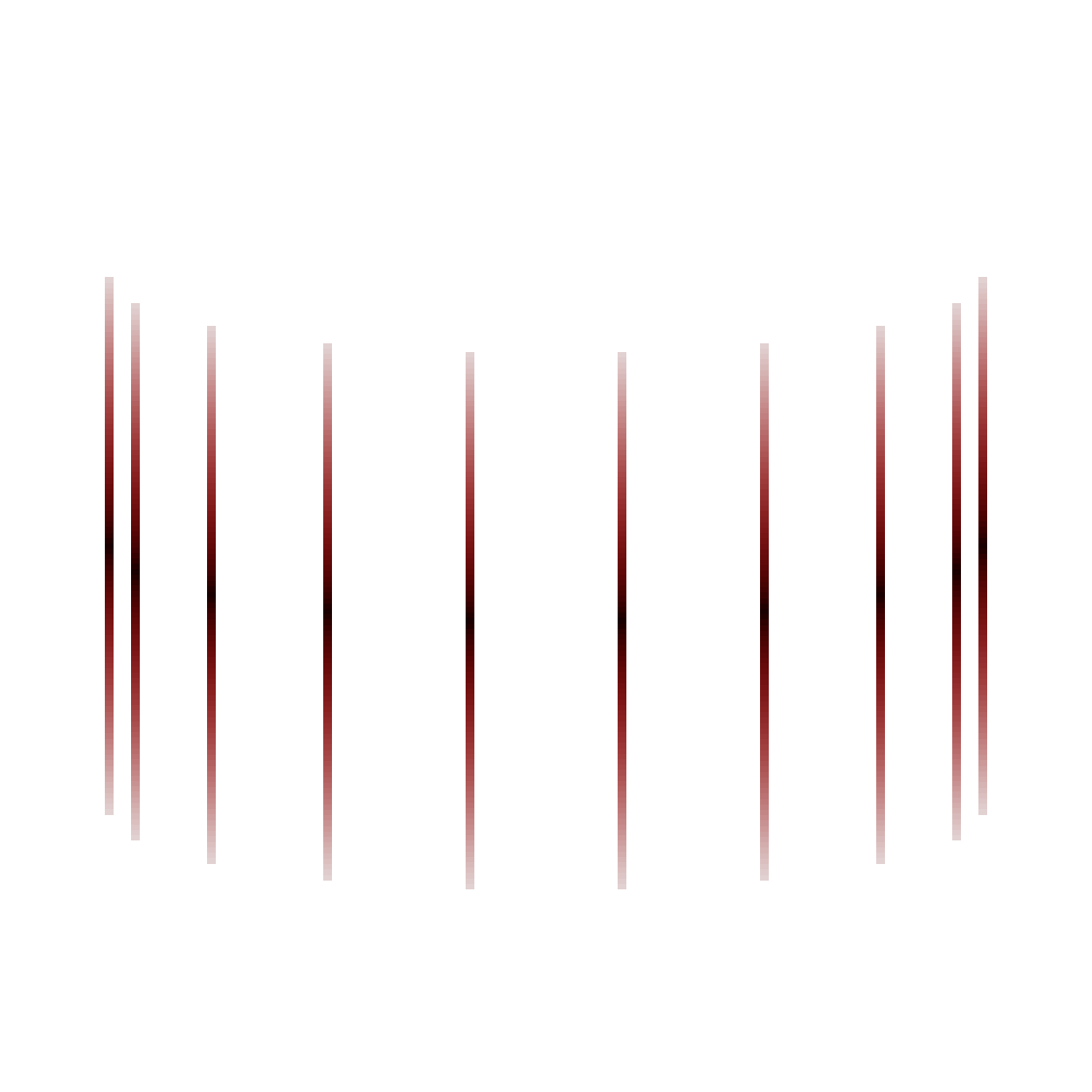}};
  
\end{tikzpicture}
}
\subfigure[]{
\begin{tikzpicture}[scale=0.18, thick]
  
\draw[color=white] (-8, 0) -- (8, 0);
  
\draw[] (-5, -8) -- (-5, 8);
\draw[] (5, -8) -- (5, 8);
\draw[style=dashed] (-5, -8) arc (180:360:5 and 1);
\draw[style=dashed] (0, 8) ellipse (5 and 1);
  
\node at (0, -13) { $T^{*}Q \approx \mathbb{S}^{1} \times \mathbb{R}$ };
 
\node[] at (0,0) {\includegraphics[width=1.78cm]{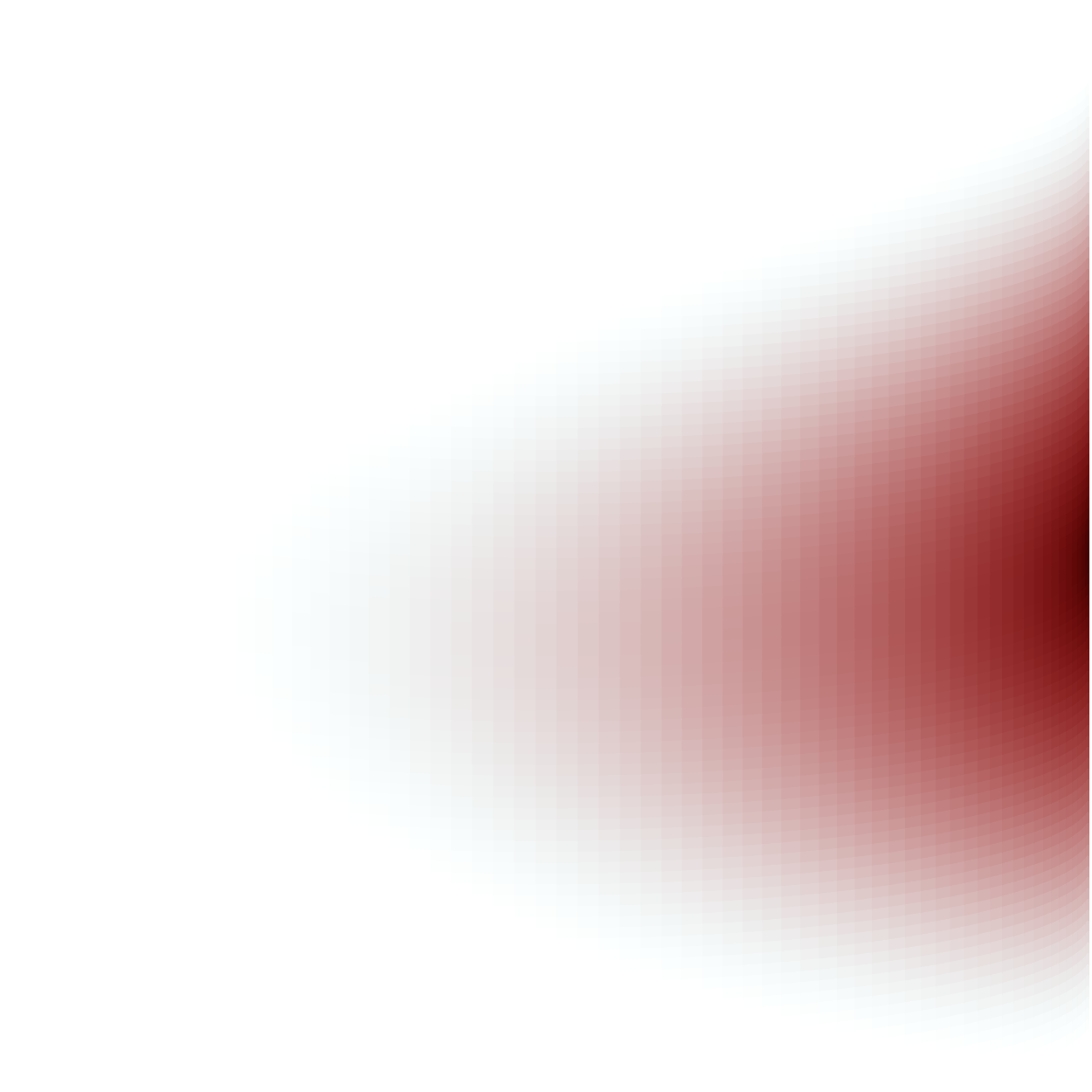}};
  
\end{tikzpicture}
}
\subfigure[]{
\begin{tikzpicture}[scale=0.18, thick]
  
\draw[color=white] (-8, 0) -- (8, 0);
  
\draw[] (-5, -8) -- (-5, 8);
\draw[] (5, -8) -- (5, 8);
\draw[style=dashed] (-5, -8) arc (180:360:5 and 1);
\draw[style=dashed] (0, 8) ellipse (5 and 1);
  
\node[] at (0,0) {\includegraphics[width=1.78cm]{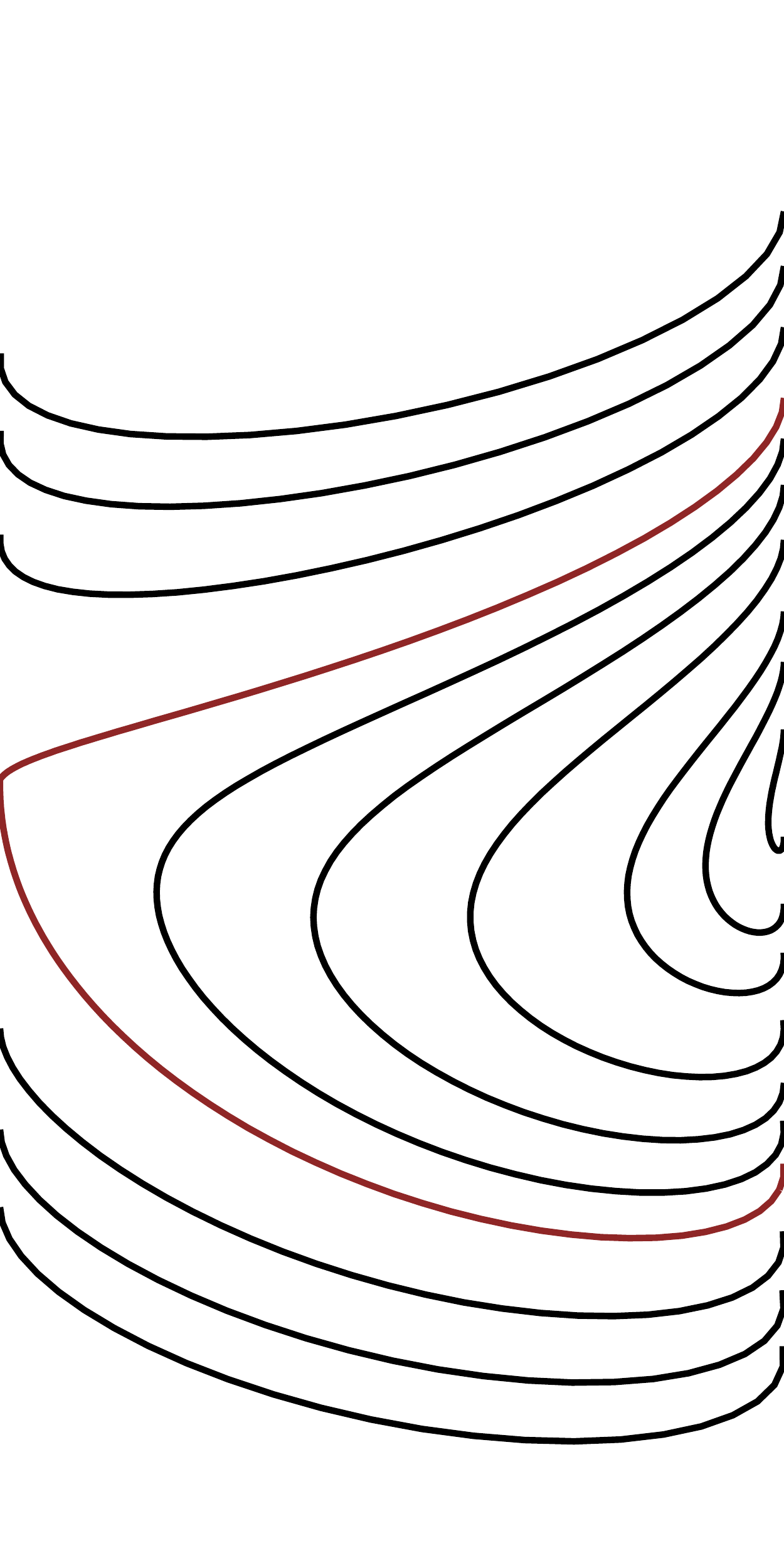}};
\fill[color=dark] (5, -0.33) circle (3pt);
  
\node at (0, -13) { $T^{*}Q \approx \mathbb{S}^{1} \times \mathbb{R}$ };
  
\end{tikzpicture}
}
\caption{Hamiltonian Monte Carlo maps a probabilistic system into a
Hamiltonian one.  Here, for example, (a) a smooth probability distribution
on the circle is lifted by (b) a disintegration on the cotangent fibers to 
define (c) a probability distribution on the cotangent bundle.  This joint 
distribution then canonically defines (d) a compatible Hamiltonian system.}
\label{fig:hmc_construction}
\end{figure*}
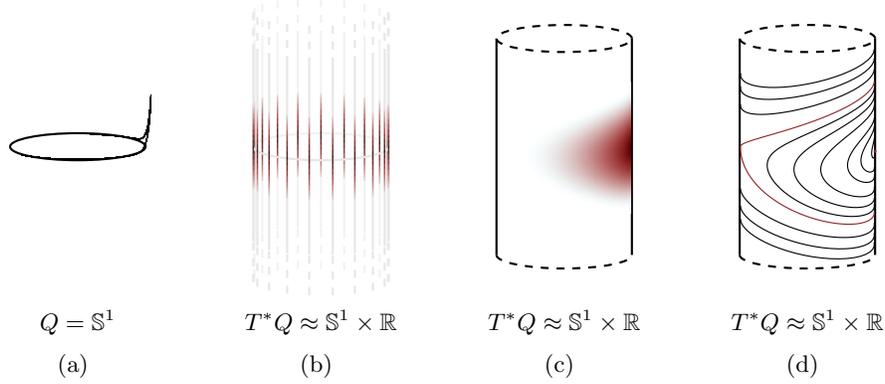

Hamiltonian Monte Carlo constructs such a kernel by mapping the
given probabilistic system into a Hamiltonian 
system~\citep{BetancourtEtAl:2014}.  Formally, any choice of a 
disintegration on the cotangent bundle, 
$\xi \in \Xi \! \left( \varpi : T^{*} Q \rightarrow Q \right)$,
immediately lifts the target distribution onto the cotangent bundle via
\begin{equation*}
\pi_{H} = \varpi^{*} \pi \wedge \xi.
\end{equation*}
Denoting $\theta$ the tautological one-form on the cotangent
bundle with $\Omega = \wedge_{n = 1}^{N} \dd \theta$ the
corresponding symplectic volume form, we can then define the
\textit{Hamiltonian},
\begin{equation*}
H = - \log \frac{ \dd \left( \varpi^{*} \pi \wedge \xi \right) }{ \dd \Omega },
\end{equation*}
and the corresponding Hamiltonian system,
$\left( T^{*} Q, \dd \theta, H \right)$ (Figure \ref{fig:hmc_construction}).
The critical feature of this construction is that the lifted target distribution 
is the canonical measure on the cotangent bundle,
\begin{equation*}
\pi_{H} = e^{-H} \Omega;
\end{equation*}
consequently the lifted distribution is preserved by the canonical Hamiltonian 
flow, which can then be used as a basis for a Markov kernel.

First, however, let's consider the local corollary of this 
construction. In a local neighborhood of the sample space,
$\mathcal{U}_{\alpha} \subset Q$, the target distribution 
decomposes as
\begin{equation*}
\pi = e^{-V} \dd q^{1} \wedge \ldots \wedge \dd q^{n},
\end{equation*}
where $V$ is known as the \textit{potential energy}.  Similarly,
in the corresponding neighborhood of the cotangent bundle, 
$\varpi^{-1} \! \left( \mathcal{U}_{\alpha} \right) \subset T^{*} Q$, 
the smooth disintegration, $\xi$ decomposes into
\begin{equation*}
\xi = e^{-K} \dd p_{1} \wedge \ldots \wedge \dd p_{n}
+ \text{horizontal} \; n\text{-forms},
\end{equation*}
with $K$ known as the \textit{kinetic energy}.  Locally the lift onto the 
cotangent bundle becomes
\begin{align*}
\pi_{H} 
&= \varpi^{*} \pi \wedge \xi
\\
&=
e^{- \left( V+ K \right) } 
\dd q^{1} \wedge \ldots \wedge \dd q^{n} \wedge 
\dd p_{1} \wedge \ldots \wedge \dd p_{n}
\\
&=
e^{-H} \Omega,
\end{align*}
with the Hamiltonian
\begin{equation*}
H = - \log \frac{ \dd \pi_{H} }{ \dd \Omega } = K + V,
\end{equation*}
taking a form familiar from classical mechanics~\citep{JoseEtAl:1998}.

We can now use the Hamiltonian flow of this engineered Hamiltonian 
system to construct a powerful Markov transition.  First we lift an initial
point from the sample space to the cotangent bundle by sampling from 
the corresponding cotangent fiber,
\begin{align*}
& p \sim \iota_{q}^{*} \xi
\\
l &: Q \rightarrow T^{*} Q 
\\
& \quad q \, \mapsto \left( q, p \right)
\\
& l_{*} \pi = \pi_{H}.
\end{align*}
We then apply the Hamiltonian flow for a random time depending
on the initial point,
\begin{align*}
t & \sim \pi_{T (q, p)}
\\
\phi^{H}_{t} &: T^{*} Q \rightarrow T^{*} Q
\\
& \left( \phi^{H}_{t} \right)_{*} \pi_{H} = \pi_{H},
\end{align*}
and finally project back down to the sample space,
\begin{align*}
\varpi : T^{*} Q & \rightarrow Q
\\
\varpi_{*} \pi_{H} &= \pi.
\end{align*}

Composing these steps together,
\begin{equation*}
g = \varpi \circ \phi^{H}_{t} \circ l,
\end{equation*}
yields a space of measure-preserving diffeomorphisms,
\begin{align*}
g \in & \, G 
\\
g : Q & \rightarrow Q
\\
g_{*} \pi &= \pi,
\end{align*}
with the corresponding semi-direct product measure, 
$\gamma_{q} =  \pi_{T (q, p)} \rtimes \iota^{*}_{q} \xi$,
that immediately defines a Hamiltonian kernel as an iterated random 
function~\citep{DiaconisEtAl:1999, Quas:1991}
\begin{equation*}
\T_{\mathrm{HMC}} \! \left( q, A\right) \equiv
\int_{G} \gamma_{q} \! \left( \dd g \right) 
\mathbb{I}_{A} \! \left( g \left( q \right) \right),
\end{equation*}
where $\mathbb{I}$ is the indicator function,
\begin{equation*}
\mathbb{I}_{A} \! \left( q \right) \propto
\left\{
\begin{array}{rr}
0, & q \notin A\\
1, & q \in A
\end{array} 
\right. , \, q \in Q, A \in \mathcal{B} \! \left( Q \right).
\end{equation*}

\subsection{Specifying an Optimal Hamiltonian Kernel from the 
Geometry of Microcanonical Systems}
\label{sec:microcanonical}

Unfortunately this construction is too general: every choice of 
cotangent disintegration, $\xi$, and distribution over integration times, 
$\pi_{T (q, p)}$, yields a different kernel, and the performance of these 
kernels can vary substantially when applied to a given target distribution.  
Consequently, a careful choice of kernel is critical to realizing the full 
potential of Hamiltonian Monte Carlo.

In this section I review how Hamiltonian systems naturally disintegrate
into microcanonical systems compatible with the Hamiltonian flow.  By 
analyzing the interaction of the Hamiltonian flow with this microcanonical 
geometry we can guide the construction of a unique kernel optimized to 
a given target distribution.

\subsubsection{The Microcanonical Disintegration}

\begin{figure*}
\centering
%
\begin{tikzpicture}[scale=0.23, thick]
  
  \draw[] (-5, -8) -- (-5, 8);
  \draw[] (5, -8) -- (5, 8);
  \draw[style=dashed] (-5, -8) arc (180:360:5 and 1);
  \draw[style=dashed] (0, 8) ellipse (5 and 1);
  
  \node[] at (0,0) {\includegraphics[width=2.3cm]
    {cylinder_hamiltonian.pdf}};
  \fill[color=dark] (5, -0.33) circle (3pt);
  
  \draw[->] (6, 0) -- +(4, 0);
  
\end{tikzpicture}
%
\begin{tikzpicture}[scale=0.23, thick]
  
  \draw[] (-5, 0) -- (-5, 8);
  \draw[] (5, 4.5) -- (5, 8);
  \draw[style=dashed] (0, 8) ellipse (5 and 1);
  
  \node[] at (0,0) {\includegraphics[width=2.3cm]
    {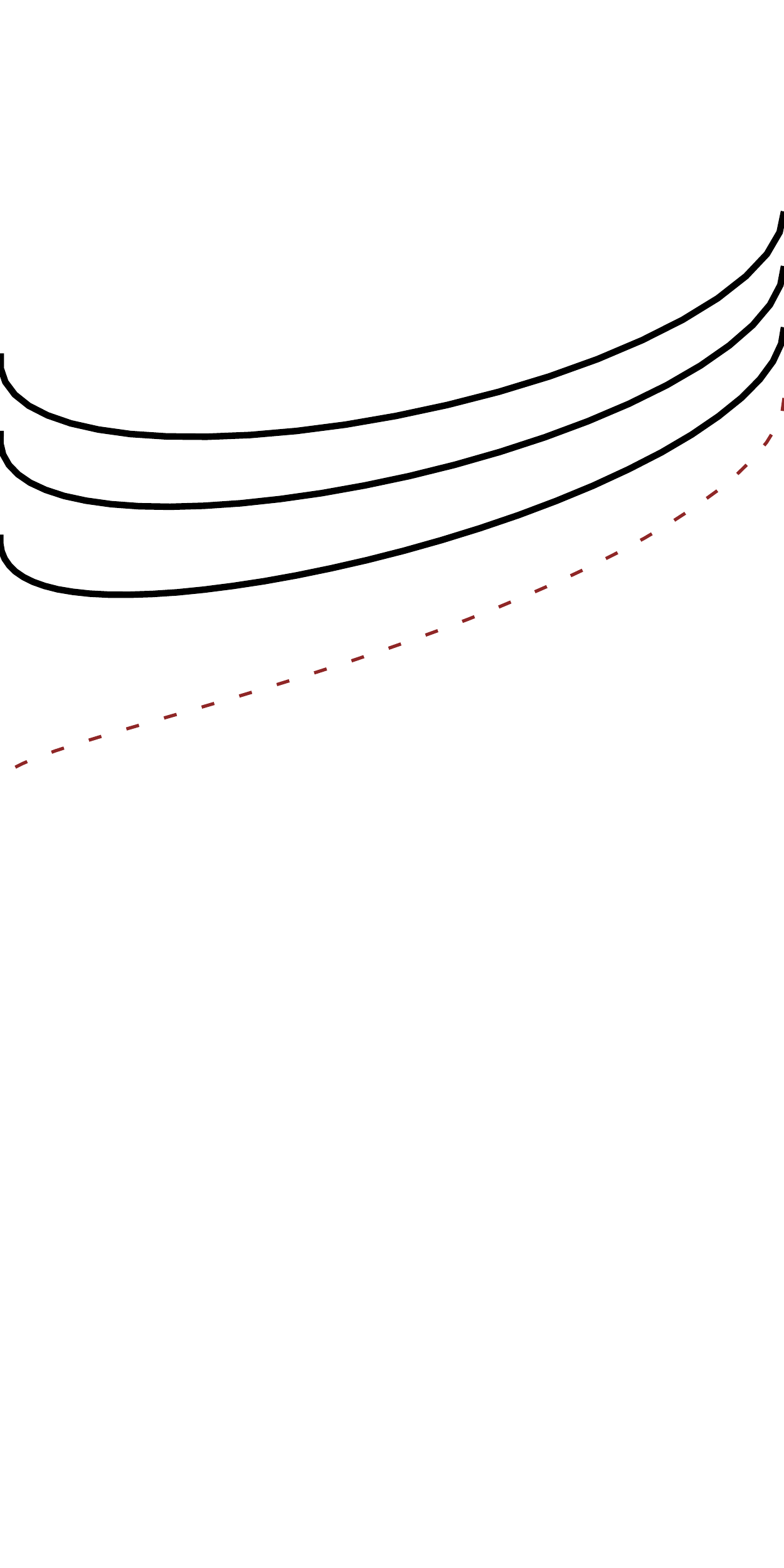}};
  
  \node[] at (7, 0) { $+$};
  
\end{tikzpicture}
%
\begin{tikzpicture}[scale=0.23, thick]
  
  \draw[] (5, -5) -- (5, 5);
  \node[] at (0,0) {\includegraphics[width=2.3cm]
    {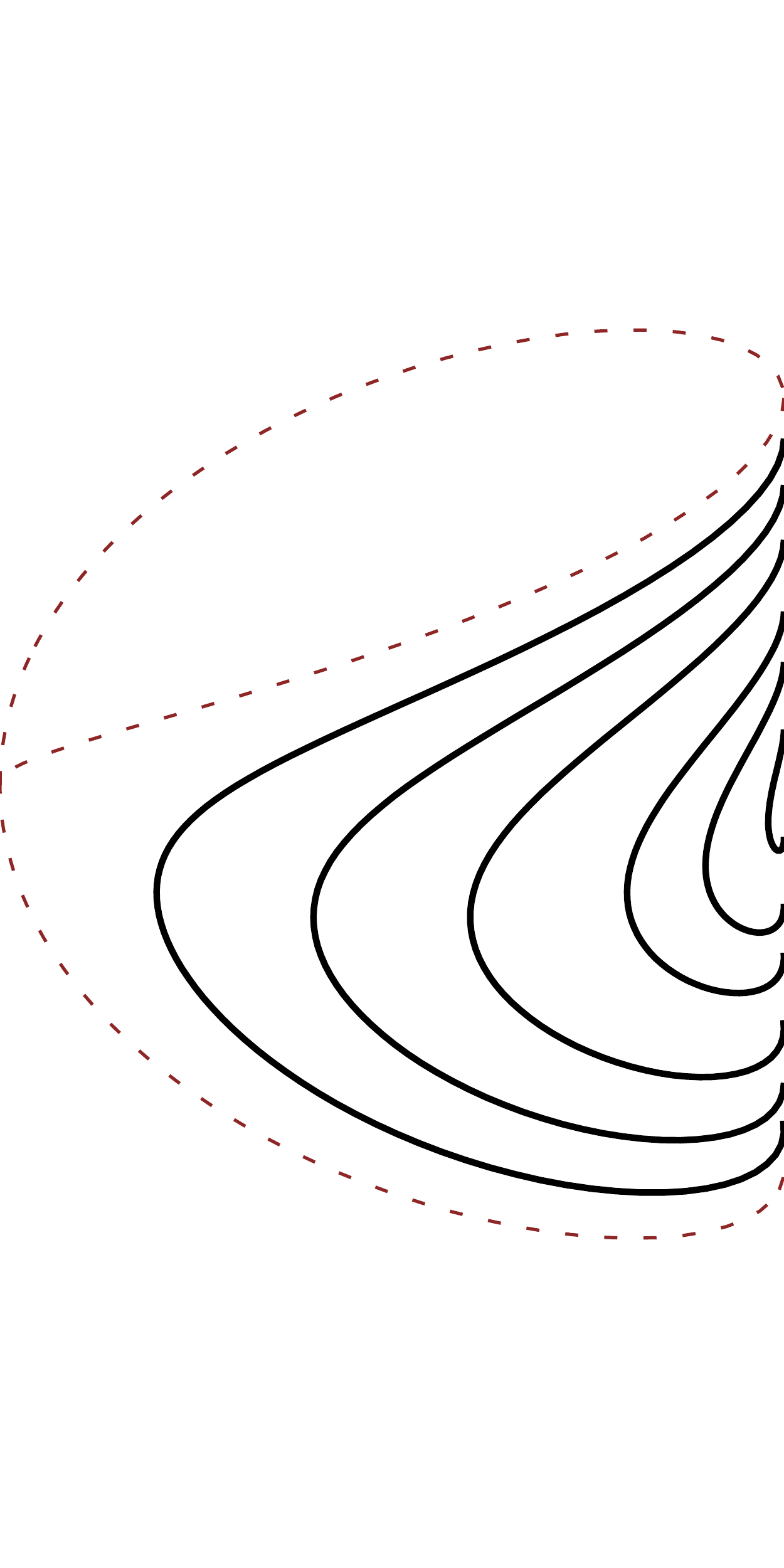}};
  
  \node[] at (7, 0) { $+$};
  
\end{tikzpicture}
%
\begin{tikzpicture}[scale=0.23, thick]
  
  \draw[] (-5, -8) -- (-5, -0.2);
  \draw[] (5, -8) -- (5, -5);
  \draw[style=dashed] (-5, -8) arc (180:360:5 and 1);
  
  \node[] at (0,0) {\includegraphics[width=2.3cm]
    {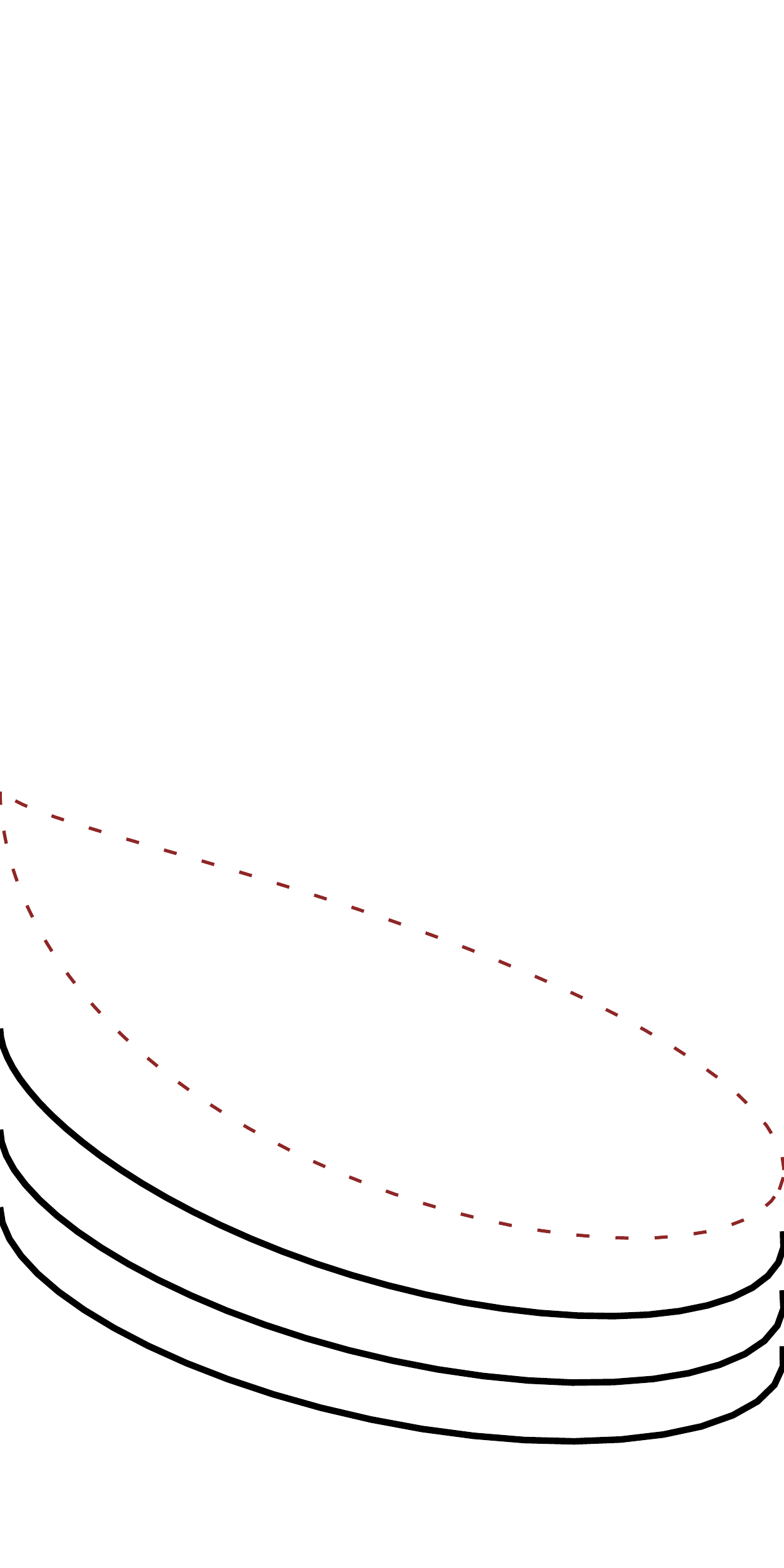}};

\end{tikzpicture}
\caption{The foliation of a Hamiltonian system into level sets naturally 
defines a fiber bundle on which we can disintegrate measures.  For 
example, once the critical level sets, here shown in red, are removed, 
a Hamiltonian system on the cylinder becomes a smooth fiber bundle 
with fiber space $F = \mathbb{S}^{1}$.  Correspondingly, the canonical 
distribution disintegrates into microcanonical distributions uniform on 
each circular fiber.}
\label{fig:cylinder_level_sets}
\end{figure*}

One of the special properties of Hamiltonian systems is that they
foliate into \textit{level sets}, or submanifolds of constant energy, $E$,
\begin{equation*}
H^{-1} \! \left( E \right) = \left\{ z \in T^{*} Q \mid H \! \left( z \right) = E \right\}.
\end{equation*}
These level sets can be \textit{regular}, in which case they contain only 
regular points of the Hamiltonian, or they can be \textit{critical}, in which 
case they contain at least one critical point of the Hamiltonian.  When the 
critical level sets are removed, the cotangent bundle decomposes into 
disconnected components, $T^{*} Q = \coprod_{i} M_{i}$, each of which 
foliates into level sets that are diffeomorphic to some common manifold
(Figure \ref{fig:cylinder_level_sets}).  Consequently each 
$H : M_{i} \rightarrow \mathbb{R}$ becomes a smooth fiber bundle with 
the level sets taking the role of the fibers.

The canonical distribution restricted to each of these components then 
disintegrates into \textit{microcanonical distributions} uniform on each 
level set,
\begin{equation*}
\pi_{H^{-1} (E) } = 
\frac{ \vec{v} \, \lrcorner \, \Omega }
{ \int_{H^{-1} (E) } \iota^{*}_{E} \left( \vec{v} \, \lrcorner \, \Omega \right) },
\end{equation*}
and a marginal \textit{energy distribution} given by
\begin{equation*}
\pi_{E} = H_{*} \pi = 
\frac{ e^{- E} }{ \int_{T^{*} Q} e^{-H} \Omega }
\frac{ \left( \int_{H^{-1} (E) } \iota_{E}^{*} 
\left( \vec{v} \, \lrcorner \, \Omega \right) \right)  }
{ \dd H \! \left( \vec{v} \right) } \dd E,
\end{equation*}
where $\vec{v}$ is any positively-oriented horizontal vector field satisfying 
$\dd H \! \left(\vec{v}\right) = c$ for some $0 < c < \infty$.  Because the 
critical level sets have zero measure with respect to the canonical 
distribution, the component disintegrations also define a valid 
disintegration of the entire cotangent bundle.  

The expectation of any smooth function $f : Q \rightarrow \mathbb{R}$, 
with respect to the target distribution, $\mathbb{E}_{\pi} \! \left[ f \right]$,
then decouples into expectations with respect to the microcanonical
distributions nested in an expectation with respect to energies,
\begin{align}
\mathbb{E}_{\pi} \! \left[ f \right]
&=
\mathbb{E}_{\pi_{H}} \! \left[ f \right]
\nonumber \\
&= \int_{T^{*} Q} f \, \pi_{H} 
\nonumber \\
&= \int_{T^{*} Q} f \pi_{E}  \wedge \pi_{ \sLS }
\nonumber \\
&= 
\int_{E} \pi_{E} \int_{\sLS} f \, \pi_{ \sLS }.
\label{expectation_decomp}
\end{align}

\begin{figure}
\centering
\includegraphics[width=4in]{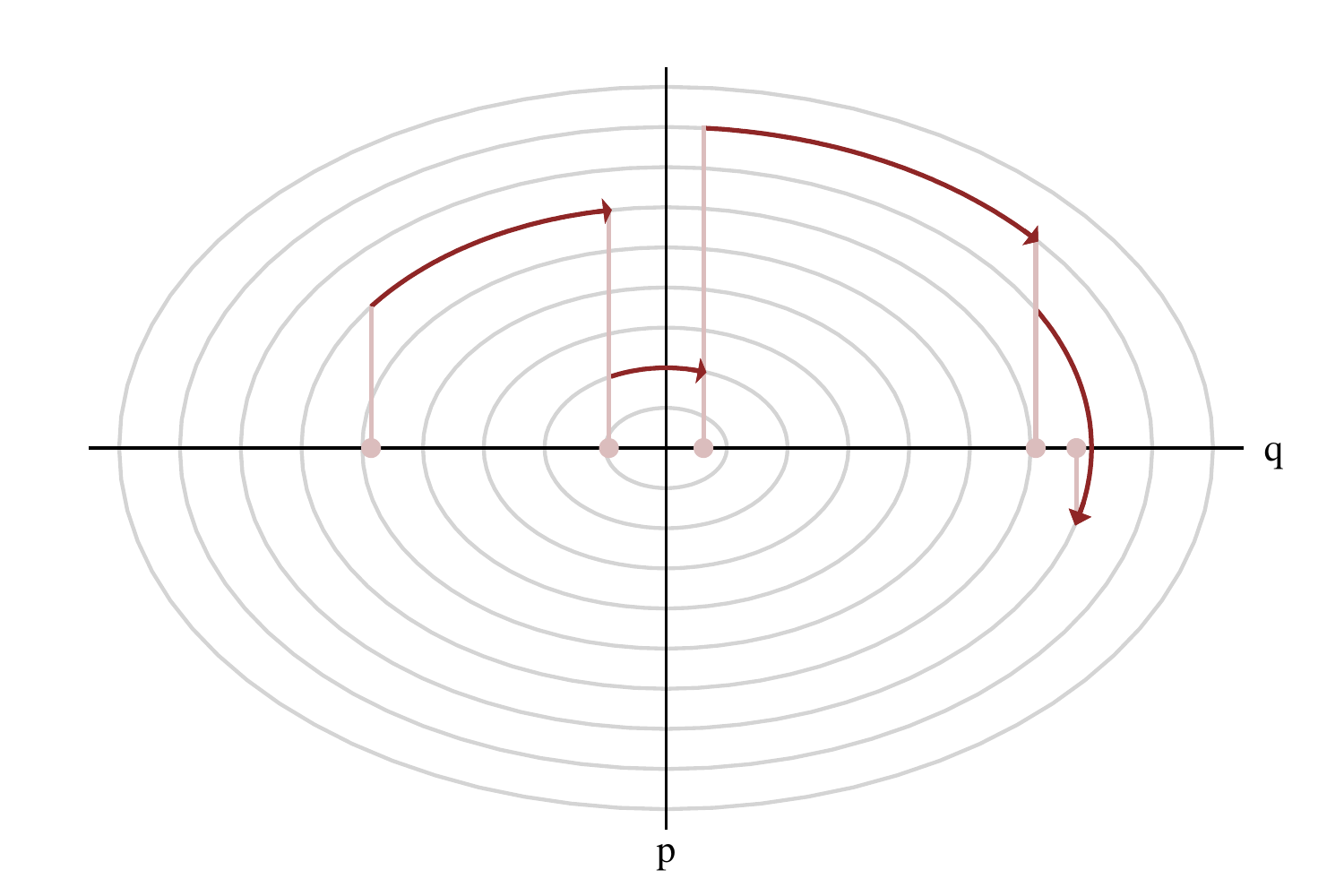}
\caption{Every Hamiltonian Markov chain alternatives between a deterministic
Hamiltonian flow that explores a single level set (dark red) and a momentum
resampling that transitions between level sets with a random walk (light red).  
The longer the flow is integrated the more efficiently the Markov chain can 
explore each level set and the smaller the autocorrelations will be.  When the 
flow is integrated for only an infinitesimally small time the Markov chain devolves 
into a Langevin diffusion.}
\label{fig:hmc_cartoon}
\end{figure}

Critically, the microcanonical disintegration is compatible with the Hamiltonian 
flow: every Hamiltonian trajectory is confined to a single level set and, because
the Hamiltonian flow restricted to a level set also preserves the corresponding 
microcanoncial distribution, these trajectories will explore the microcanonical 
distribution if integrated long enough.  Consequently a Hamiltonian Markov chain 
decouples into a deterministic flow along levels sets, with the projection and 
subsequent lift, $\lambda \circ \varpi : T^{*} Q \rightarrow T^{*} Q$, resampling 
the momentum and inducing a random walk between level sets (Figure 
\ref{fig:hmc_cartoon}).  The autocorrelation of the Markov chain then depends on 
both how effectively the Hamiltonian flow explores each microcanonical distribution 
and how effectively the momentum resampling explores the marginal energy 
distribution.  

\begin{figure}
\centering
\subfigure[]{ \includegraphics[width=2.5in]{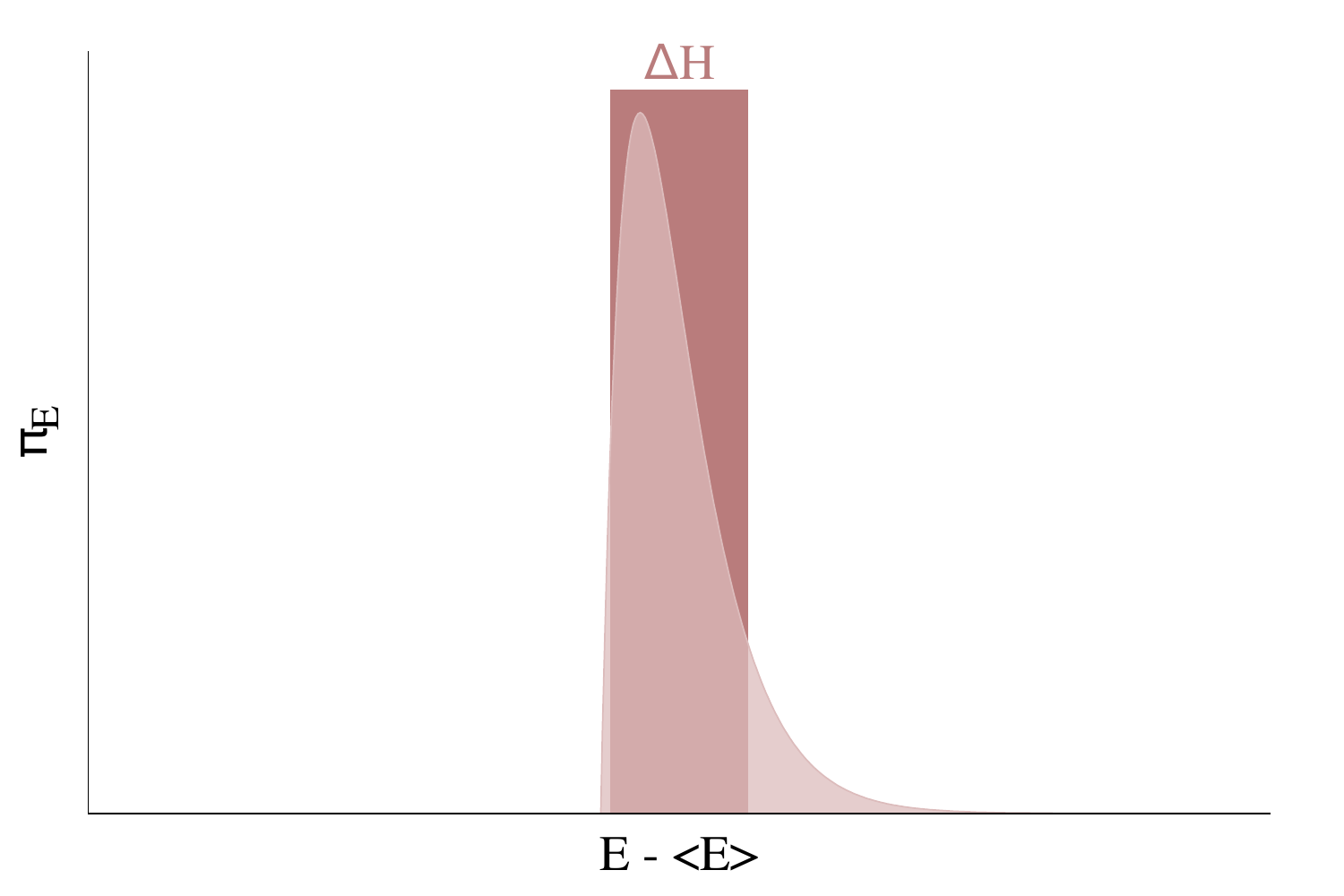} }
\subfigure[]{ \includegraphics[width=2.5in]{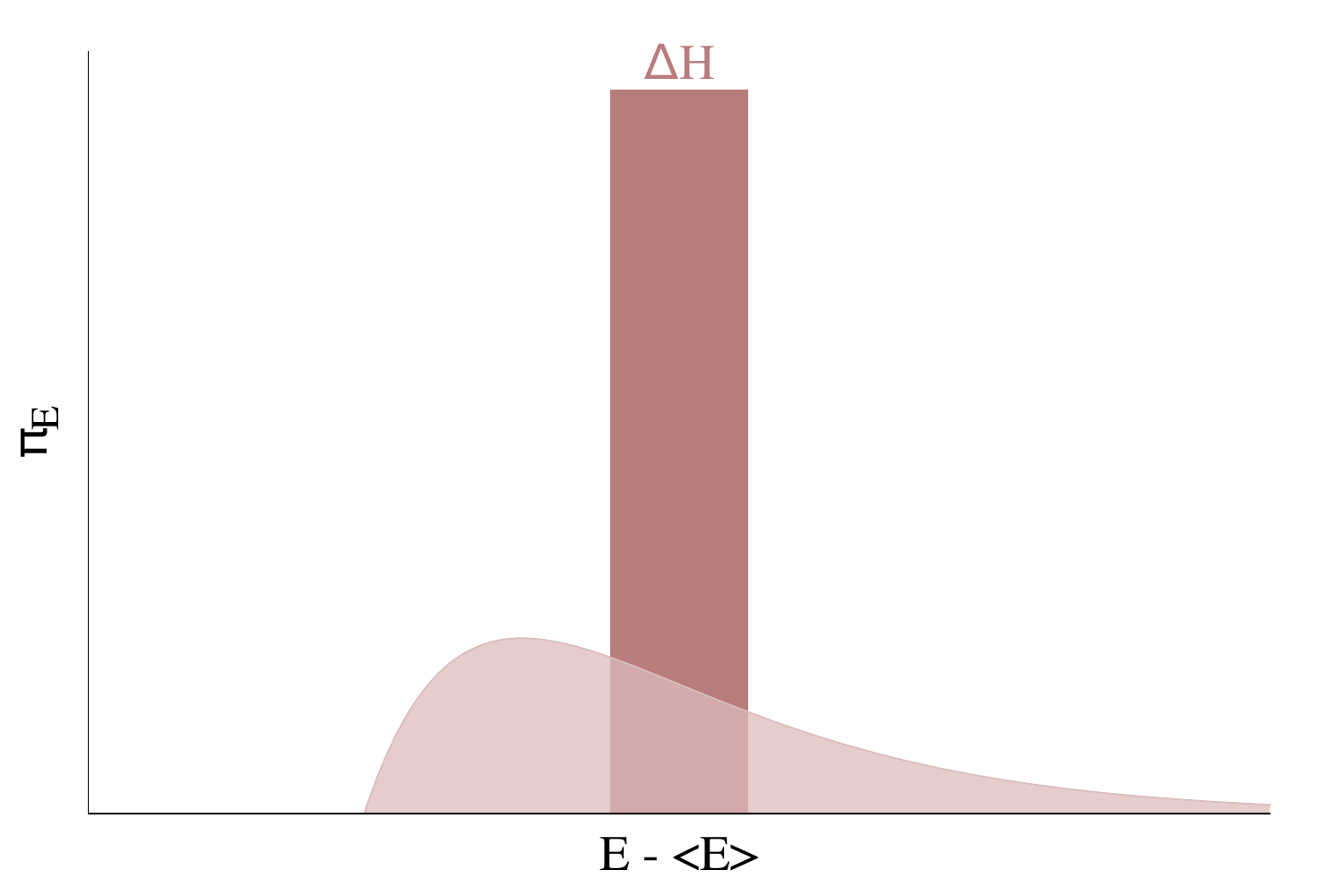} }
\caption{Every momentum resampling induces a change in the Hamiltonian,
which allows a Hamiltonian Markov chain to randomly walk amongst energy
level sets.  (a) When the expected variation, $\Delta H$, is similar to the width 
of the marginal energy distribution this random walk will rapidly explore this 
distribution, but (b) when the expected variation is small the exploration will
suffer from large autocorrelations.  Optimizing the exploration of the marginal
energy distribution provides an implicit criteria for selecting an optimal cotangent 
disintegration, and the energy autocorrelations define a constructive diagnostic 
for a poorly chosen cotangent disintegration.}
\label{fig:energy_marginals}
\end{figure}

The exploration of the energy distribution depends on how much the Hamiltonian
varies under a momentum resampling, $\Delta H$, relative to the width of the 
marginal energy distribution: the less the energy can vary in each transition the
fewer level sets can be reached and the larger the autocorrelations will be 
(Figure \ref{fig:energy_marginals}).  Because this ratio is fully determined by the 
interaction of the cotangent disintegration and the target distribution, it provides 
the foundations for the optimal choice of the cotangent disintegration itself.  
Formalizing this approach will be the subject of future work.

When the cotangent disintegration is well-chosen, the performance of the 
resulting Markov chain is then determined by how effectively the Hamiltonian 
flow explores each microcanonical distribution, which in turn depends
on for how long the flow is integrated along each level set.  If the flow
is integrated for only a short time then the transition will examine only
a small neighborhood of each level set and the Markov chain will suffer
from large autocorrelations.  As the integration time grows the flow more
completely explores each level set and reduces the autocorrelation of the
chain.  The additional exploration given by increasing the integration time, 
however, will eventually suffer from diminishing returns and ultimately not
be worth the additional cost.  Formalizing this intuition into a criterion for 
selecting an optimal compromise requires a more careful investigating of 
how Hamiltonian flow explores each microcanonical distributions.

\subsubsection{The Ergodicity of Hamiltonian Flow}
\label{sec:ergodicity}

The ideal circumstance for exploration is \textit{dynamical ergodicity}, 
where almost every trajectory eventually passes through almost every point 
on the corresponding level set, at least in the limit of an infinite integration time.  
Under these conditions the Birkhoff ergodic theorem~\citep{Petersen:1989} 
states that the temporal average of any function along a trajectory converges 
to the the spatial average with respect to the microcanonical distribution,
\begin{equation*}
\left< f \right>_{\phi^{H}} \! \left( z, T \right)
\equiv
\lim_{T \rightarrow \infty}
\frac{1}{T} \int_{0}^{T} \mathrm{d} t \,  f \circ \phi^{H}_{t} \! \left( z \right)
=
\mathbb{E}_{\pi_{\sLS}} \! \left[ f \right],
\end{equation*}
for $\pi_{\sLS}$ almost all initial $z \in \LS \subset T^{*} Q$.  In particular, a 
uniform sample from any trajectory will converge in distribution to a sample 
from the corresponding microcanonical distribution as the integration time 
grows, suggesting that we take $\pi_{T (z)} = U \! \left(0, T(z) \right)$ for
some appropriating chosen $T(z)$.

Unfortunately Hamiltonian systems are not always dynamically ergodic.
Depending on the topology of the level sets and the nonlinearity of the Hamiltonian, 
for example, trajectories may be confined to only subspaces within a level 
set~\citep{HoferEtAl:2011}, and identifying those systems that are dynamically 
ergodic is challenging if not outright infeasible.  The only guarantee that we have 
for a generic Hamiltonian system is that the time average along the flow 
converges to the spatial expectation along the domain of the trajectory,
$\pi_{\phi^{H} ( z )}$,
\begin{align*}
\lim_{T \rightarrow \infty}
\frac{1}{T} \int_{0}^{T} \mathrm{d} t \,  f \circ \phi^{H}_{t} \! \left( z \right)
&=
\mathbb{E}_{\pi_{\phi^{H} ( z )}} \! \left[ f \right]
\\
&=
\frac{ \int_{\phi^{H} ( z )} \pi_{H} f }{ \int_{\phi^{H} (z)} \pi_{H} }
\\
&=
\frac{ \int_{\phi^{H} ( z )} \pi_{H^{-1}(H(z))} f }
{ \int_{\phi^{H} (z)} \pi_{H^{-1}(H(z))} },
\end{align*}
where the domain, $\phi^{H} \! \left(z \right)$, is also known as
an \textit{orbit} of the Hamiltonian flow,
\begin{equation*}
\phi^{H} \! \left(z \right) 
= 
\left\{ \phi^{H}_{t} \! \left( z \right), \forall t \in \mathbb{R} \right\}
\subset H^{-1} \! \left( H \! \left( z \right) \right).
\end{equation*}
Although the trajectory may not explore the entire level set, it will at least 
explore the entire orbit, and a uniform sample from the trajectory will 
converge in distribution to a sample from $\pi_{\phi^{H} ( z )}$ as the 
integration time grows.

\begin{figure}
\centering
\subfigure[]{ \includegraphics[width=2.5in]{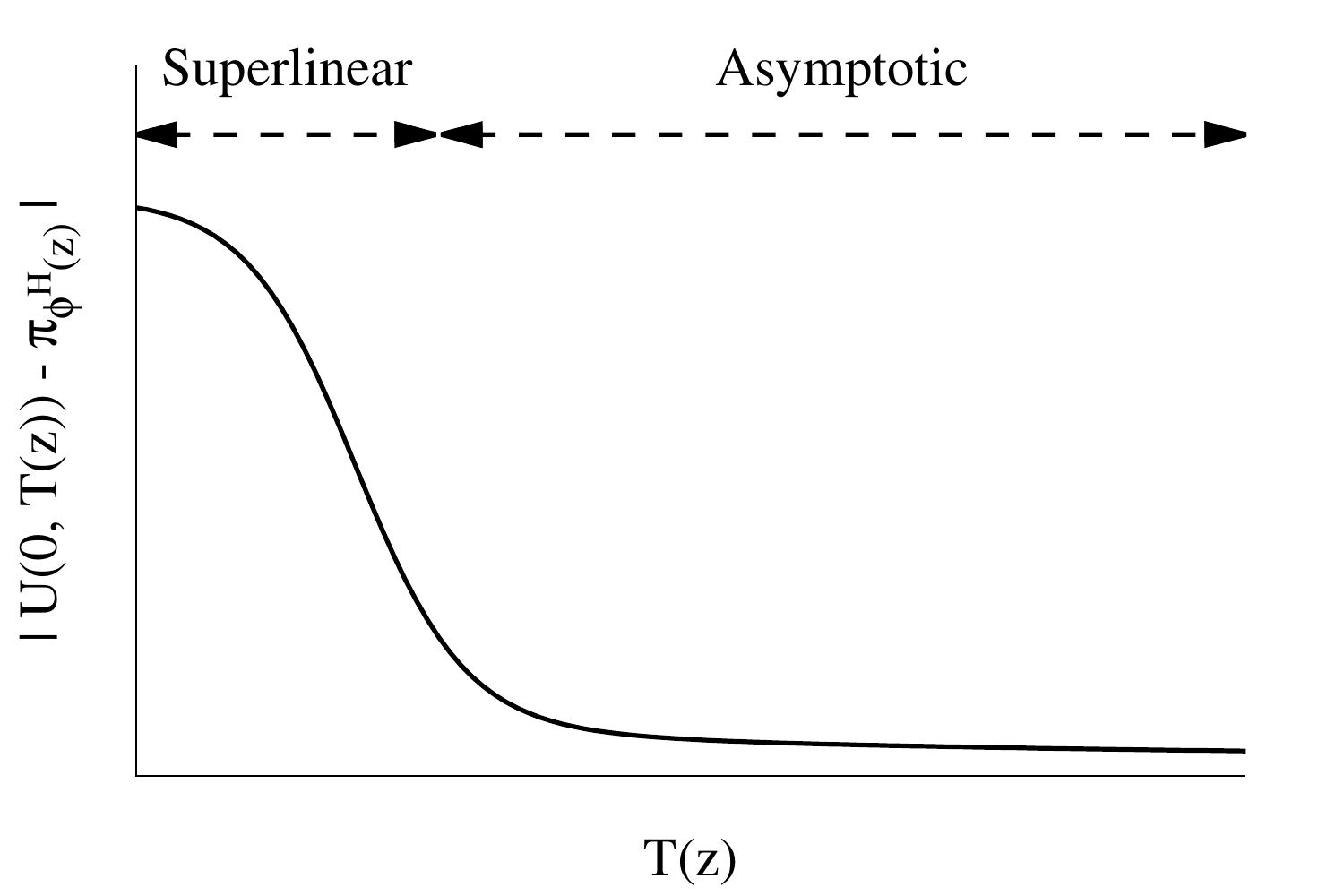} } \\
\subfigure[]{ \includegraphics[width=2.5in]{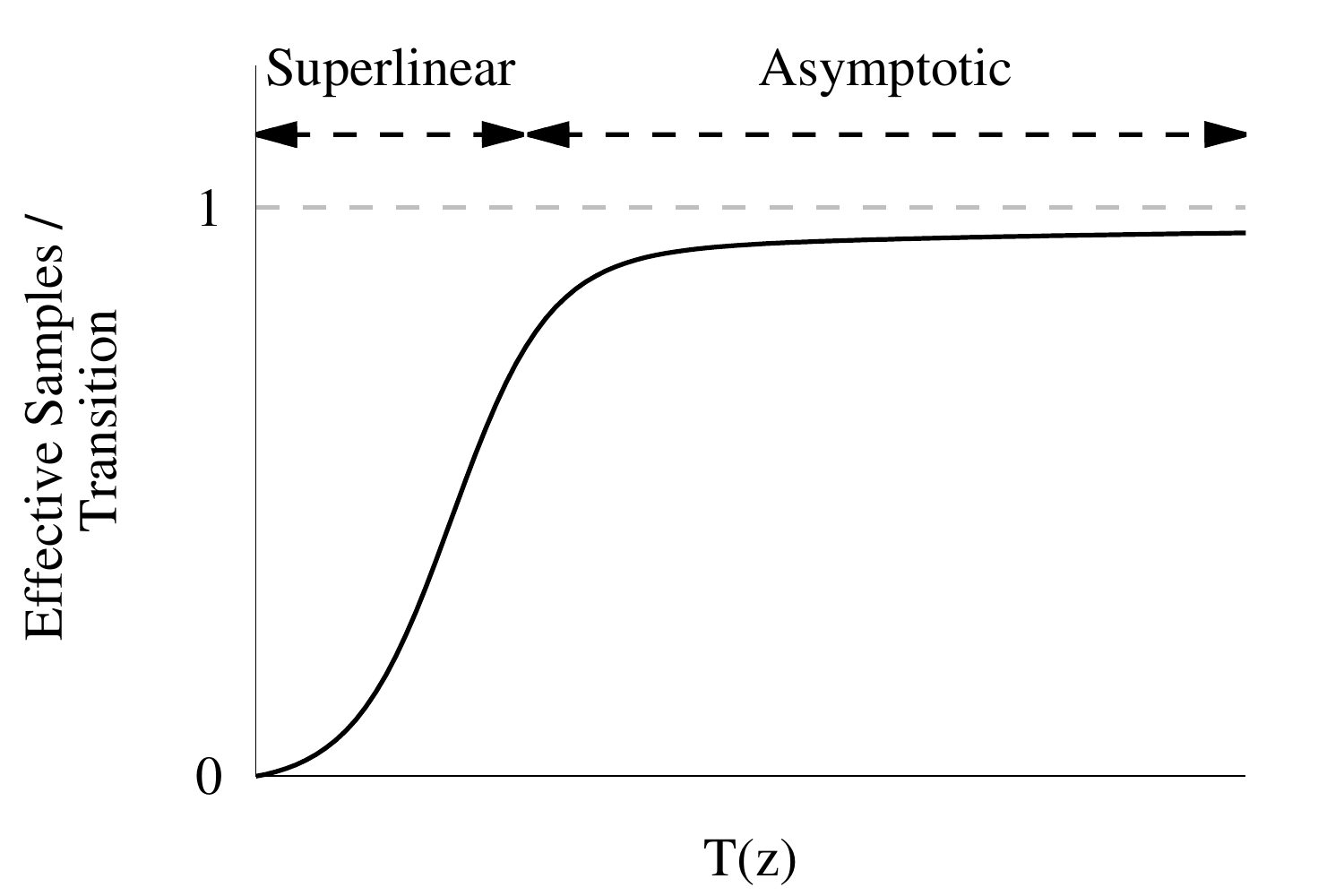} }
\subfigure[]{ \includegraphics[width=2.5in]{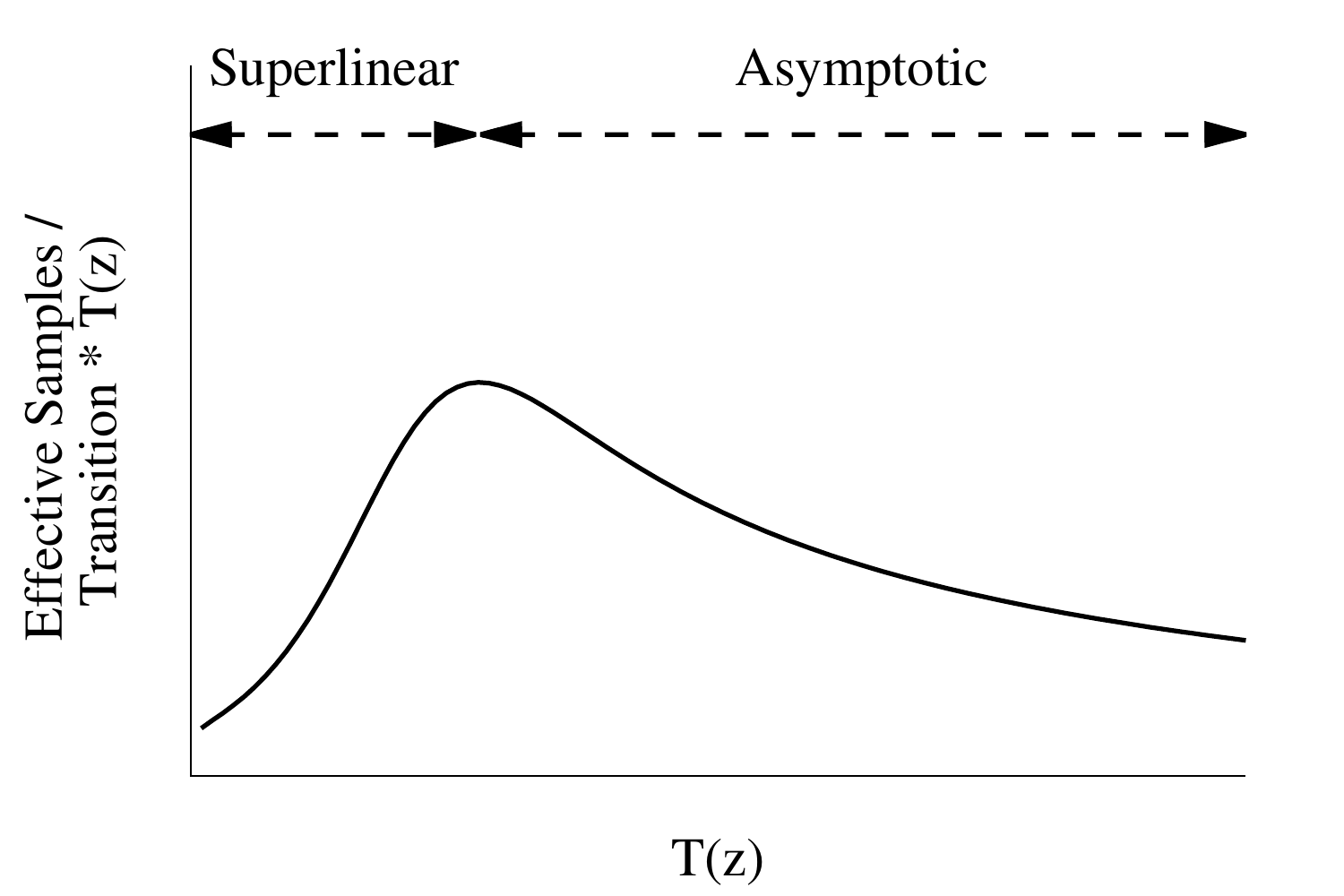} }
\caption{(a) Temporal averages along a Hamiltonian trajectory converge
to the corresponding spatial expectation, $\pi_{\phi^{H}( z )}$, as the
integration time grows, (b) inducing convergence of any Monte Carlo 
estimator, here represented by the number of effective samples per
transition.  Typically this convergence is initially rapid and superlinear 
before settling into an asymptotic regime where the convergence continues 
only with the square of the integration time.  (c) Because cost of simulating
each trajectory scales with the integration time, those integration times, 
$T \! \left( z \right)$, that identify the transition between these two regimes 
uniformly for all $z \in T^{*} Q$, yields optimal performance.}
\label{fig:microcanonical_convergence}
\end{figure}

Even though integrating for ever longer times will improve convergence,
yielding more accurate samples and reducing the autocorrelations of the 
resulting Hamiltonian Markov chain, longer integration times may not be
worth the additional cost.  When the integration time is small and the trajectory 
is just beginning to explore its orbit, for example, the convergence to the 
corresponding spatial expectation can be superlinear, justifying the linear 
cost of increasing the integration time.  For long integration times, however, 
the temporal expectations typically converge with only the square root of the 
integration time~\citep{CancesEtAl:2005}, and the cost of additional integration 
begins to undermine the performance of the chain 
(Figure \ref{fig:microcanonical_convergence}).  

\begin{figure}
\centering
\subfigure[]{ \includegraphics[width=2.5in]{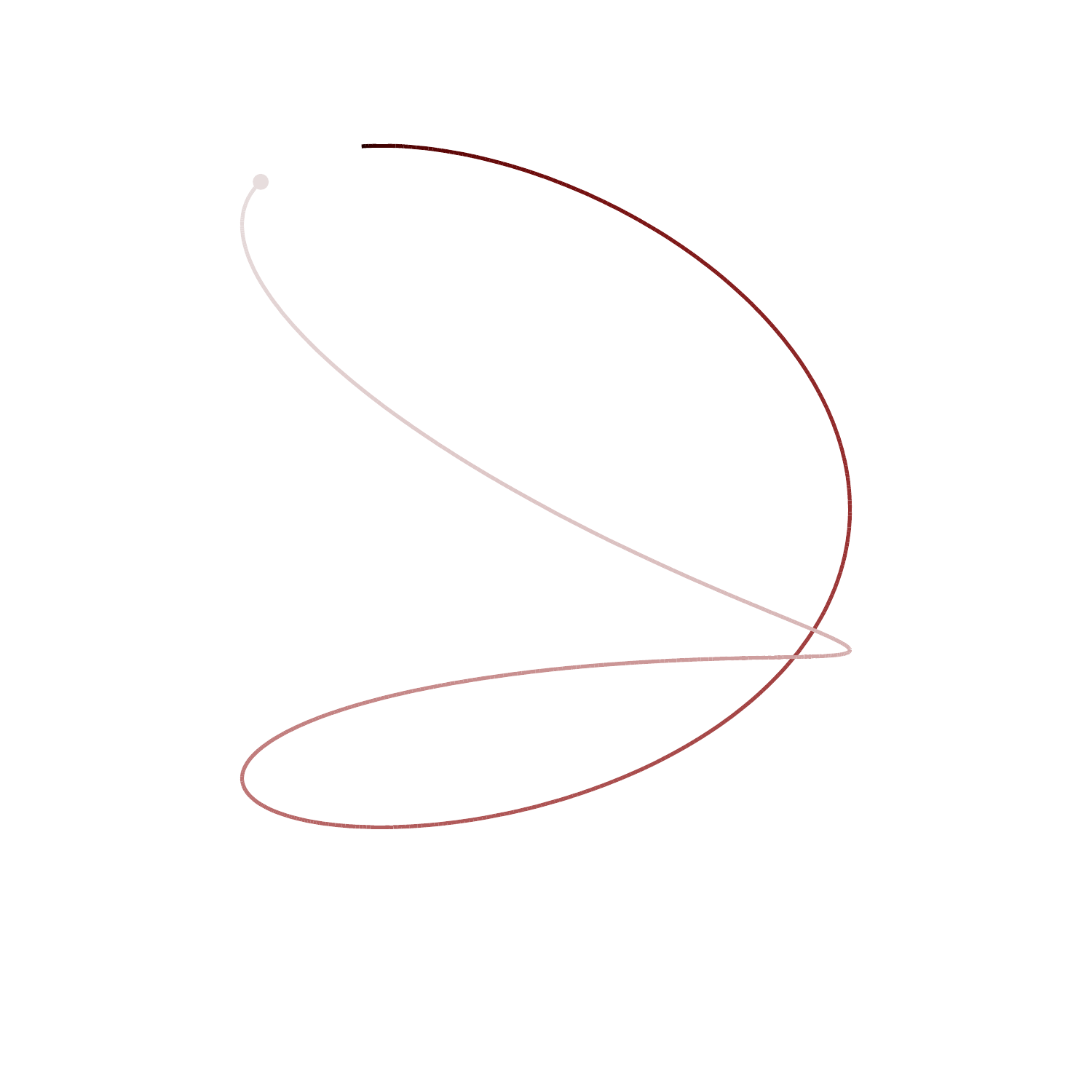} } 
\subfigure[]{ \includegraphics[width=2.5in]{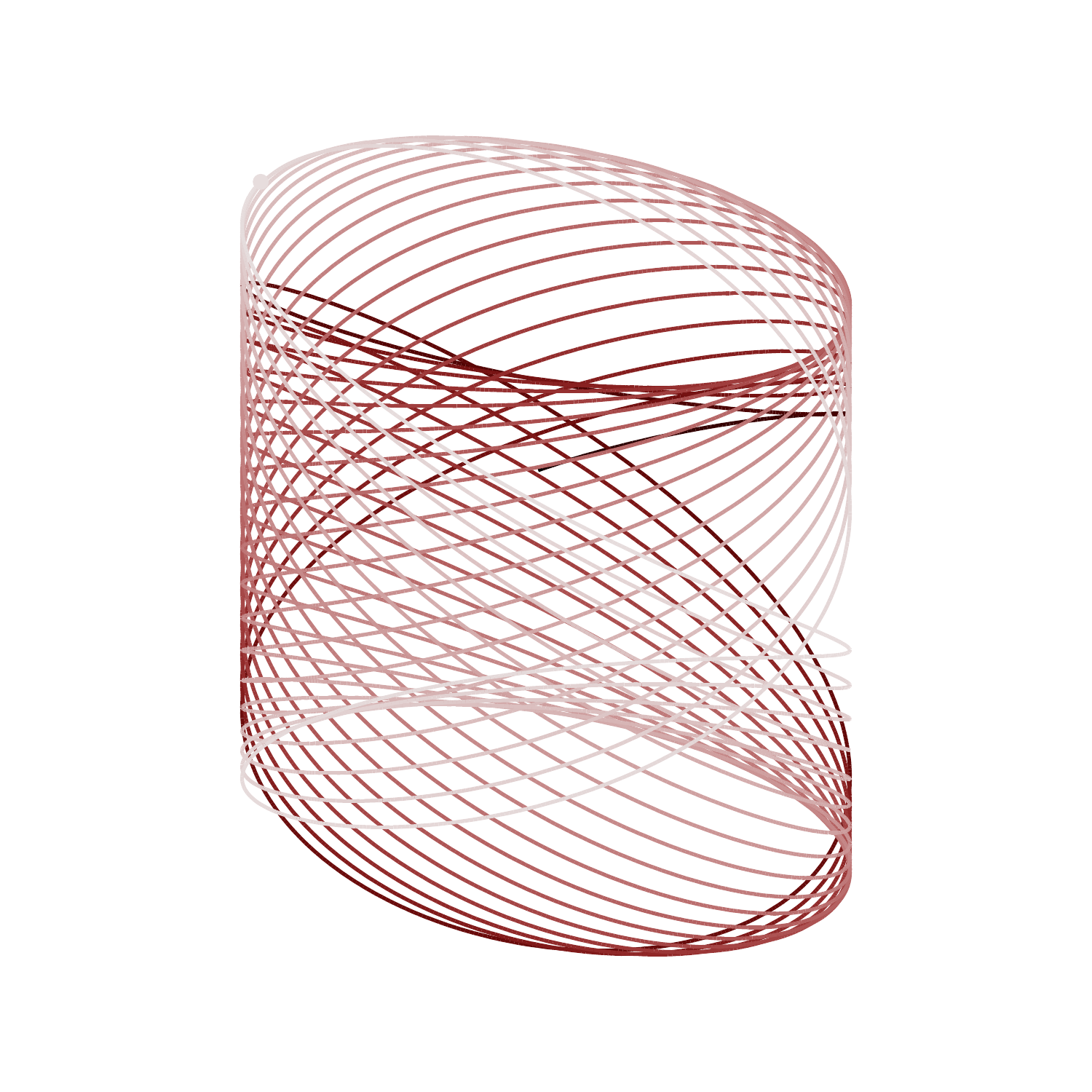} }
\caption{(a) Intuitively, the temporal average along a Hamiltonian trajectory 
rapidly converges to the spatial expectation over its orbit as the trajectory 
first spans the orbit.  (b) Longer trajectories simply refine this initial exploration, 
yielding better but slower convergence.}
\label{fig:convergence_cartoon}
\end{figure}

Consequently, optimal performance requires identifying a maximal 
integration time for each trajectory, $T \! \left( z \right)$, with the resulting 
uniform measure, $\pi_{T (z)} = U \! \left(0, T(z) \right)$, that identifies the 
transition between these two regimes uniformly across all level sets.  
Intuitively this transition should occur after the trajectory has 
first traversed the extent of its orbit (Figure \ref{fig:convergence_cartoon}a), 
with the asymptotic behavior corresponding to the trajectory exploring finer 
and finer details of the orbit (Figure \ref{fig:convergence_cartoon}b).  
Formalizing this intuition into an explicit optimization criterion, however, is 
not straightforward.

\subsubsection{Poincar\'{e} Recurrence and Autocorrelation Functions}

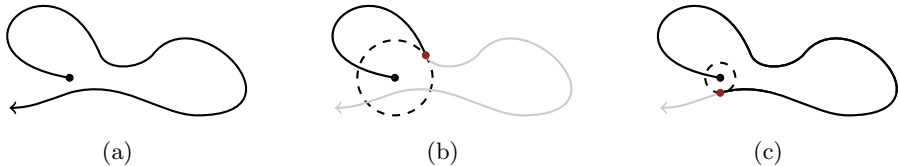
\begin{figure*}
\centering
\subfigure[]{
\begin{tikzpicture}[scale=0.1, thick]
  \fill[color=black] (-7, -3) circle (15pt);
 
  \begin{scope}
    \clip (-20, -10) rectangle (20, 10);
    \draw[->, color=black, thick] (-7, -3)
    .. controls (-25, 0) and (-9, 15) .. (-2.9, 0)
    .. controls (-2, -2) and (2, -2) .. (4, 0)
    .. controls (10, 8) and (25, -8) .. (10, -8)
    .. controls (5, -8) and (0, -3) .. (-7, -5)
    .. controls (-10, -6) and (-12, -7) .. (-15, -7);
  \end{scope}
\end{tikzpicture}
}
\subfigure[]{
\begin{tikzpicture}[scale=0.1, thick]
  \fill[color=black] (-7, -3) circle (15pt);
  \draw[style=dashed] (-7, -3) circle (5);
 
  \begin{scope}
    \clip (-20, -10) rectangle (20, 10);
    \draw[->, color=gray80, thick] (-7, -3)
    .. controls (-25, 0) and (-9, 15) .. (-2.9, 0)
    .. controls (-2, -2) and (2, -2) .. (4, 0)
    .. controls (10, 8) and (25, -8) .. (10, -8)
    .. controls (5, -8) and (0, -3) .. (-7, -5)
    .. controls (-10, -6) and (-12, -7) .. (-15, -7);
  \end{scope}
  
   \begin{scope}
    \clip (-20, -4) rectangle (-2.9, 7);
    \draw[->, color=black, thick] (-7, -3)
    .. controls (-25, 0) and (-9, 15) .. (-2.9, 0)
    .. controls (-2, -2) and (2, -2) .. (4, 0)
    .. controls (10, 8) and (25, -8) .. (10, -8)
    .. controls (5, -8) and (0, -3) .. (-7, -5)
    .. controls (-10, -6) and (-12, -7) .. (-15, -7);
  \end{scope}
  
  \fill[color=dark] (-2.9, 0) circle (15pt);
\end{tikzpicture}
}
\subfigure[]{
\begin{tikzpicture}[scale=0.1, thick]
  \fill[color=black] (-7, -3) circle (15pt);
  \draw[style=dashed] (-7, -3) circle (2); 
 
  \begin{scope}
    \clip (-20, -10) rectangle (20, 10);
    \draw[->, color=gray80, thick] (-7, -3)
    .. controls (-25, 0) and (-9, 15) .. (-2.9, 0)
    .. controls (-2, -2) and (2, -2) .. (4, 0)
    .. controls (10, 8) and (25, -8) .. (10, -8)
    .. controls (5, -8) and (0, -3) .. (-7, -5)
    .. controls (-10, -6) and (-12, -7) .. (-15, -7);
  \end{scope}
  
   \begin{scope}
    \clip (-20, -5) rectangle (17, 7);
    \draw[->, color=black, thick] (-7, -3)
    .. controls (-25, 0) and (-9, 15) .. (-2.9, 0)
    .. controls (-2, -2) and (2, -2) .. (4, 0)
    .. controls (10, 8) and (25, -8) .. (10, -8)
    .. controls (5, -8) and (0, -3) .. (-7, -5)
    .. controls (-10, -6) and (-12, -7) .. (-15, -7);
  \end{scope}
  
   \begin{scope}
    \clip (-7, -10) rectangle (17, 7);
    \draw[->, color=black, thick] (-7, -3)
    .. controls (-25, 0) and (-9, 15) .. (-2.9, 0)
    .. controls (-2, -2) and (2, -2) .. (4, 0)
    .. controls (10, 8) and (25, -8) .. (10, -8)
    .. controls (5, -8) and (0, -3) .. (-7, -5)
    .. controls (-10, -6) and (-12, -7) .. (-15, -7);
  \end{scope}
  
  \fill[color=dark] (-7, -5) circle (15pt);
\end{tikzpicture}
}
\caption{When its orbit is compact, (a) a Hamiltonian trajectory
(b) will return to any neighborhood of almost every initial point
within some finite recurrence time.  (c) The smaller the neighborhood,
the longer the recurrence time, and the more thoroughly the trajectory 
will explore its orbit and converge to the corresponding spatial
expectation.}
\label{fig:poincare_recurrence}
\end{figure*}

One natural strategy for identifying optimal integration times is to appeal 
to \textit{Poincar\'{e} recurrence}.  If the Hamiltonian is proper and its level 
sets compact~\citep{Lee:2011} then all Hamiltonian orbits will be bounded 
and the Poincar\'{e} recurrence theorem~\citep{Zaslavsky:2008} states that 
the trajectories originating from almost any point will explore the corresponding 
orbit and then return to any neighborhood of that point within some finite 
\textit{recurrence time} (Figure \ref{fig:poincare_recurrence}).  

The recurrence corresponding to well-behaved neighborhoods then 
immediately formalizes and implements our above intuitions.  At least
it would if we could define the necessary behavior and then explicitly
define the corresponding neighborhoods and identify recurrence
exactly.  Unfortunately none of these are particularly practical for
most Hamiltonian systems.

An alternative strategy that is both general and easily implemented, 
if less inspired, is to use an auxiliary \textit{autocorrelation function},
\begin{equation*}
\kappa \! \left(T, z \right) =
\kappa \! \left(\phi^{H}_{T} \! \left( z \right), z \right),
\end{equation*}
that monotonically converges to zero for all initial initial points,
\begin{equation*}
\lim_{T \rightarrow \infty} \left| \kappa \! \left(T, z \right) \right| = 0,
\, \forall z \in T^{*} Q.
\end{equation*}
Relaxing this to a uniform bound gives an \emph{termination criterion},
\begin{equation} \label{eqn:autocorr_bound}
\left| \kappa \! \left(T, z \right) \right| \leq \delta, \delta \in \mathbb{R}^{+},
\end{equation}
which implicitly defines a set of integration times,
\begin{equation*} \label{eqn:implicit_times}
T_{\kappa} \! \left( z \right) = 
\min \left\{ t \mid \left| \kappa \! \left(t, z \right) \right| \leq \delta \right\},
\end{equation*}
with $\delta$ providing some control over the amount of convergence.
If $\kappa \! \left(T, z\right)$ is not monotonic, for example if it oscillates 
around zero, then this interpretation becomes more complicated;  although 
this is not ideal, the resulting integration times may still provide some 
uniformity of exploration over each level set and hence identify useful
integration times.

Additionally, these two strategies are not mutually exclusive.  Because 
there always exists a compact neighborhood containing $z$ with 
$\phi^{H}_{T_{\kappa} ( z ) }$ on its boundary, when the level sets
are compact $T_{\kappa} ( z )$ can always be interpreted as a 
recurrence time for some implicit recurrence neighborhood.
Consequently, for some geometries Poincar\'{e} recurrence may 
be useful in motivating useful autocorrelation functions

\section{Hamiltonian Monte Carlo in Practice}
\label{sec:hmc_in_practice}

Regardless of how a Hamiltonian kernel is chosen, any implementation
of the underlying Hamiltonian flow requires solving a system of $2n$ 
first-order ordinary differential equations.  For all but the simplest systems 
analytical solutions are unfeasible and we must instead resort to simulating 
the flow numerically. Fortunately, there exist a family of numerical integrators 
that employ the underlying symplectic geometry to conserve many of the 
properties of the exact flow~\citep{HairerEtAl:2006, LeimkuhlerEtAl:2004}.  
These \textit{symplectic integrators} exactly preserve the symplectic volume 
form with only small variations in the Hamiltonian along the simulated flow.

In fact, symplectic integrators simulate some flow exactly, just not the flow 
corresponding to $H$.  Backwards error analysis shows that the discrete time 
steps of a $k$-th order symmetric symplectic integrator exactly fall onto the flow 
for some \textit{modified Hamiltonian}, given by an even, asymptotic expansion 
with respect to the integrator step size, $\epsilon$,
\begin{equation*}
\widetilde{H} = H 
+ \sum_{n=k / 2}^{N} \epsilon^{2 n} H_{\left( n \right) } 
+ \mathcal{O} \! \left( e^{- c / \epsilon } \right).
\end{equation*}
Because it is exponentially small in the step size, the asymptotic error is 
typically neglected and the leading-order behavior of the modified 
Hamiltonian is given by
\begin{equation*}
\widetilde{H} = H + \epsilon^{k} G + \mathcal{O} ( \epsilon^{k + 2} ).
\end{equation*}

This discretized, approximate flow then generates a series of states, 
\begin{equation*}
z_{L} \equiv \Phi^{\widetilde{H}}_{\epsilon, L \cdot \epsilon} ( z_{0} ) \in T^{*} Q,
\, L \in \mathbb{Z},
\end{equation*}
that tracks the true flow for exponentially long times.  The symplectic integrator 
will still introduce some error, however, and, while that error can be managed by 
the choice of step size, it will still bias the resulting Markov chain if left uncorrected.  
Correcting this error is a delicate problem that depends crucially on how the 
numerical trajectories are used, and hence the distribution of integration times,
$\pi_{T(z)}$.
 
\subsection{Static Implementations}
\label{sec:static_impl}

The simplest implementation of Hamiltonian Monte Carlo uses a single,
\emph{static} integration time, $T \! \left( z \right) = T$, or, equivalently,
a static number of symplectic integrator steps, $L = T / \epsilon$.

When using only the final point of each trajectory, in other words taking
a Dirac measure on integration times, $\pi_{T(z)} = \delta_{L \cdot \epsilon}$,
we might naively consider treating the numerical trajectory as a Metropolis
proposal, accepting the final state with only probability,
\begin{align*}
a \! \left( z_{0},  z_{L} \right)
&= 
\min \! \left[ 1, 
\frac{\dd \pi_{H} }{ \dd \Omega } \! \left( z_{L} \right)
\frac{\dd \Omega }{ \dd \pi_{H} } \! \left( z_{0} \right)
\right]
\\
&= 
\min \! \left[ 1, 
\exp \! \left( H \! \left( z_{L} \right) - H \! \left( z_{0} \right) \right) 
\right].
\end{align*}
Unfortunately the non-reversible nature of the flow renders it an invalid 
Metropolis proposal unless augmented. 

The numerical trajectory becomes a valid Metropolis proposal only when 
manipulated into an involution~\citep{Tierney:1998}, for example, by 
composing the flow with any operator, $R$, satisfying
\begin{equation*}
\Phi^{\tilde{H}}_{\epsilon, L \cdot \epsilon} 
\circ R \circ 
\Phi^{\tilde{H}}_{\epsilon, L \cdot \epsilon} 
= \mathrm{Id}_{T^{*}Q}.
\end{equation*}
The probability of accepting the final state is then given by
\begin{align*}
a \! \left( z_{0}, R \! \left( z_{L} \right) \right)
&= 
\min \! \left[ 1, 
\frac{\dd \pi_{H} }{ \dd \Omega } \! \left( R \! \left( z_{L} \right) \right)
\frac{\dd \Omega }{ \dd \pi_{H} } \! \left( z_{0} \right)
\right]
\\
&= 
\min \! \left[ 1, 
\exp \! \left( H \circ R \circ \! \left( z_{L} \right) - H \! \left( z_{0} \right) \right) 
\right].
\end{align*}

\begin{figure*}
\centering
\begin{tikzpicture}[scale=0.5, thick]

\draw[->, color=gray80] (0, 2) -- +(7,0);
\foreach \i in {0, 1, 2, 3} {
 \fill[color=dark] (2 * \i, 2) circle (4pt);
}

\fill[color=dark] (0, 2) circle (0pt)
node[below, color=black] { $z_{0}$ };

\fill[color=dark] (4, 2) circle (0pt)
node[below, color=black] { $z_{l}$ };

\fill[color=dark] (6, 2) circle (0pt)
node[below, color=black] { $z_{L}$ };

\fill[] (12, 2) circle (0pt)
node[right, color=black] { $\mathbb{P} \! \left[ z_{l} | z_{0} \right] > 0$ };

\draw[->, color=gray80] (0, 0) -- +(11,0);
\foreach \i in {0, 1} {
 \fill[color=gray80] (2 * \i, 0) circle (4pt);
}
\foreach \i in {2, 3, 4, 5} {
 \fill[color=dark] (2 * \i, 0) circle (4pt);
}

\fill[color=dark] (0, 0) circle (0pt)
node[below, color=black] { $z_{0}$ };

\fill[color=dark] (4, 0) circle (0pt)
node[below, color=black] { $z_{l}$ };

\fill[color=dark] (10, 0) circle (0pt)
node[below, color=black] { $z_{L + l}$ };

\fill[] (12, 0) circle (0pt)
node[right, color=black] { $\mathbb{P} \! \left[ z_{0} | z_{l} \right] = 0$ };

\end{tikzpicture}
\caption{If numerical trajectories are generated by integrating only forwards in time
then a state cannot be sampled from the entire trajectory without destroying
the invariance of the target distribution.  Sampling from trajectories while
maintaining the correct invariant distribution requires considering trajectories that 
integrate both forwards and backwards in time from the initial state, $z_{0}$.}
\label{fig:nonreversible_trajectories}
\end{figure*}
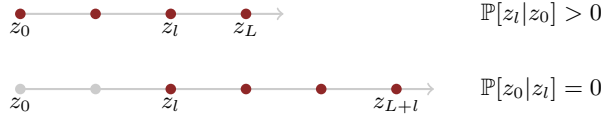

Our analysis of the microcanonical geometry, however, motivated not
a Dirac measure on integration times but rather sampling uniformly 
from the entire trajectory, $\pi_{T} = U \! \left(0, T \right)$.  Unfortunately, 
sampling from a numerical trajectory while also correcting for the error in
the symplectic integrator is a not straightforward given that Metropolis 
sampling from states generated by integrating only forwards in time 
breaks detailed balance (Figure \ref{fig:nonreversible_trajectories}),
\begin{align*}
\mathbb{P} \! \left[ z_{l} \mid z_{0} \right]
\frac{\dd \pi_{H} }{ \dd \Omega } \! \left( z_{0} \right) &= 
\frac{ 
\frac{\dd \pi_{H} }{ \dd \Omega } \! \left( z_{l} \right) 
\frac{\dd \pi_{H} }{ \dd \Omega } \! \left( z_{0} \right)
}
{ \sum_{m = 0}^{L} \frac{ \dd \pi_{H} }{ \dd \Omega } \! \left( z_{m} \right) }
\\
\mathbb{P} \! \left[ z_{0} \mid z_{l} \right] 
\frac{\dd \pi_{H} }{ \dd \Omega } \! \left( z_{l} \right)
&= 0,
\end{align*}
and obstructs the invariance of the target distribution.

In order to guarantee detailed balance and hence maintain the invariance of 
the target distribution, we need to consider not just those numerical trajectories 
that begin at the initial point but also those trajectories that only contain the initial point.  
Defining $\mathfrak{T}_{z, L}$ as the set of all numerical trajectories 
of length $L$ that contains the state $z$, we need to consider transitions that 
first sample a trajectory $\mathfrak{t} \in \mathfrak{T}_{z_{0}, L}$, with probability 
$\mathbb{P} \! \left[ \mathfrak{t} | z_{0} \right]$ and then sample a state from that
trajectory with probabilities $\mathbb{P} \! \left[ z | \mathfrak{t} \right]$.

Provided that the states within each trajectory are appropriately weighted
with the Metropolis probabilities,
\begin{equation} \label{eqn:state_probs}
\mathbb{P} \! \left[ z | \mathfrak{t} \right]
=
\frac{ \frac{\dd \pi_{H} }{ \dd \Omega } \! \left( z \right) }
{ \sum_{z' \in \mathfrak{t} } \frac{ \dd \pi_{H} }{ \dd \Omega } \! \left( z' \right) }
=
\frac{ e^{-H ( z ) } }
{ \sum_{z' \in \mathfrak{t}} e^{- H ( z' ) } },
\end{equation}
then the equality of trajectory probabilities,
\begin{equation} \label{eqn:trajectory_probs}
\mathbb{P} \! \left[ \mathfrak{t} | z_{1} \right]
= \mathbb{P} \! \left[ \mathfrak{t} | z_{2} \right],
\, \forall \mathfrak{t} \in \mathfrak{T}_{z_{1}, L} \cap \mathfrak{T}_{z_{2}, L}
\equiv \mathfrak{T}_{(z_{1}, z_{2}), L},
\end{equation}
is sufficient to ensure detailed balance,
\begin{align*}
\mathbb{P} \! \left[ z_{1} | z_{2} \right] 
\frac{\dd \pi_{H} }{ \dd \Omega } \! \left( z_{2} \right)
&=
\sum_{\mathfrak{t} \in \mathfrak{T}_{(z_{1}, z_{2}), L} }
\mathbb{P} \! \left[ z_{1} | \mathfrak{t} \right] 
\mathbb{P} \! \left[ \mathfrak{t} | z_{2} \right] 
\frac{\dd \pi_{H} }{ \dd \Omega } \! \left( z_{2} \right)
\\
&=
\sum_{\mathfrak{t} \in \mathfrak{T}_{(z_{1}, z_{2}), L} }
\frac{ \frac{\dd \pi_{H} }{ \dd \Omega } \! \left( z_{1} \right) }
{ \sum_{z' \in \mathfrak{t} } \frac{ \dd \pi_{H} }{ \dd \Omega } \! \left( z' \right) }
\mathbb{P} \! \left[ \mathfrak{t} | z_{2} \right] 
\frac{\dd \pi_{H} }{ \dd \Omega } \! \left( z_{2} \right)
\\
&=
\left(
\sum_{\mathfrak{t} \in \mathfrak{T}_{(z_{1}, z_{2}), L} }
\frac{ \frac{\dd \pi_{H} }{ \dd \Omega } \! \left( z_{2} \right) }
{ \sum_{z' \in \mathfrak{t} } \frac{ \dd \pi_{H} }{ \dd \Omega } \! \left( z' \right) }
\mathbb{P} \! \left[ \mathfrak{t} | z_{2} \right] 
\right)
\frac{\dd \pi_{H} }{ \dd \Omega } \! \left( z_{1} \right)
\\
&=
\left(
\sum_{\mathfrak{t} \in \mathfrak{T}_{(z_{1}, z_{2}), L} }
\frac{ \frac{\dd \pi_{H} }{ \dd \Omega } \! \left( z_{2} \right) }
{ \sum_{z' \in \mathfrak{t} } \frac{ \dd \pi_{H} }{ \dd \Omega } \! \left( z' \right) }
\mathbb{P} \! \left[ \mathfrak{t} | z_{1} \right] 
\right) \frac{\dd \pi_{H} }{ \dd \Omega } \! \left( z_{1} \right)
\\
&=
\left(
\sum_{\mathfrak{t} \in \mathfrak{T}_{(z_{1}, z_{2}), L} }
\mathbb{P} \! \left[ z_{2} | \mathfrak{t} \right] 
\mathbb{P} \! \left[ \mathfrak{t} | z_{1} \right] 
\right) \frac{\dd \pi_{H} }{ \dd \Omega } \! \left( z_{1} \right)
\\
&=
\mathbb{P} \! \left[ z_{2} | z_{1} \right]
\frac{\dd \pi_{H} }{ \dd \Omega } \! \left( z_{1} \right).
\end{align*}

There are various methods for appropriately weighting the states in a 
numerical trajectory according to \eqref{eqn:state_probs}.  For example,
we could simply sample from the multinomial distribution defined by the 
Metropolis probabilities directly, or even apply a slice sampler that first samples  
$u \sim U \! \left( 0, 1 \right)$ and then uniformly samples from those points 
on the trajectory satisfying
\begin{equation*}
\frac{ e^{-H ( z ) } }
{ \sum_{z' \in \mathfrak{t}} e^{- H ( z' ) } }
> u.
\end{equation*}
Designing a transition from an initial state to a numerical trajectory satisfying 
\eqref{eqn:trajectory_probs} is a more subtle challenge.  

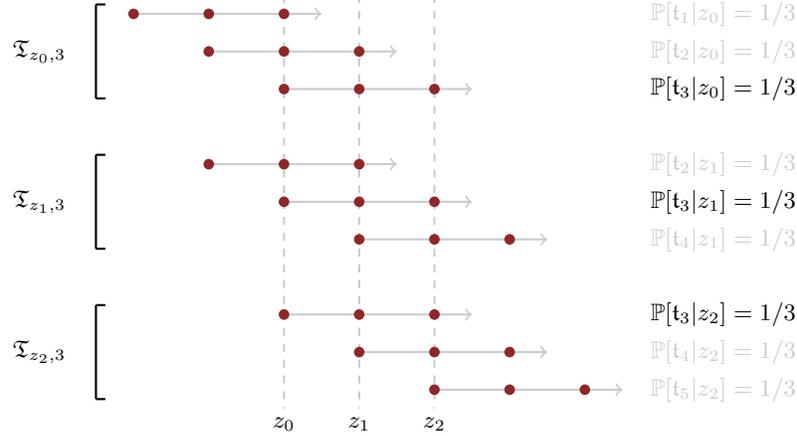
\begin{figure*}
\centering
\begin{tikzpicture}[scale=0.5, thick]

\draw[color=gray80, dashed] (8, 11.5) -- +(0, -11)
node[below, color=black] { $z_{2}$ };

\draw[color=gray80, dashed] (6, 11.5) -- +(0, -11)
node[below, color=black] { $z_{1}$ };

\draw[color=gray80, dashed] (4, 11.5) -- +(0, -11)
node[below, color=black] { $z_{0}$ };

\draw[-] (-1, 11.25) -- +(0, -2.5);
\draw[-] (-1, 11.25) -- +(0.25, 0);
\draw[-] (-1, 8.75) -- +(0.25, 0);
\node[color=black] at (-2.5, 10) { $\mathfrak{T}_{z_{0}, 3}$ };

\draw[-] (-1, 7.25) -- +(0, -2.5);
\draw[-] (-1, 7.25) -- +(0.25, 0);
\draw[-] (-1, 4.75) -- +(0.25, 0);
\node[color=black] at (-2.5, 6) { $\mathfrak{T}_{z_{1}, 3}$ };

\draw[-] (-1, 3.25) -- +(0, -2.5);
\draw[-] (-1, 3.25) -- +(0.25, 0);
\draw[-] (-1, 0.75) -- +(0.25, 0);
\node[color=black] at (-2.5, 2) { $\mathfrak{T}_{z_{2}, 3}$ };

\foreach \i in {0, 1, 2} {
  \draw[->, color=gray80] (2 * \i + 4, 3 - \i) -- + (5, 0);
  \foreach \j in {0, 1, 2} {
    \fill[color=dark] (2 * \j + 2 * \i + 4, 3 - \i) circle (4pt);
  }
}

\fill[] (13.5, 11) circle (0pt)
node[right, color=gray80] { $\mathbb{P} \! \left[ \mathfrak{t}_{1} | z_{0} \right] = 1 / 3$ };

\fill[] (13.5, 10) circle (0pt)
node[right, color=gray80] { $\mathbb{P} \! \left[ \mathfrak{t}_{2} | z_{0} \right] = 1 / 3$ };

\fill[] (13.5, 9) circle (0pt)
node[right, color=black] { $\mathbb{P} \! \left[ \mathfrak{t}_{3} | z_{0} \right] = 1 / 3$ };

\foreach \i in {0, 1, 2} {
  \draw[->, color=gray80] (2 * \i + 2, 7 - \i) -- + (5, 0);
  \foreach \j in {0, 1, 2} {
    \fill[color=dark] (2 * \j + 2 * \i + 2, 7 - \i) circle (4pt);
  }
}

\fill[] (13.5, 7) circle (0pt)
node[right, color=gray80] { $\mathbb{P} \! \left[ \mathfrak{t}_{2} | z_{1} \right] = 1 / 3$ };

\fill[] (13.5, 6) circle (0pt)
node[right, color=black] { $\mathbb{P} \! \left[ \mathfrak{t}_{3} | z_{1} \right] = 1 / 3$ };

\fill[] (13.5, 5) circle (0pt)
node[right, color=gray80] { $\mathbb{P} \! \left[ \mathfrak{t}_{4} | z_{1} \right] = 1 / 3$ };

\foreach \i in {0, 1, 2} {
  \draw[->, color=gray80] (2 * \i, 11 - \i) -- + (5, 0);
  \foreach \j in {0, 1, 2} {
    \fill[color=dark] (2 * \j + 2 * \i, 11 - \i) circle (4pt);
  }
}

\fill[] (13.5, 3) circle (0pt)
node[right, color=black] { $\mathbb{P} \! \left[ \mathfrak{t}_{3} | z_{2} \right] = 1 / 3$ };

\fill[] (13.5, 2) circle (0pt)
node[right, color=gray80] { $\mathbb{P} \! \left[ \mathfrak{t}_{4} | z_{2} \right] = 1 / 3$ };

\fill[] (13.5, 1) circle (0pt)
node[right, color=gray80] { $\mathbb{P} \! \left[ \mathfrak{t}_{5} | z_{2} \right] = 1 / 3$ };

\end{tikzpicture}
\caption{By uniformly sampling all numerical trajectories in 
$\mathfrak{T}_{z, L}$ regardless of the initial $z$, we ensure 
that $\mathbb{P} \! \left[ \mathfrak{t} | z \right] = 1 / L, \forall z \in \mathfrak{t}$, 
which immediate guarantees \eqref{eqn:trajectory_probs} and 
hence detailed balance of the resulting Markov chain.}
\label{fig:reversible_trajectories}
\end{figure*}

One immediate solution is to simply sample trajectories in 
$\mathfrak{T}_{z_{0}, L}$ uniformly,
\begin{align*}
\mathbb{P} \! \left[ \mathfrak{t} | z_{0} \right] 
&=
\left\{
\begin{array}{rr}
0, & \mathfrak{t} \notin \mathfrak{T}_{z_{0}, L} \\
1 / \left| \mathfrak{T}_{z_{0}, L} \right|, & \mathfrak{t} \in \mathfrak{T}_{z_{0}, L}
\end{array} 
\right. 
\\
&=
\left\{
\begin{array}{rr}
0, & \mathfrak{t} \notin \mathfrak{T}_{z_{0}, L} \\
\hspace{9mm} 1 / L, & \mathfrak{t} \in \mathfrak{T}_{z_{0}, L}
\end{array} 
\right. .
\end{align*}
Because each trajectory is equally likely regardless of the initial point,
\eqref{eqn:trajectory_probs} holds trivially (Figure \ref{fig:reversible_trajectories}).  
Moreover, sampling trajectories uniformly is straightforward to implement,
for example by sampling $L' \sim U \! \left[ 0, L \right]$ and integrating 
backwards for $L'$ steps and forwards for $L - L'$ steps.  When sampling
a final state using the Metropolis probabilities directly, this is equivalent to Neal's
Windowed State Algorithm with $W = L$~\citep{Neal:1994}.

Still, all of this effort lets us uniformly sample only from static trajectories of constant
length, $L$, and not the dynamic trajectories capable of expanding to ensure uniform
exploration of the underlying level sets.  Sampling from trajectories that integrate for a 
dynamic number of steps determined by a termination criterion such as 
\eqref{eqn:autocorr_bound} requires a careful extension of the static implementation.  

\subsection{Dynamic Implementations}

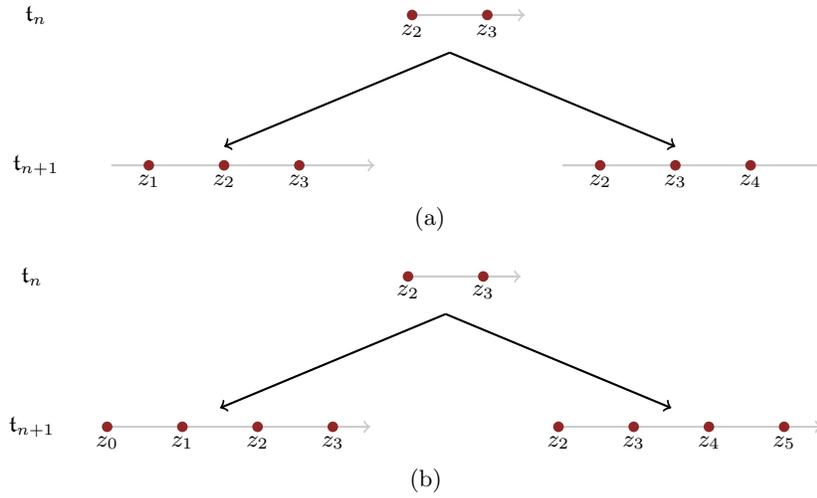
\begin{figure*}
\centering
\subfigure[]{
\begin{tikzpicture}[scale=0.5, thick]

\draw[-, color=white] (1, 1) -- + (21, 0);

\node[color=black] at (0, 5) { $\mathfrak{t}_{n}$ };
\node[color=black] at (0, 1) { $\mathfrak{t}_{n + 1}$ };

\draw[->, color=gray80] (10, 5) -- + (3, 0);
\foreach \i in {0, 1} {
  \pgfmathsetmacro{\ind}{\i + 2}
  \fill[color=dark] (2 * \i + 10, 5) circle (4pt)
  node[below, color=black] { $z_{\pgfmathprintnumber[fixed, precision=1]{\ind}}$ };
}

\draw[->, color=black] (11, 4) -- (5, 1.5);

\draw[->, color=gray80] (2, 1) -- + (7, 0);
\foreach \i in {0, 1, 2} {
  \pgfmathsetmacro{\ind}{\i + 1}
  \fill[color=dark] (2 * \i + 3, 1) circle (4pt)
  node[below, color=black] { $z_{\pgfmathprintnumber[fixed, precision=1]{\ind}}$ };
}

\draw[->, color=black] (11, 4) -- (17, 1.5);

\draw[->, color=gray80] (14, 1) -- + (7, 0);
\foreach \i in {0, 1, 2} {
  \pgfmathsetmacro{\ind}{\i + 2}
  \fill[color=dark] (2 * \i + 15, 1) circle (4pt)
  node[below, color=black] { $z_{\pgfmathprintnumber[fixed, precision=1]{\ind}}$ };
}
\end{tikzpicture}
}
\subfigure[]{
\begin{tikzpicture}[scale=0.5, thick]

\draw[-, color=white] (1, 1) -- + (21, 0);

\node[color=black] at (0, 5) { $\mathfrak{t}_{n}$ };
\node[color=black] at (0, 1) { $\mathfrak{t}_{n + 1}$ };

\draw[->, color=gray80] (10, 5) -- + (3, 0);
\foreach \i in {0, 1} {
  \pgfmathsetmacro{\ind}{\i + 2}
  \fill[color=dark] (2 * \i + 10, 5) circle (4pt)
  node[below, color=black] { $z_{\pgfmathprintnumber[fixed, precision=1]{\ind}}$ };
}

\draw[->, color=black] (11, 4) -- (5, 1.5);

\draw[->, color=gray80] (2, 1) -- + (7, 0);
\foreach \i in {0, 1, 2, 3} {
  \fill[color=dark] (2 * \i + 2, 1) circle (4pt)
  node[below, color=black] { $z_{\i}$ };
}

\draw[->, color=black] (11, 4) -- (17, 1.5);

\draw[->, color=gray80] (14, 1) -- + (7, 0);
\foreach \i in {0, 1, 2, 3} {
  \pgfmathsetmacro{\ind}{\i + 2}
  \fill[color=dark] (2 * \i + 14, 1) circle (4pt)
  node[below, color=black] { $z_{\pgfmathprintnumber[fixed, precision=1]{\ind}}$ };
}
\end{tikzpicture}
}
\caption{Uniformly sampling numerical trajectories of a dynamic length requires 
that the trajectories are generated incrementally.  Trajectories can be generated
recursively, either with (a) additive increments that randomly integrate the 
trajectory forwards or backward a single step or (b) multiplicative increments 
that double a trajectory of length $L$ by randomly integrating forward or 
backwards $L$ additional steps.}
\label{fig:dynamic_trajectories}
\end{figure*}

\begin{figure*}
\centering
\begin{tikzpicture}[scale=0.5, thick]
\draw[color=gray80, dashed] (7, 6) -- +(-4, -2);
\draw[color=gray80, dashed] (7, 6) -- +(4, -2);
\foreach \i in {0, 1} {
  \draw[color=gray80, dashed] (8 * \i + 3, 4) -- +(-2, -2);
  \draw[color=gray80, dashed] (8 * \i + 3, 4) -- +(2, -2);
}
\foreach \i in {0, 1, ..., 3} {
  \draw[color=gray80, dashed] (4 * \i + 1, 2) -- +(-1, -2);
  \draw[color=gray80, dashed] (4 * \i + 1, 2) -- +(1, -2);
}
\draw[->, color=gray80] (0, 0) -- +(15,0);
\foreach \i in {0, 1, ..., 7} { 
  \fill[color=dark] (2 * \i, 0) circle (4pt); 
}
\fill[color=dark] (0, 0) circle (0pt)
node[below, color=black] { $z_{-}$ };
\fill[color=dark] (14, 0) circle (0pt)
node[below, color=black] { $z_{+}$ };
\end{tikzpicture}
\caption{A numerical trajectory of length $L = 2^{D}$ can be
represented as the leaves of a perfect, ordered binary tree of depth $D$.  
The initial and final points of the trajectory, labeled $z_{-}$ and $z_{+}$ 
respectively, serve as the tree boundaries.
}
\label{fig:trajectory_as_tree}
\end{figure*}
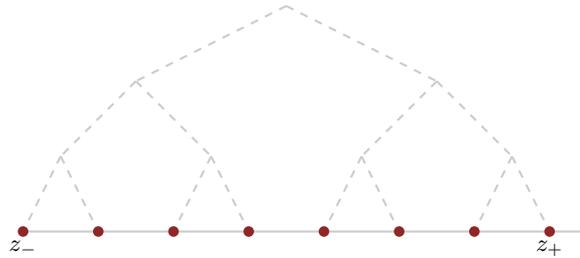

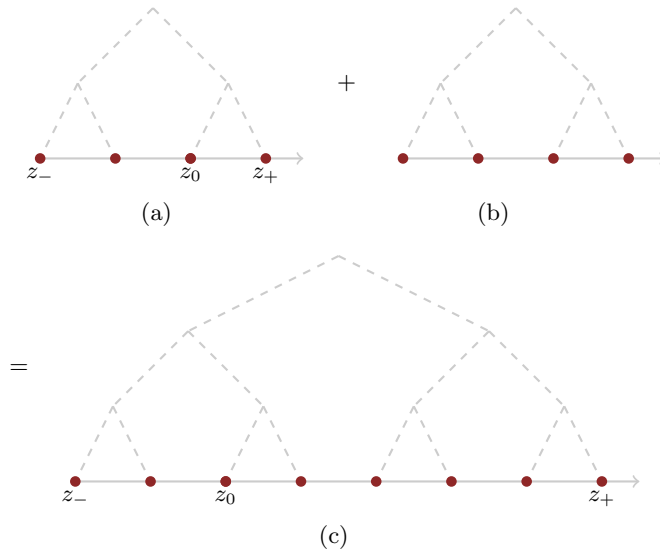
\begin{figure*}
\centering
\subfigure[]{
\begin{tikzpicture}[scale=0.5, thick]

\foreach \i in {0} {
  \draw[color=gray80, dashed] (8 * \i + 3, 4) -- +(-2, -2);
  \draw[color=gray80, dashed] (8 * \i + 3, 4) -- +(2, -2);
}

\foreach \i in {0, 1} {
  \draw[color=gray80, dashed] (4 * \i + 1, 2) -- +(-1, -2);
  \draw[color=gray80, dashed] (4 * \i + 1, 2) -- +(1, -2);
}

\draw[->, color=gray80] (0, 0) -- +(7,0);
\foreach \i in {0, 1, 2, 3} {
 \fill[color=dark] (2 * \i, 0) circle (4pt);
}

\fill[color=dark] (0, 0) circle (0pt)
node[below, color=black] { $z_{-}$ };

\fill[color=dark] (4, 0) circle (4pt)
node[below, color=black] { $z_{0}$ };

\fill[color=dark] (6, 0) circle (0pt)
node[below, color=black] { $z_{+}$ };

\end{tikzpicture}
}
\subfigure[]{
\begin{tikzpicture}[scale=0.5, thick]

\node[color=black] at (-1.5, 2) { $+$ };

\foreach \i in {0} {
  \draw[color=gray80, dashed] (8 * \i + 3, 4) -- +(-2, -2);
  \draw[color=gray80, dashed] (8 * \i + 3, 4) -- +(2, -2);
}

\foreach \i in {0, 1} {
  \draw[color=gray80, dashed] (4 * \i + 1, 2) -- +(-1, -2);
  \draw[color=gray80, dashed] (4 * \i + 1, 2) -- +(1, -2);
}

\draw[->, color=gray80] (0, 0) -- +(7,0);
\foreach \i in {0, 1, 2, 3} {
 \fill[color=dark] (2 * \i, 0) circle (4pt);
}

\fill[color=dark] (6, 0) circle (0pt)
node[below, color=white] { $z_{+}$ };

\end{tikzpicture}
}
\subfigure[]{
\begin{tikzpicture}[scale=0.5, thick]

\node[color=black] at (-1.5, 3) { $=$ };
\node[color=white] at (15.5, 3) { $=$ };

\draw[color=gray80, dashed] (7, 6) -- +(-4, -2);
\draw[color=gray80, dashed] (7, 6) -- +(4, -2);

\foreach \i in {0, 1} {
  \draw[color=gray80, dashed] (8 * \i + 3, 4) -- +(-2, -2);
  \draw[color=gray80, dashed] (8 * \i + 3, 4) -- +(2, -2);
}

\foreach \i in {0, 1, 2, 3} {
  \draw[color=gray80, dashed] (4 * \i + 1, 2) -- +(-1, -2);
  \draw[color=gray80, dashed] (4 * \i + 1, 2) -- +(1, -2);
}

\draw[->, color=gray80] (0, 0) -- +(15,0);
\foreach \i in {0, 1, ..., 7} {
 \fill[color=dark] (2 * \i, 0) circle (4pt);
}

\fill[color=dark] (0, 0) circle (0pt)
node[below, color=black] { $z_{-}$ };

\fill[color=dark] (4, 0) circle (4pt)
node[below, color=black] { $z_{0}$ };

\fill[color=dark] (14, 0) circle (0pt)
node[below, color=black] { $z_{+}$ };

\end{tikzpicture}
}
\caption{In the tree representation, a multiplicative expansion of a numerical 
trajectory of length $L = 2^{D}$ is given by randomly selecting a boundary, here
$z_{+}$, and integrating away from the tree $L$ additional steps.  This process 
can also be considered as appending (b) a new tree of depth $D$ to (a) the original 
tree of depth $D$ to give (c) an expanded tree of depth $D + 1$.}
\label{fig:tree_building}
\end{figure*}

In order to maintain a uniform distribution over numerical trajectories when their
length, $L$, is dynamic we have to build up each trajectory incrementally, checking 
if a termination criterion like \eqref{eqn:implicit_times} has been satisfied after each
expansion (Algorithm \ref{algo:naive_generation}).  For example, a uniformly 
sampled trajectory can be build up additively by iteratively expanding the trajectory
one step at a time in a random direction (Figure \ref{fig:dynamic_trajectories}a).  
If we consider trajectories of only lengths $2^{D}$ then we can also build up 
uniformly sampled trajectories multiplicatively, expanding a trajectory of length 
$L$ by integrating $L$ additional steps in a random direction (Figure 
\ref{fig:dynamic_trajectories}b).  In this multiplicative scheme each intermediate
trajectory can also be interpreted as a balanced binary tree 
(Figures \ref{fig:trajectory_as_tree}, \ref{fig:tree_building}).

\begin{algorithm}
\caption{Given a means to expand a trajectory, such as the additive or multiplicative
schemes discussed in the text, and a termination criterion that implicitly identifies
the optimal integration time, a uniformly sampled trajectory can be built up 
recursively.}
\label{algo:naive_generation}
\begin{algorithmic}[0]
\Function{expand\_trajectory}{$\mathfrak{t}$}
\EndFunction
\Function{check\_termination}{$\mathfrak{t}$}
\EndFunction
\State
\Function{naive\_build\_trajectory}{$\mathfrak{t}$}
   \State $\mathfrak{t}_{\mathrm{new}} \gets $ \Call{expand\_trajectory}{$\mathfrak{t}$}
   \If{ \Call{check\_termination}{$\mathfrak{t}_{\mathrm{new}}$} }
     \State \textbf{return} $\mathfrak{t}_{\mathrm{new}}$
   \Else
     \State \Call{naive\_build\_trajectory}{$\mathfrak{t}_{\mathrm{new}}$}
   \EndIf
\EndFunction
\end{algorithmic}
\end{algorithm}

\begin{figure*}
\centering
\begin{tikzpicture}[scale=0.5, thick]

\draw[color=gray80, dashed] (7, 6) -- +(-4, -2);
\draw[color=gray80, dashed] (7, 6) -- +(4, -2);

\foreach \i in {0, 1} {
  \draw[color=gray80, dashed] (8 * \i + 3, 4) -- +(-2, -2);
  \draw[color=gray80, dashed] (8 * \i + 3, 4) -- +(2, -2);
}

\foreach \i in {0, 1, ..., 3} {
  \draw[color=gray80, dashed] (4 * \i + 1, 2) -- +(-1, -2);
  \draw[color=gray80, dashed] (4 * \i + 1, 2) -- +(1, -2);
}

\draw[->, color=gray80] (0, 0) -- +(15,0);
\foreach \i in {0, 1, ..., 7} {
 \fill[color=dark] (2 * \i, 0) circle (4pt);
}

\fill[color=dark] (0, 0) circle (4pt)
node[below, color=black] { $z_{-}$ };

\fill[color=dark] (4, 0) circle (4pt)
node[below, color=black] { $z_{0}$ };

\fill[color=dark] (8, 0) circle (4pt)
node[below, color=black] { $z'_{-}$ };

\fill[color=dark] (10, 0) circle (4pt)
node[below, color=black] { $z'_{+}$ };

\fill[color=dark] (14, 0) circle (4pt)
node[below, color=black] { $z_{+}$ };

\draw[color=dark] (9, 2) -- +(-1, -2);
\draw[color=dark] (9, 2) -- +(1, -2);

\end{tikzpicture}
\caption{Uniformly sampling a numerical trajectory around the initial state, 
$z_{0}$, is not suffient to ensure detailed balance when the trajectory
length is dynamic.  The problem is that if the termination criterion is
satisfied in the interior of the trajectory, here between $z_{-}'$ and
$z_{+}'$ then both 
$\mathbb{P} \! \left[ \mathfrak{t} | z_{-}' \right] = 0$ and
$\mathbb{P} \! \left[ \mathfrak{t} | z_{+}' \right] = 0$
despite $\mathbb{P} \! \left[ \mathfrak{t} | z_{0} \right] \neq 0$.
}
\label{fig:premature_termination}
\end{figure*}
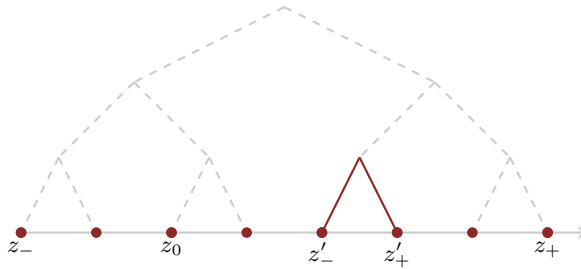

Unfortunately, when the length is chosen dynamically uniformly sampling trajectories
is no longer sufficient to ensure \eqref{eqn:trajectory_probs}, as different initial states
may lead to different terminal lengths (Figure \ref{fig:premature_termination}).  In order 
to guarantee detailed balance we have to treat each increment as a proposal, rejecting 
any extensions which include states from which 
$\mathbb{P} \! \left[ \mathfrak{t}_{\mathrm{new}} | z \right] = 0$ (Algorithm
\ref{algo:robust_generation}).  When the trajectory length is limited by a failed proposal 
the resulting trajectory will not satisfy the termination criterion exactly, a price we have 
to pay to ensure uniform samples that target the correct distribution. 
 
\begin{algorithm}
\caption{Ensuring detailed balance with dynamic trajectories requires not just
uniformly sampling a trajectory for an initial point, but also ensuring that the final
trajectory can be reached from all points in that trajectory.  Given a means of
validating each intermediate trajectory the final algorithm is a straightforward 
modification of Algorithm \ref{algo:naive_generation}.}
\label{algo:robust_generation}
\begin{algorithmic}[0]
\Function{expand\_trajectory}{$\mathfrak{t}$}
\EndFunction
\Function{validate\_trajectory}{$\mathfrak{t}$}
\EndFunction
\Function{check\_termination}{$\mathfrak{t}$}
\EndFunction
\State
\Function{build\_trajectory}{$\mathfrak{t}$}
   \State $\mathfrak{t}_{\mathrm{new}} \gets $ \Call{expand\_trajectory}{$\mathfrak{t}$}
   \If {\Call{validate\_trajectory}{$\mathfrak{t}_{\mathrm{new}}$} }
     \If{ \Call{check\_termination}{$\mathfrak{t}_{\mathrm{new}}$} }
       \State \textbf{return} $\mathfrak{t}_{\mathrm{new}}$
     \Else
       \State \Call{naive\_build\_trajectory}{$\mathfrak{t}_{\mathrm{new}}$}
     \EndIf
   \Else
     \State \textbf{return} $\mathfrak{t}$
   \EndIf
\EndFunction
\end{algorithmic}
\end{algorithm}

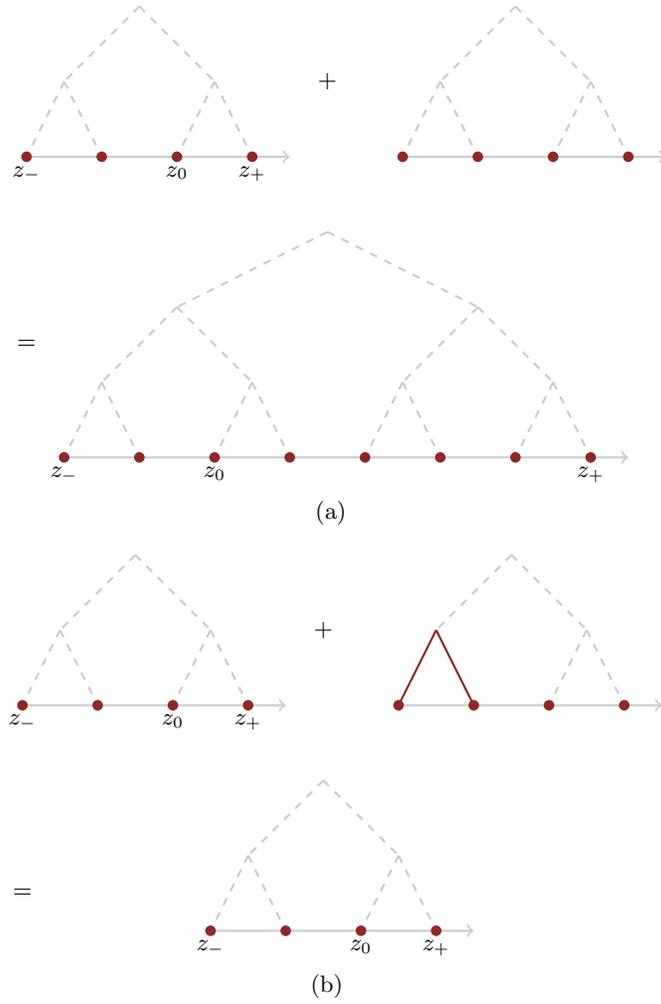
\begin{figure*}
\centering
\subfigure[]{
\begin{tikzpicture}[scale=0.5, thick]

\draw[color=gray80, dashed] (3, 4) -- +(-2, -2);
\draw[color=gray80, dashed] (3, 4) -- +(2, -2);

\foreach \i in {0, 1} {
  \draw[color=gray80, dashed] (4 * \i + 1, 2) -- +(-1, -2);
  \draw[color=gray80, dashed] (4 * \i + 1, 2) -- +(1, -2);
}

\draw[->, color=gray80] (0, 0) -- +(7,0);
\foreach \i in {0, 1, 2, 3} {
 \fill[color=dark] (2 * \i, 0) circle (4pt);
}

\fill[color=dark] (0, 0) circle (0pt)
node[below, color=black] { $z_{-}$ };

\fill[color=dark] (4, 0) circle (0pt)
node[below, color=black] { $z_{0}$ };

\fill[color=dark] (6, 0) circle (0pt)
node[below, color=black] { $z_{+}$ };

\node[color=black] at (8, 2) { $+$ };

\draw[color=gray80, dashed] (13, 4) -- +(-2, -2);
\draw[color=gray80, dashed] (13, 4) -- +(2, -2);

\foreach \i in {2, 3} {
  \draw[color=gray80, dashed] (4 * \i + 3, 2) -- +(-1, -2);
  \draw[color=gray80, dashed] (4 * \i + 3, 2) -- +(1, -2);
}

\draw[->, color=gray80] (10, 0) -- +(7,0);
\foreach \i in {0, 1, 2, 3} {
 \fill[color=dark] (2 * \i + 10, 0) circle (4pt);
}

\node[color=black] at (0, -5) { $=$ };

\draw[color=gray80, dashed] (8, -2) -- +(-4, -2);
\draw[color=gray80, dashed] (8, -2) -- +(4, -2);

\foreach \i in {0, 1} {
  \draw[color=gray80, dashed] (8 * \i + 4, -4) -- +(-2, -2);
  \draw[color=gray80, dashed] (8 * \i + 4, -4) -- +(2, -2);
}

\foreach \i in {0, 1, ..., 3} {
  \draw[color=gray80, dashed] (4 * \i + 2, -6) -- +(-1, -2);
  \draw[color=gray80, dashed] (4 * \i + 2, -6) -- +(1, -2);
}

\draw[->, color=gray80] (1, -8) -- +(15,0);
\foreach \i in {0, 1, ..., 7} {
 \fill[color=dark] (2 * \i + 1, -8) circle (4pt);
}

\fill[color=dark] (1, -8) circle (0pt)
node[below, color=black] { $z_{-}$ };

\fill[color=dark] (5, -8) circle (0pt)
node[below, color=black] { $z_{0}$ };

\fill[color=dark] (15, -8) circle (0pt)
node[below, color=black] { $z_{+}$ };

\end{tikzpicture}
}
\subfigure[]{
\begin{tikzpicture}[scale=0.5, thick]

\draw[color=gray80, dashed] (3, 4) -- +(-2, -2);
\draw[color=gray80, dashed] (3, 4) -- +(2, -2);

\foreach \i in {0, 1} {
  \draw[color=gray80, dashed] (4 * \i + 1, 2) -- +(-1, -2);
  \draw[color=gray80, dashed] (4 * \i + 1, 2) -- +(1, -2);
}

\draw[->, color=gray80] (0, 0) -- +(7,0);
\foreach \i in {0, 1, 2, 3} {
 \fill[color=dark] (2 * \i, 0) circle (4pt);
}

\fill[color=dark] (0, 0) circle (0pt)
node[below, color=black] { $z_{-}$ };

\fill[color=dark] (4, 0) circle (0pt)
node[below, color=black] { $z_{0}$ };

\fill[color=dark] (6, 0) circle (0pt)
node[below, color=black] { $z_{+}$ };

\node[color=black] at (8, 2) { $+$ };

\draw[color=gray80, dashed] (13, 4) -- +(-2, -2);
\draw[color=gray80, dashed] (13, 4) -- +(2, -2);

\foreach \i in {2, 3} {
  \draw[color=gray80, dashed] (4 * \i + 3, 2) -- +(-1, -2);
  \draw[color=gray80, dashed] (4 * \i + 3, 2) -- +(1, -2);
}

\draw[->, color=gray80] (10, 0) -- +(7,0);
\foreach \i in {0, 1, 2, 3} {
 \fill[color=dark] (2 * \i + 10, 0) circle (4pt);
}

\draw[color=dark] (11, 2) -- +(-1, -2);
\draw[color=dark] (11, 2) -- +(1, -2);

\node[color=black] at (0, -5) { $=$ };

\draw[color=gray80, dashed] (8, -2) -- +(-2, -2);
\draw[color=gray80, dashed] (8, -2) -- +(2, -2);

\foreach \i in {0, 1} {
  \draw[color=gray80, dashed] (4 * \i + 6, -4) -- +(-1, -2);
  \draw[color=gray80, dashed] (4 * \i + 6, -4) -- +(1, -2);
}

\draw[->, color=gray80] (5, -6) -- +(7,0);
\foreach \i in {0, 1, 2, 3} {
 \fill[color=dark] (2 * \i + 5, -6) circle (4pt);
}

\fill[color=dark] (5, -6) circle (0pt)
node[below, color=black] { $z_{-}$ };

\fill[color=dark] (9, -6) circle (0pt)
node[below, color=black] { $z_{0}$ };

\fill[color=dark] (11, -6) circle (0pt)
node[below, color=black] { $z_{+}$ };

\end{tikzpicture}
}
\caption{Ensuring detailed balance requires validating each trajectory expansion
before accepting the new trajectory.  For multiplicative expansion this requires
that no internal subtree of the proposal satisfies the termination criterion.  (a) If 
no subtree satisfies the termination criterion then it can appended to the trajectory,
resulting in an expanded trajectory that can then be checked for termination and
further expanded as necessary.  (b) Conversely, when a subtree does satisfy the 
termination criterion then the proposal must be rejected and the trajectory construction 
immediately terminated.
}
\label{fig:dynamic_tree_building}
\end{figure*}

How a given trajectory is validated to ensure that there are no states with 
$\mathbb{P} \! \left[ \mathfrak{t}_{\mathrm{new}} | z \right] = 0$ depends
on the expansion method.  Additive expansions, for example, require that
the termination criterion not be satisfied for every pair of states in the
trajectory.  If checked recursively, this requires $L$ checks after proposing
a new trajectory of length $L$, as well as the local storage of each state in 
the trajectory.  Multiplicative expansions have the advantage that the termination 
criterion needs to be checked for only the subtrees 
(Figure \ref{fig:dynamic_tree_building}), requiring only $\log(L)$ checks for a 
proposed trajectory of length $L$ and only $\log(L)$ states in memory at any 
given time.

\subsection{Alternative Schemes}

Before considering explicit termination criteria, let us briefly pause to discuss 
alternative schemes for constructing Markov chains using Hamiltonian flow.  
Ultimately, the cause of poor performance when using a poorly chosen integration 
time is the momentum resampling induced by the projection and lifting needed 
to map from the cotangent bundle down to the target space and back.  
Constructing a Markov chain on the cotangent bundle directly, however, could 
invalidate the need for the momentum resampling and conceivably yield improved 
performance even with a suboptimal integration time.

The Horowitz scheme~\citep{Horowitz:1991}, for example, uses Hamiltonian flow
mixed with only partial momentum resampling to move between the level sets.
After integrating for the prescribed integration time, a Metropolis correction is 
applied: if the state is accepted then the final momentum is mixed with newly 
sampled momenta, maintaining some coherency in the exploration.  The cost of 
this approach, however, is that in order to preserve the target distribution the 
momentum must be completely negated after a rejection, causing the next 
trajectory to return to a neighborhood that has already been explored.  Extra-chance 
schemes~\citep{SohlEtAl:2014, CamposEtAl:2015} take this idea even further,
applying a fixed number of proposals which do not modify the momentum at all 
after an acceptance while continuing to negate after rejections to ensure the 
correct stationary distribution. 

\begin{figure*}
\centering
\begin{tikzpicture}[scale=0.5, thick]
  \draw [mid, thick] plot[smooth, tension=1.1] coordinates {(0,0) (8, 4) (15, 2) (20.5, 2)}
  node[right] { $\phi^{H}$ };
  
  \draw[color=gray80] (0, 0) -- (1, 1);
  \draw[color=gray80] (1, 1) -- (2, 2.25);
  \draw[color=gray80] (2, 2.25) -- (3, 3.5);
  \draw[color=gray80] (3, 3.5) -- (4, 4);
  \draw[color=gray80] (4, 4) -- (5, 4);
  \draw[color=gray80] (5, 4) -- (6, 3.75);
  \draw[color=gray80] (6, 3.75) -- (7, 3.8);
  \draw[color=gray80] (7, 3.8) -- (8, 4);
  \draw[color=gray80] (8, 4) -- (9, 4.5);
  \draw[color=gray80] (9, 4.5) -- (10, 5);
  \draw[color=gray80] (10, 5) -- (11, 4.9);
  \draw[color=gray80] (11, 4.9) -- (12, 3.96);
  \draw[color=gray80] (12, 3.96) -- (13, 2.73);
  \draw[color=gray80] (13, 2.73) -- (14, 2.6);
  \draw[color=gray80] (14, 2.6) -- (15, 3);
  \draw[color=gray80] (15, 3) -- (16, 3.38);
  \draw[color=gray80] (16, 3.38) -- (17, 3.3);
  \draw[color=gray80] (17, 3.3) -- (18, 2.98);
  \draw[color=gray80] (18, 2.98) -- (19, 2.75);
  \draw[color=gray80] (19, 2.75) -- (20, 2.9);
  \draw[->, color=gray80] (20, 2.9) -- +(0.5, 0.1)
    node [right, color=dark] { $\Phi^{\widetilde{H}}_{\epsilon}$ };
  
  \fill[color=dark] (0, 0) circle (4pt);
  \fill[color=dark] (1, 1) circle (4pt);
  \fill[color=dark] (2, 2.25) circle (4pt);
  \fill[color=dark] (3, 3.5) circle (4pt);
  \fill[color=dark] (4, 4) circle (4pt);
  \fill[color=dark] (5, 4) circle (4pt);
  \fill[color=dark] (6, 3.75) circle (4pt);
  \fill[color=dark] (7, 3.8) circle (4pt);
  \fill[color=dark] (8, 4) circle (4pt);
  \fill[color=dark] (9, 4.5) circle (4pt);
  \fill[color=dark] (10, 5) circle (4pt);
  \fill[color=dark] (11, 4.9) circle (4pt);
  \fill[color=dark] (12, 3.96) circle (4pt);
  \fill[color=dark] (13, 2.73) circle (4pt);
  \fill[color=dark] (14, 2.6) circle (4pt);
  \fill[color=dark] (15, 3) circle (4pt);
  \fill[color=dark] (16, 3.38) circle (4pt);
  \fill[color=dark] (17, 3.3) circle (4pt);
  \fill[color=dark] (18, 2.98) circle (4pt);
  \fill[color=dark] (19, 2.75) circle (4pt);
  \fill[color=dark] (20, 2.9) circle (4pt);
\end{tikzpicture}
\caption{Because they are volume preserving, symplectic integrators do not drift
from the true Hamiltonian flow even over exponentially long integration times.
The numerical trajectory effectively oscillates near the true trajectory, and 
any increasing error is only temporary.
}
\label{fig:integrator_oscillations}
\end{figure*}
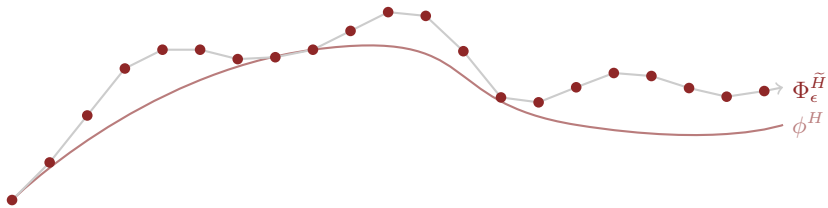

Both schemes, however, can maintain the coherency of the exploration only while
the symplectic integrator is near the true flow, devolving into diffusive exploration
as the symplectic integrator strays and the proposals are rejected.  Optimal
performance is then achieved when the total integration time, for one proposal in a
Horowitz scheme or the many proposals of an extra-chance scheme, is matched
to the first excursion of the symplectic integrator.  This almost always results in
premature termination, however, as the volume preservation of the symplectic 
integrator ensures that it does not drift and that these excursions are only temporary
(Figure \ref{fig:integrator_oscillations}).  To truly exploit the exploratory power of
Hamiltonian flow with symplectic integrators, we need to be able integrate far past
these temporary excursions.

\section{Explicit Termination Criteria}

Now that we know how to implement Hamiltonian Monte Carlo with dynamic
integration times we just need to select an explicit termination criterion capable 
of identifying optimal, or at least approximately optimal, integration times.  
Fortunately, the underlying geometry proves ever fruitful, naturally motivating a
canonical autocorrelation function that yields a set of integration times known 
as an \textit{exhaustion}.  After constructing these objects for both exact and 
numerical trajectories I also consider termination criteria that arise naturally
when the target space is equipped with a Riemannian metric. 

\subsection{Theoretical Exhaustions}

A particularly natural way to define autocorrelation functions on a Hamiltonian 
system is through the temporal expectation of the temporal derivative of any 
scalar function, $u$,
\begin{equation*}
\kappa_{u} \! \left(T, z \right)
\equiv
\frac{1}{T} \int_{0}^{T} \mathrm{d} t \,  \frac{\dd u}{\dd t} \circ \phi^{H}_{t} \! \left( z \right)
=
\frac{ u \circ \phi^{H}_{T} \! \left( z \right) - u \! \left( z \right) }{ T }.
\end{equation*}
Provided that the scalar function is bounded,
\begin{equation*}
\left| u \circ \phi^{H}_{t} \! \left(z \right) - u \! \left( z \right) \right| 
< \infty, \, \forall t \in \mathbb{R},
\end{equation*}
then every such expectation vanishes asymptotically,
\begin{equation*}
\lim_{t \rightarrow \infty} \kappa_{u} \! \left(t , z\right)
=
\lim_{t \rightarrow \infty} 
\frac{ u \circ \phi^{H}_{t} \! \left(z \right) - u \! \left( z \right) }{ t }
=
0,
\end{equation*}
making it a potential termination criterion.  Care must be taken, however,
as the scalar function may recur, $u \circ \phi^{H}_{t} \! \left(z \right) = 
u \! \left( z \right)$, preventing $\kappa_{u}$ from being monotonic and 
possibly resulting in premature integration times.

There aren't many scalar functions available to construct such an autocorrelation
function for a generic Hamiltonian system.  One canonical scalar function is the 
Hamiltonian itself, but, because the Hamiltonian is conserved by the Hamiltonian 
flow, its time rate of change vanishes trivially making it unsuitable for tracking
convergence.  The only other scalar function canonical to a general Hamiltonian 
system is the \textit{virial}, $G = q^{i} p_{i}$.  When the Hamiltonian is proper and 
all trajectories bounded, the virial itself is always bounded and provides a
potential candidate.

Collecting the resulting integration times together defines an 
\textit{exhaustion}.
\begin{definition}
An exhaustion, $T_{\delta} \! \left( z \right)$, is the family of integration times 
at each point in the cotangent bundle such that the temporal average of the 
rate of change of the virial along the resulting Hamiltonian flow is uniformly bounded,
\begin{equation*}
\left| \frac{1}{T_{\delta}} \int_{0}^{T_{\delta}} \mathrm{d} t \,  
\frac{\dd G}{\dd t} \circ \phi^{H}_{t} \! \left( z \right) \right|
= 
\left| 
\frac{ G \circ \phi^{H}_{T_{\delta}} \! \left( z \right) 
- G \! \left( z \right) }{T_{\delta}} \right|
< \delta, 
\, \forall z \in T^{*} Q.
\end{equation*}

\end{definition}
Provided that the Hamiltonian is proper a valid exhaustion can always
be constructed.  

Although exhaustions are canonical to any Hamiltonian system, they will not,
in general, identify optimal integration times for any choice of $\delta$.  The 
real utility of an exhaustive termination criterion is that it ensures uniform
convergence across all level sets and reduces the tuning problem to the
single exhaustion threshold, $\delta$.  How to identify an optimal threshold 
for a given problem remains an open problem.

\subsection{Numerical Exhaustions}

Because the exact flow is approximated with a symplectic integrator, exhaustions 
defined using exact expectations are not quite applicable to any practical 
implementation of Hamiltonian Monte Carlo.  Instead we can replace the exact 
expectation with a Metropolis-corrected expectation over the numerical trajectory,
\begin{equation*}
\frac{1}{\left| \mathfrak{t} \right|} \sum_{z \in \mathfrak{t}} 
\mathbb{P} \! \left[ z | \mathfrak{t} \right]
\frac{\mathrm{d}G}{\mathrm{d}t} \! \left (z \right),
\end{equation*}
with
\begin{equation*}
\mathbb{P} \! \left[ z | \mathfrak{t} \right]
=
\frac{ \frac{\dd \pi_{H} }{ \dd \Omega } \! \left( z \right) }
{ \sum_{z' \in \mathfrak{t} } \frac{ \dd \pi_{H} }{ \dd \Omega } \! \left( z' \right) }
=
\frac{ e^{-H ( z ) } }
{ \sum_{z' \in \mathfrak{t}} e^{- H ( z' ) } }.
\end{equation*}
Because this expectation converges to the continuous expectation in the limit of 
infinite steps,
\begin{equation*}
\lim_{\left| \mathfrak{k} \right| \rightarrow \infty}
\frac{1}{\left| \mathfrak{t} \right|} \sum_{z \in \mathfrak{t}} 
\mathbb{P} \! \left[ z | \mathfrak{t} \right]
\frac{\mathrm{d}G}{\mathrm{d}t} \! \left (z \right)
=
\lim_{T \rightarrow \infty}\frac{1}{T} \int^{T}_{0} 
\mathrm{d}t \, \frac{\mathrm{d}G}{\mathrm{d}t} 
\circ \phi^{H}_{t} \! \left (z \right)
= 0,
\end{equation*}
the numerical expectation will converge to the true expectation and provide
similar termination behavior. Hence we can define an equivalent 
\textit{numerical exhaustion}.
\begin{definition}
A numerical exhaustion, $\mathfrak{T}_{\delta}$, is the set of numerical 
trajectories such that the Metropolis-corrected expectation of the rate of change 
of the virial along any element, $\mathfrak{t}$ is uniformly bounded,
\begin{equation*}
\left| \frac{1}{\left| \mathfrak{t} \right|} \sum_{z \in \mathfrak{t}} 
\mathbb{P} \! \left[ z | \mathfrak{t} \right]
\frac{\mathrm{d}G}{\mathrm{d}t} \! \left (z \right) \right|
< \delta.
\end{equation*}

\end{definition}

As in the exact case, the modified Hamiltonian foliates the manifold 
and we can define corresponding modified level sets,
\begin{equation*}
\MLS = \left\{ q, p \in M \, | \, \widetilde{H} \! \left( q, p \right) = E \right \}.
\end{equation*}
Provided that the asymptotic error is negligible and the symplectic 
integrator is \textit{topologically stable}~\citep{McLachlanEtAl:2004}, 
the modified level sets will have the same topology as the exact 
level sets.  In particular, when the exact Hamiltonian is proper and its 
level sets compact, then negligible asymptotic error and topological 
stability imply that the modified Hamiltonian is also proper and its 
level sets also compact.  Consequently the virial remains bounded 
on the numerical trajectories and the Poincar\'{e} recurrence theorem
still applies, guaranteeing that numerical exhaustions are nonempty.  
When the topological stability and negligible asymptotic error do not hold,
the numerical trajectories will rapidly diverge; these numerical divergences 
then serve as immediate diagnostics of an ill-posed numerical exhaustion.

Hence we can define a trajectory termination criterion by checking if 
$\mathfrak{t} \in \mathfrak{T}_{\delta}$, with the resulting implementation
of Hamiltonian Monte Carlo denoted \textit{Exhaustive Hamiltonian Monte
Carlo}. 

\subsection{Riemannian Termination Criteria}

Although the virial is the only candidate scalar function canonical to every
Hamiltonian system, there are additional candidates once we endow the 
sample space with additional structure, such as a Riemannian metric.  In 
particular, a Riemannian metric, $g$, allows us to define an entire family of 
disintegrations given in local coordinates by the kinetic energy
\begin{equation*}
K \! \left( q, p \right) = A \cdot f \! \left( g^{-1}_{q} \! \left( p, p \right) \right)
+ \frac{1}{2} \log \left| g_{q} \right| + \mathrm{const},
\end{equation*}
for some constant $A$ and function $f: \mathbb{R} \rightarrow \mathbb{R}$.  
Given such a Riemannian disintegration we can then define two new scalars: 
the \textit{effective potential energy},
\begin{equation*}
\widecheck{V} \! \left( q \right) 
= 
V \! \left( q \right) +  \frac{1}{2} \log \left| g_{q} \right| + \mathrm{const}.
\end{equation*}
and the \textit{effective kinetic energy},
\begin{equation*}
\widecheck{K} \! \left(q, p \right) 
= 
A \cdot f \! \left( g^{-1}_{q} \! \left( p, p \right) \right).
\end{equation*}
Because the Hamiltonian is conserved, the autocorrelation functions
induced by these two functions are simply negations of each other
and the resulting integration times identical.  The difficulty with
these functions is that they recur quickly, long before any reasonable
recurrence of the trajectory.  More formally, if the disintegration is 
Gaussian then the functions will recur at turning points of the 
orbits~\citep{HoferEtAl:2011}, which are rampant in the Hamiltonian 
systems resulting from strongly-correlated target distributions.

A Riemannian metric also admits the construction of a completely
different termination criterion.  Instead of considering the temporal
expectation of a scalar function we can appeal to the generalized 
No-U-Turn criterion~\citep{Betancourt:2013a}, which terminates 
when 
\begin{equation*}
\kappa_{\mathrm{NUTS}} \! \left(T \right) 
= 
g^{-1}_{q} \! \left( p, \rho_{T} \right) < 0
\end{equation*}
where
\begin{equation*}
\rho_{T} =
\frac{1}{T} \int_{0}^{T} \mathrm{d} t \, \left( \phi^{H}_{t} \right)_{*} \theta.
\end{equation*}
Note that when the metric is Euclidean the generalized No-U-Turn 
criterion reduces to the usual No-U-Turn criterion~\citep{HoffmanEtAl:2014}.
In fact, the use of the No-U-Turn criterion with multiplicative trajectory 
expansion and a slice sampler to draw a state from the final trajectory 
is exactly Hoffman and Gelman's No-U-Turn sampler.

For simple level set geometries the generalized No-U-Turn criterion is
satisfied when a trajectory has traveled from one side of a level set to
to the other, matching of the intuition we developed for an optimal
integration time in Section \ref{sec:ergodicity}.  Although there is no guarantee 
that the generalized No-U-Turn criterion always identifies the optimal integration 
time, its impressive empirically success suggests that it applies even in
when targeting complex distributions.  

One weakness that has arisen in some applications is that, because the criterion 
is always small in a neighborhood around the initial point, small oscillations in a 
trajectory can cause the criterion to vanish prematurely.  Additionally, evaluating 
the No-U-Turn criterion is more computationally expensive than checking a numerical 
exhaustion, especially in the general Riemannian case.

\section{Experiments}

In this section I present a series of illustrative experiments to corroborate 
the theory and intuition developed above.  I begin first with a graphical study 
of the exhaustive termination criteria and then follow with performance studies 
of various Hamiltonian Monte Carlo implementations targeting various
distributions.

\subsection{Graphical Experiments}

To illuminate the qualitative behavior of the exhaustive termination criterion
relative to the No-U-Turn criterion, consider a two-dimensional Gaussian 
distribution with a Euclidean-Gaussian disintegration, given in local coordinates
by the effective potential energy
\begin{equation*}
\widecheck{V} \! \left( q \right) 
=
\frac{1}{2} q^{i} q^{j} 
\frac{ \delta_{ij} - \left(1 - \delta_{ij} \right) \rho}
{1 - \rho^{2}}
+ \mathrm{const}
\end{equation*}
and the effective kinetic energy
\begin{equation*}
\widecheck{K} \! \left( q, p \right) 
=
\frac{1}{2} p_{i} p_{j} \delta^{ij},
\end{equation*}
where $\delta_{ij}$ is the discrete Dirac-delta function not to be confused
with the exhaustive termination threshold, $\delta$.

\begin{figure}
\centering
\subfigure[]{ \includegraphics[width=2.5in]{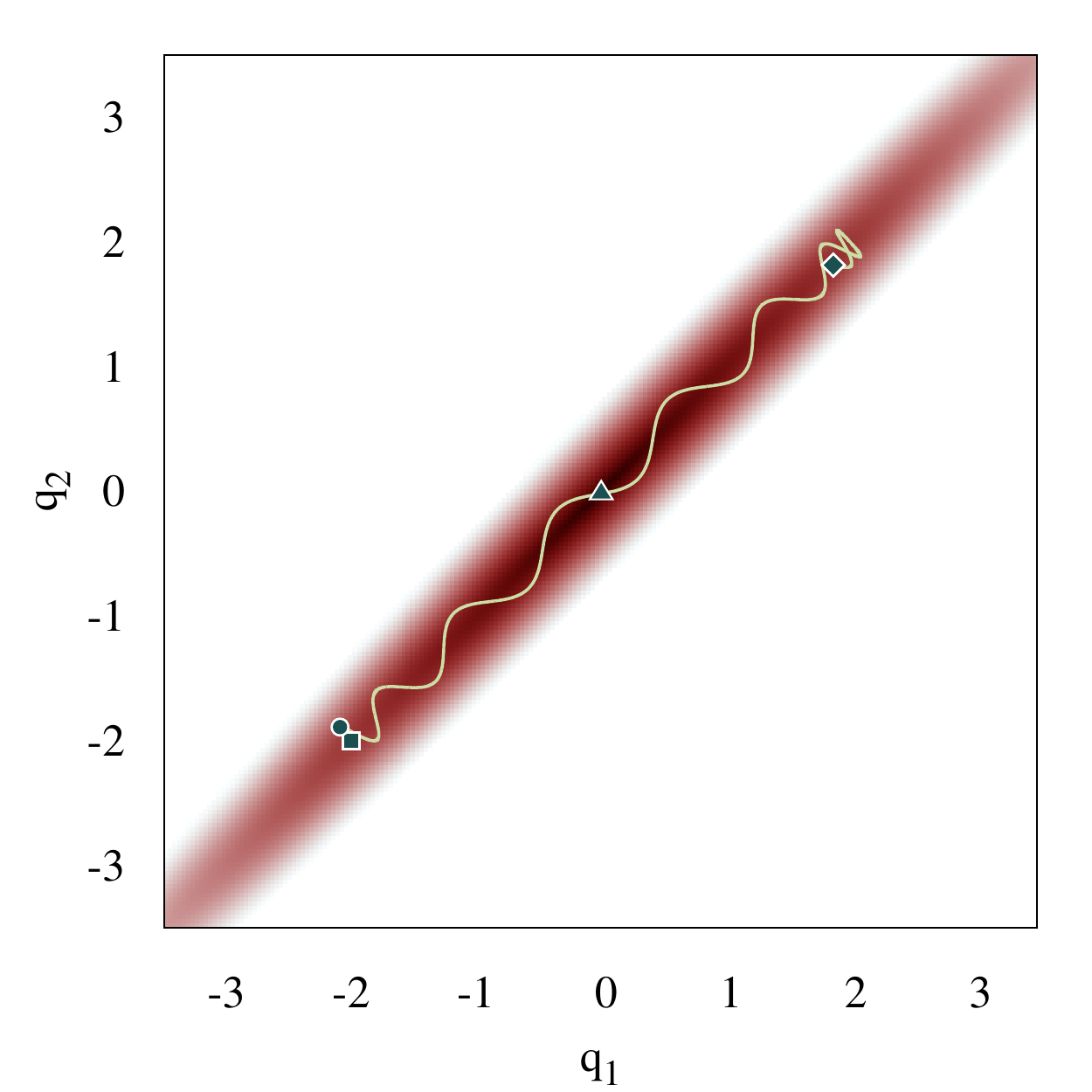} } \\
\subfigure[]{ \includegraphics[width=2.5in]{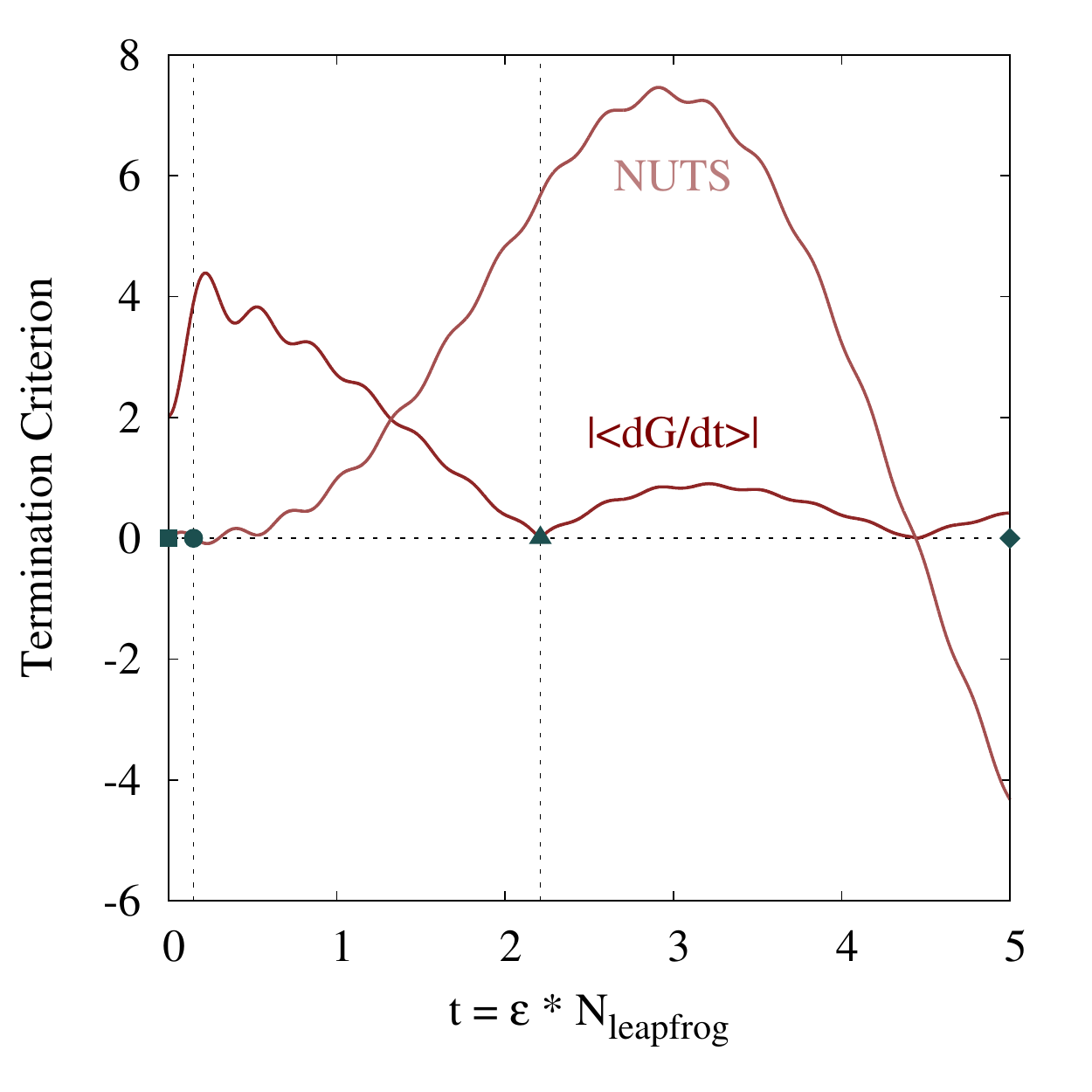} }
\subfigure[]{ \includegraphics[width=2.5in]{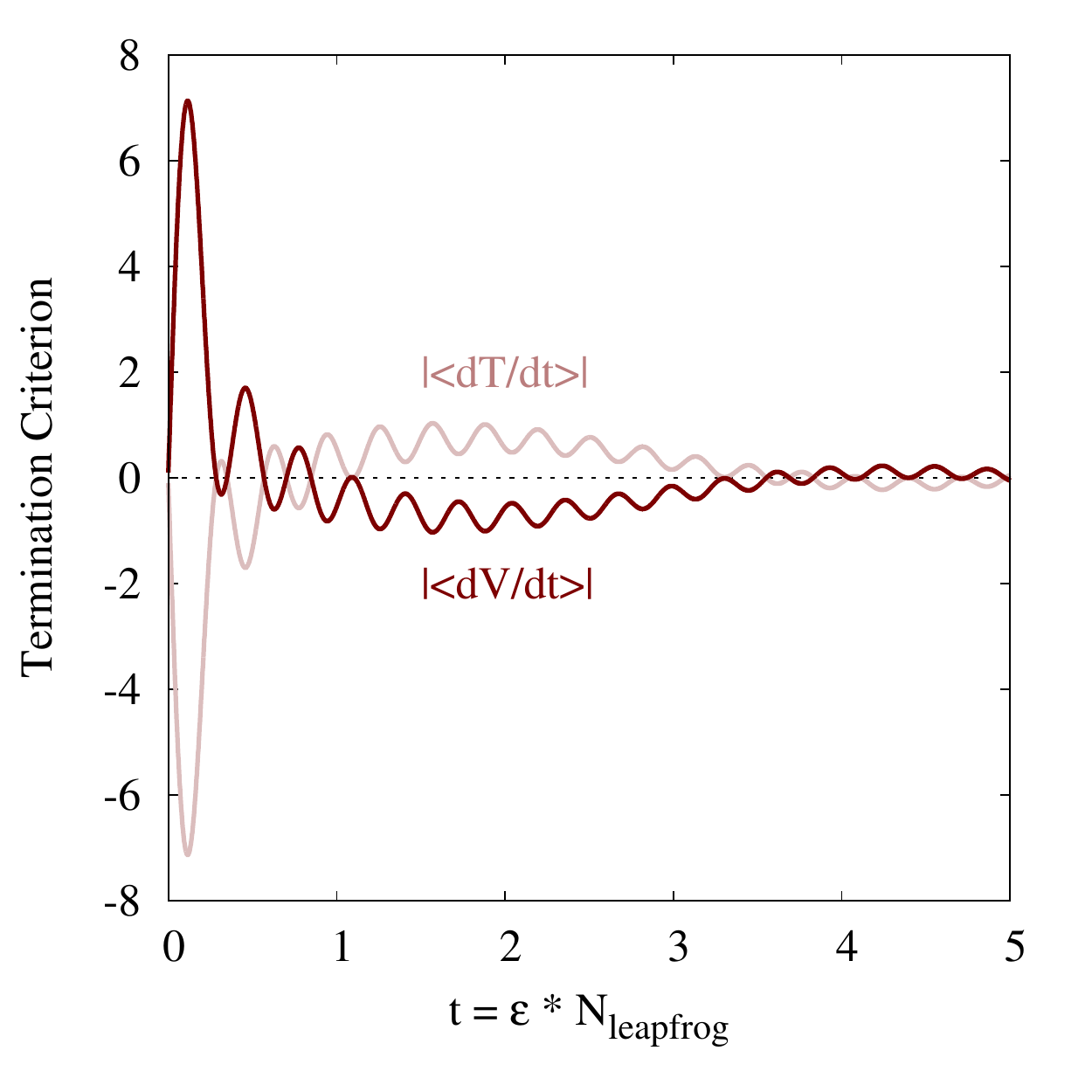} }
\caption{(a) Strong correlations in a two-dimensional Gaussian target
distribution, (b) cause the No-U-Turn criterion to terminate (circle) long before 
the exhaustive criterion for any $\delta$ (triangle). Similarly, (c) the temporal
expectations of the effective kinetic energy and effective potential energy 
vanish after only an incredibly short integration time, making them 
poor criteria for identifying optimal integration times.}
\label{fig:gauss_high}
\end{figure}

For $\rho = 0.99$ the target distribution is highly correlated 
(Figure \ref{fig:gauss_high}a).  The strong correlations induce 
premature termination of the No-U-Turn criterion but the exhaustive
termination criterion yields substantially longer integration times 
for any choice of $\delta$ (Figure \ref{fig:gauss_high}b).  Like the 
No-U-Turn criterion, the temporal expectations of the effective kinetic 
energy and effective potential energy vanish long before the exhaustive
termination criterion is satisfied (Figure \ref{fig:gauss_high}c).  

\begin{figure}
\centering
\subfigure{ \includegraphics[width=2.5in]{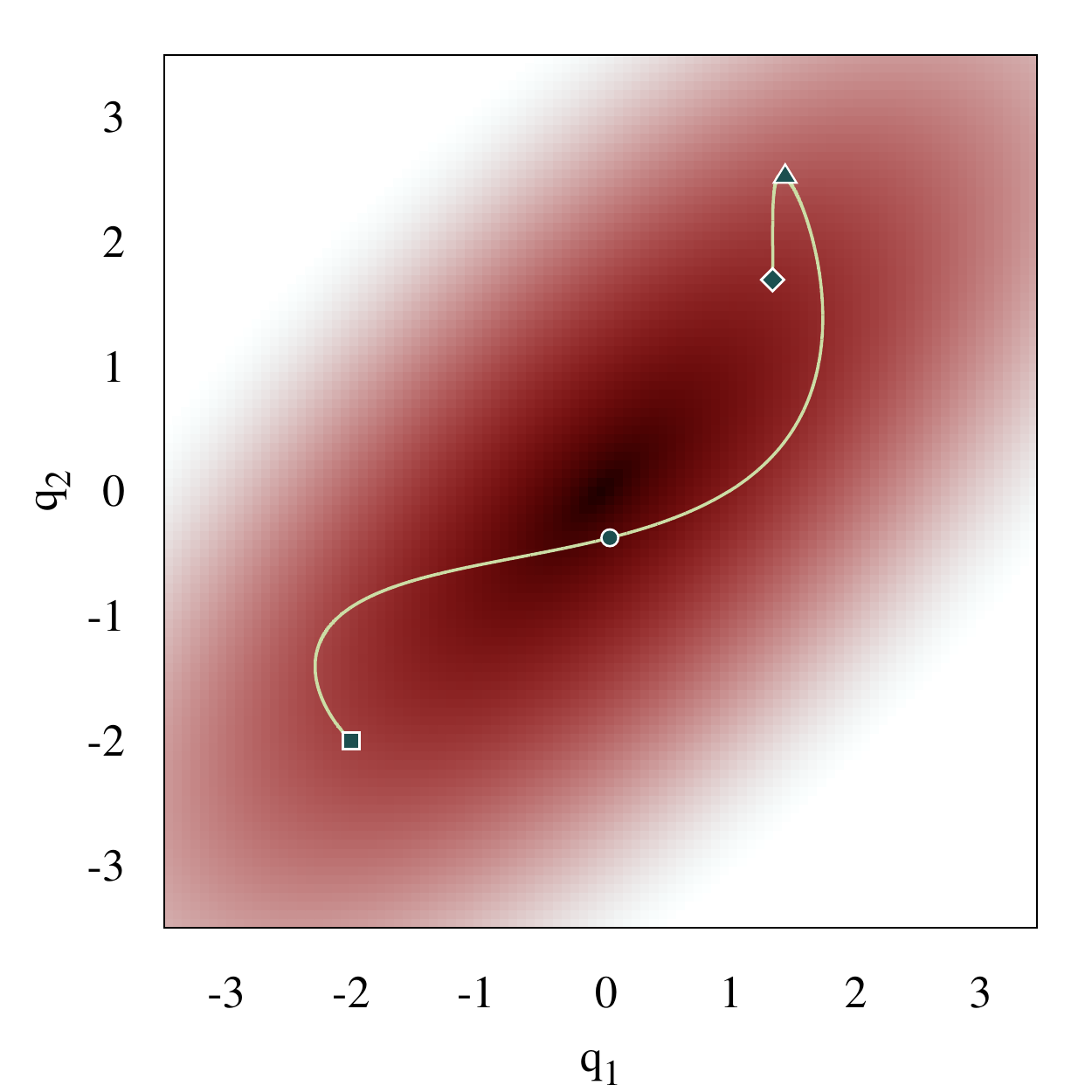} }
\subfigure{ \includegraphics[width=2.5in]{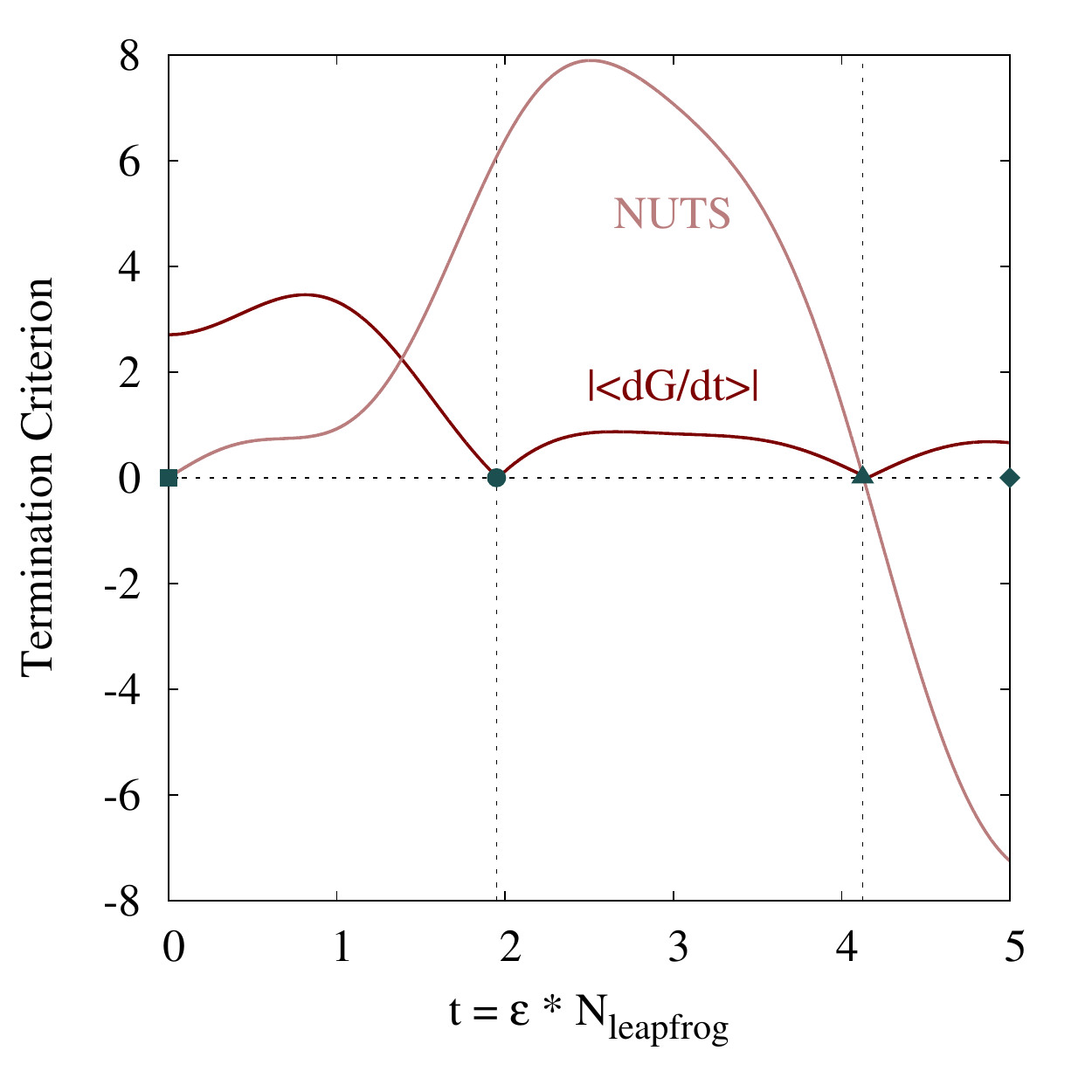} }
\caption{When targeting a two-dimensional Gaussian distribution with 
weaker correlations the No-U-Turn criterion terminates (triangle) after
traversing the level set once, and long after the exhaustion for any 
exhaustion threshold, $\delta$, (circle), ultimately yielding more effective 
exploration.}
\label{fig:gauss_low}
\end{figure}

When the correlations are relaxed to $\rho = 0.7$, however, the 
No-U-Turn criterion no longer suffers from premature termination
and provides integration times that are far more optimal than those
given by the exhaustive termination criterion (Figure \ref{fig:gauss_low}).

\subsection{Performance Experiments}

More quantitative evaluations of the termination criteria require comparing
the resulting Hamiltonian Markov chains. Both the No-U-Turn Sampler (NUTS) 
and Exhaustive Hamiltonian Monte Carlo (XHMC) were implemented using 
a second-order symplectic \textit{leapfrog} integrator with multiplicative 
trajectory expansion.  NUTS is implemented with a slice sampler over the
final trajectory while XHMC utilizes multinomial sampling.  Following the original 
implementation of the No-U-Turn Sampler, I also added an integrator error cutoff 
which rejects any trajectory $\mathfrak{t}$ sampled around the initial state, $z_{0}$, 
satisfying
\begin{equation*}
H \! \left( z_{0} \right) - H \! \left( z \right) > 1000, \, \forall z \in \mathfrak{t}.
\end{equation*}
Without any guidance on how to tune the exhaustion threshold, $\delta$, in 
all experiments XHMC is run with two nominal thresholds, $\delta = 0.1$ 
and $\delta = 0.01$.

All implementations were implemented in \textsc{Stan}~\citep{Stan:2015} and 
run with  \textsc{CmdStan}~\citep{CmdStan:2015} using the \verb+exhaustions+ 
branch (commit: \verb+c04d34ee77d831a2817cf3c7671aebc50a3bf825+).

Here I consider the performance of both samplers on an identically and 
independently distributed Gaussian target, a correlated Gaussian target, 
and a more realistic item response theory model.

\subsubsection{IID Gaussian Target}

\begin{figure}
\centering
\subfigure[]{ \includegraphics[width=2.5in]{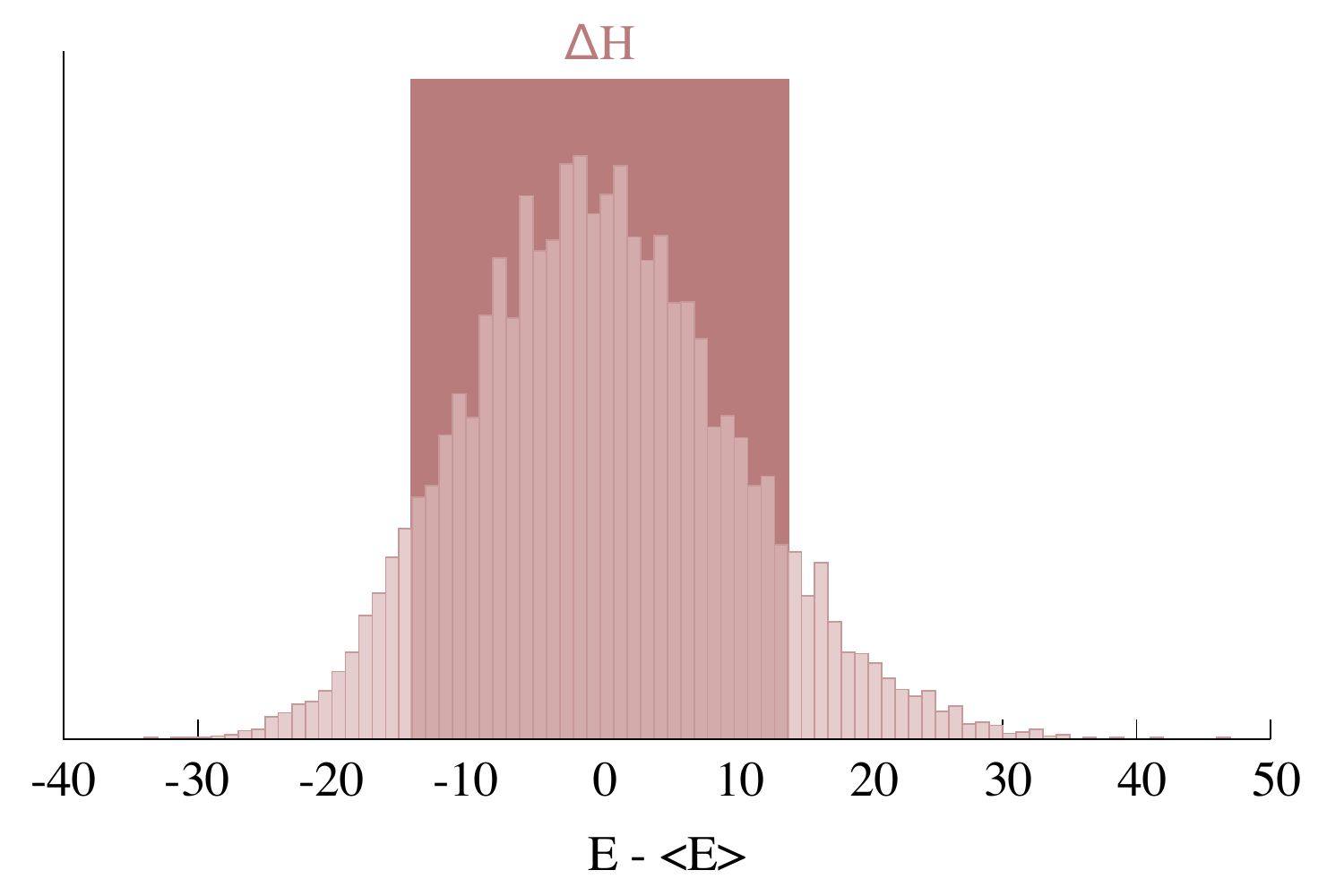} }
\subfigure[]{ \includegraphics[width=2.5in]{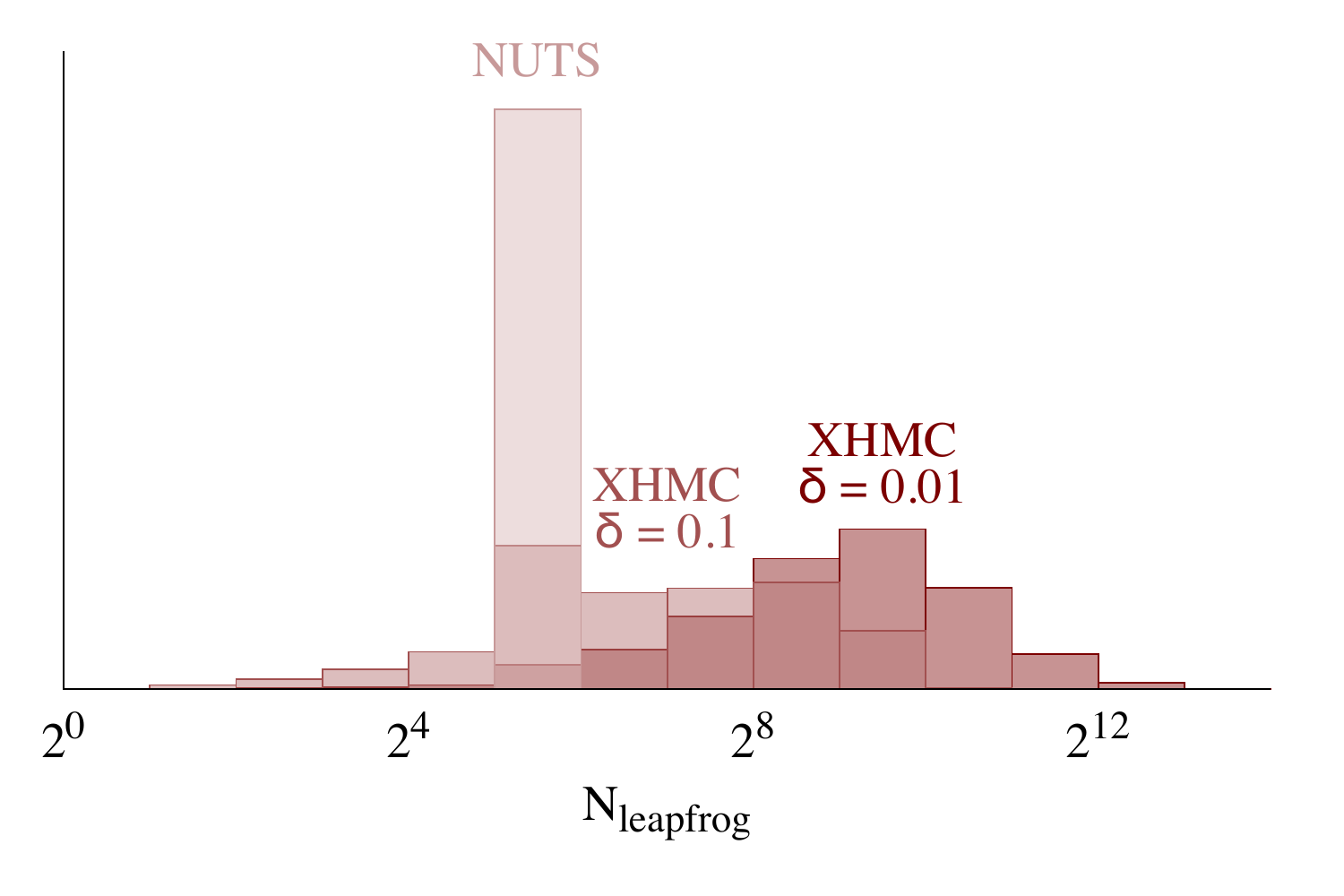} }
\subfigure[]{ \includegraphics[width=2.5in]{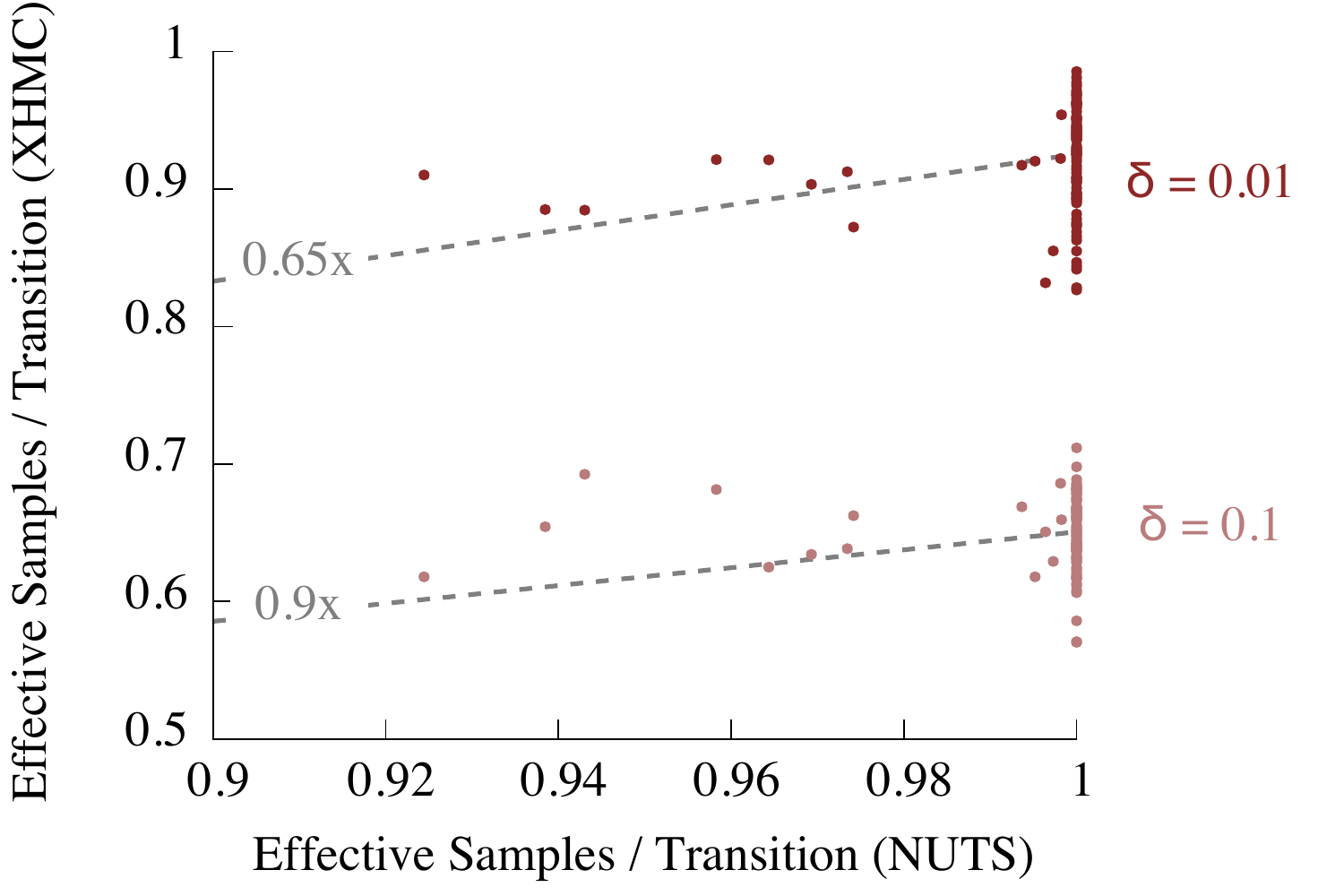} }
\subfigure[]{ \includegraphics[width=2.5in]{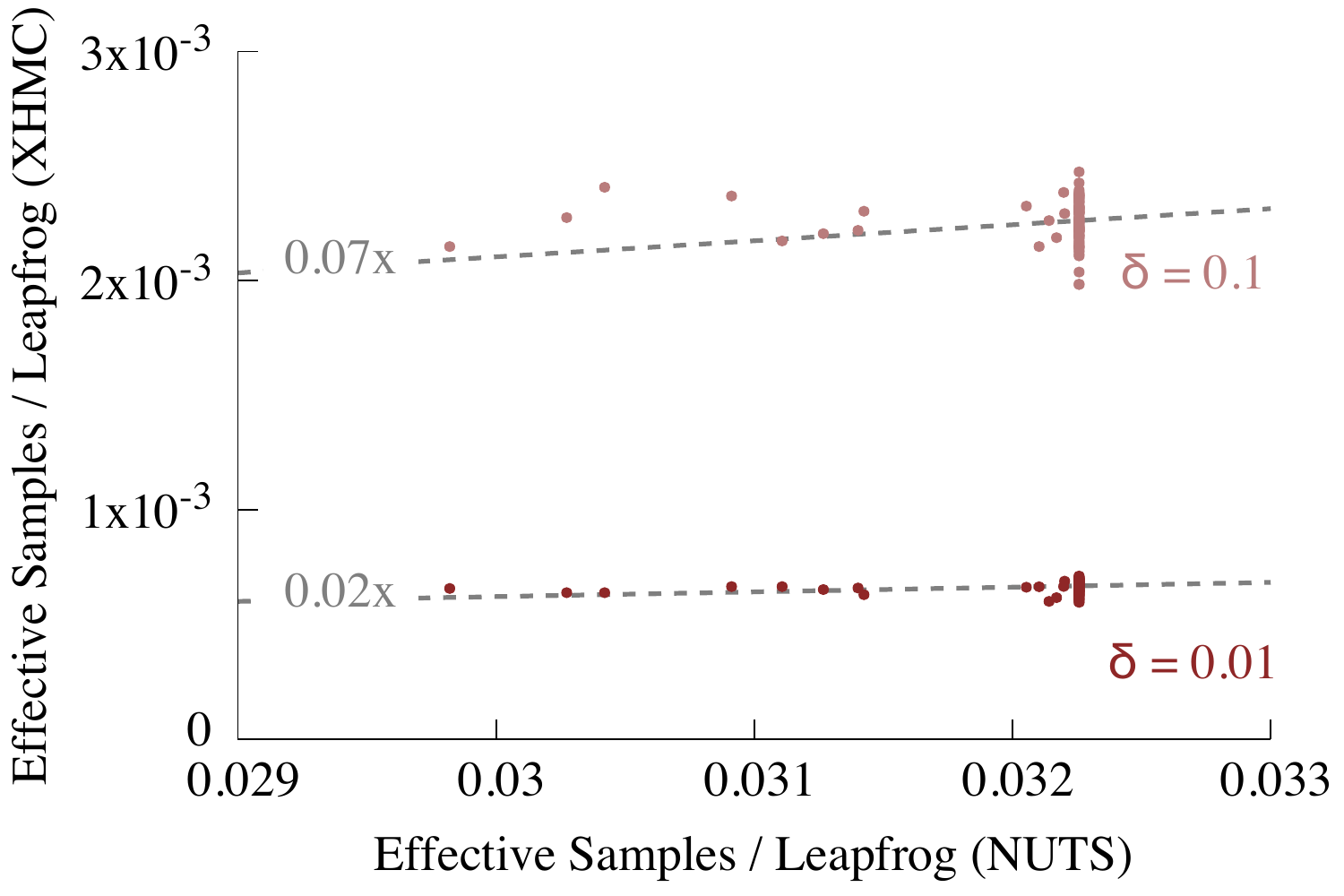} }
\caption{(a) A Euclidean-Gaussian disintegration is well-suited to an IID
Gaussian target distribution, but the ultimate sampling efficiency is
sensitive to the choice of integration time.  (b) Both XHMC tunes identify
integration times that are too long, resulting in (c) larger autocorrelations
and (d) substantially worse computational performance.}
\label{fig:gauss_iid}
\end{figure}

The $100$-dimensional IID Gaussian target with $\rho = 0$ is particularly 
nice because the optimal implementation can be identified analytically.  For 
example, the marginal energy distribution is $\chi^{2}$ with $100$ degrees 
of freedom while the variation in the momentum resampling is given by a 
$\chi^{2}$ with $50$ degrees of freedom, sufficiently wide to ensure rapid 
mixing between the level sets (Figure \ref{fig:gauss_iid}a).

Similarly, because every trajectory oscillates with the period $2\pi$ independent 
of the level set or the initial point, the optimal maximal integration time is
given by $T(z) = 2 \pi$ with the corresponding optimal trajectory length
given by $L = 2 \pi / \epsilon \sim 64$ leapfrog steps.  NUTS integrates to half of 
this time, but XHMC tends to integrate for much longer (Figure \ref{fig:gauss_iid}b),
resulting in worse effective samples per transition (Figure \ref{fig:gauss_iid}c) 
and even worse effective samples per leapfrog step (Figure \ref{fig:gauss_iid}d).

The redundant exploration in XHMC is a result of the non-stationarity of the 
exhaustive termination criterion.  For IID targets the time rate of the chance 
of the virial decomposes into a contribution from each dimension,
\begin{equation*}
\frac{ \dd G }{ \dd t} = 2 \sum_{n = 1}^{N} ( T_{n} - V_{n} ),
\end{equation*}
which oscillate to zero at different times depending on the initial state.
These contributions add incoherently and the exhaustive termination 
criterion isn't satisfied until after each dimension has oscillated through 
a full period, biasing the final samples towards the initial state and actually 
increasing the autocorrelations.  NUTS, on the other hand, is approximately 
stationary here and is able to identify near optimal integration times.

\subsubsection{Correlated Gaussian Target}

\begin{figure}
\centering
\subfigure[]{ \includegraphics[width=2.5in]{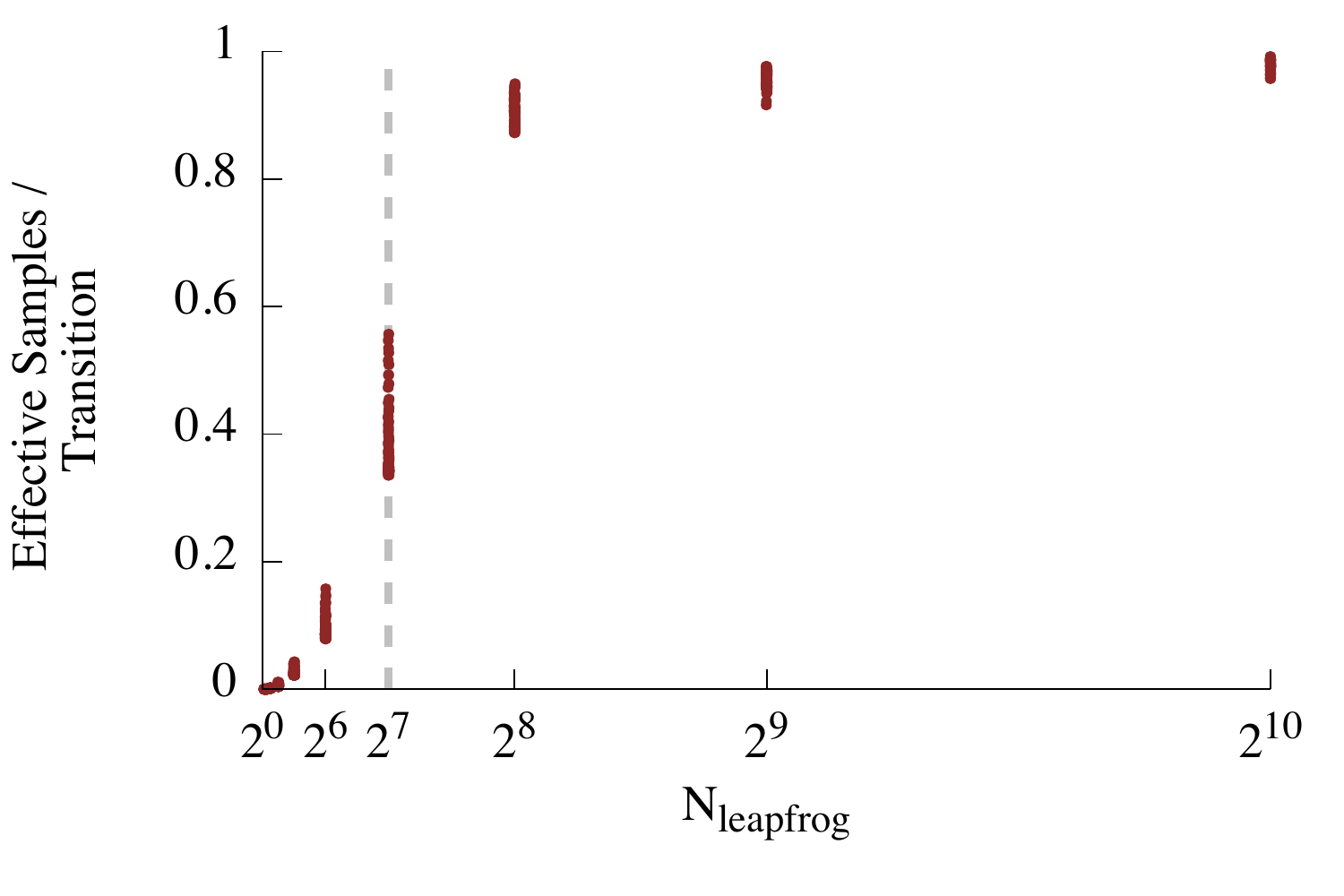} }
\subfigure[]{ \includegraphics[width=2.5in]{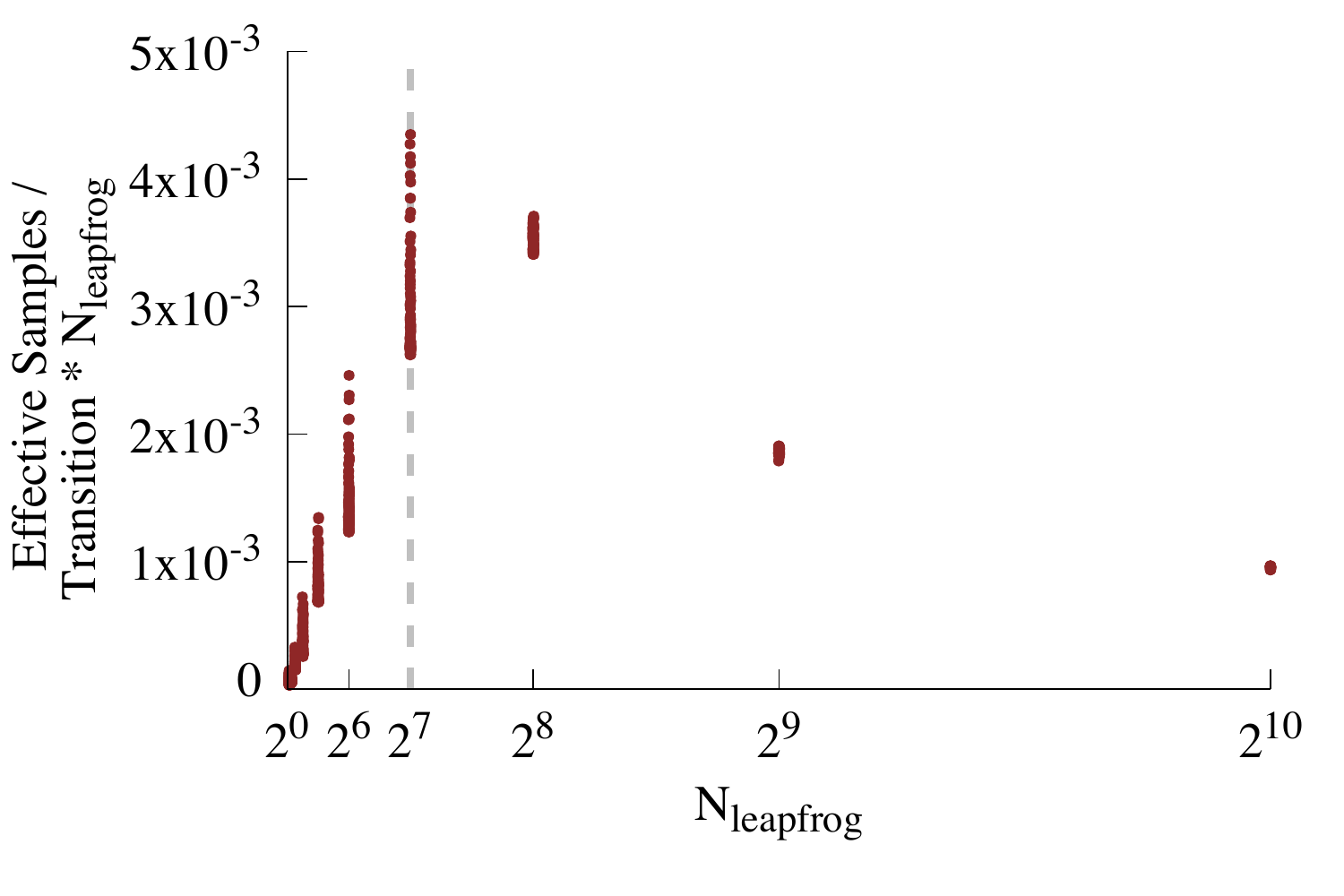} }
\caption{Two phases of convergence are evident in the autocorrelations
of a Hamiltonian Markov chain targeting a correlation Gaussian target
distribution.  For trajectory lengths below approximately $2^{7} = 128$ leapfrog 
steps the effective sample sizes grows superlinearly, but past that initial window 
the effective sample sizes grow only with the square root of the number 
of leapfrog steps.  Compare to Figure \ref{fig:microcanonical_convergence}.}
\label{fig:gauss_corr_scan}
\end{figure}

Now consider correlating the independent Gaussian components with
the covariance
\begin{equation*}
\Sigma^{ij} = \rho^{\left| i - j \right| }, \rho = 0.95.
\end{equation*}
In this case the trajectories are no longer periodic but they are dynamically
ergodic, and the rate of convergence to the microcanonical distribution
is uniform across all level sets.  To see the various phases of convergence
I sampled uniformly from static trajectories of varying lengths as described
in Section \ref{sec:static_impl}.  Up to lengths of around $2^{7} = 128$ 
leapfrog steps the trajectories converge superlinearly, but afterwards 
the convergence slows to the expected $\sqrt{t}$ asymptotic rate (Figure 
\ref{fig:gauss_corr_scan}).  Per intuition, optimal performance is achieved 
when trajectories do not grow into the asymptotic regime.

\begin{figure}
\centering
\subfigure[]{ \includegraphics[width=2.5in]{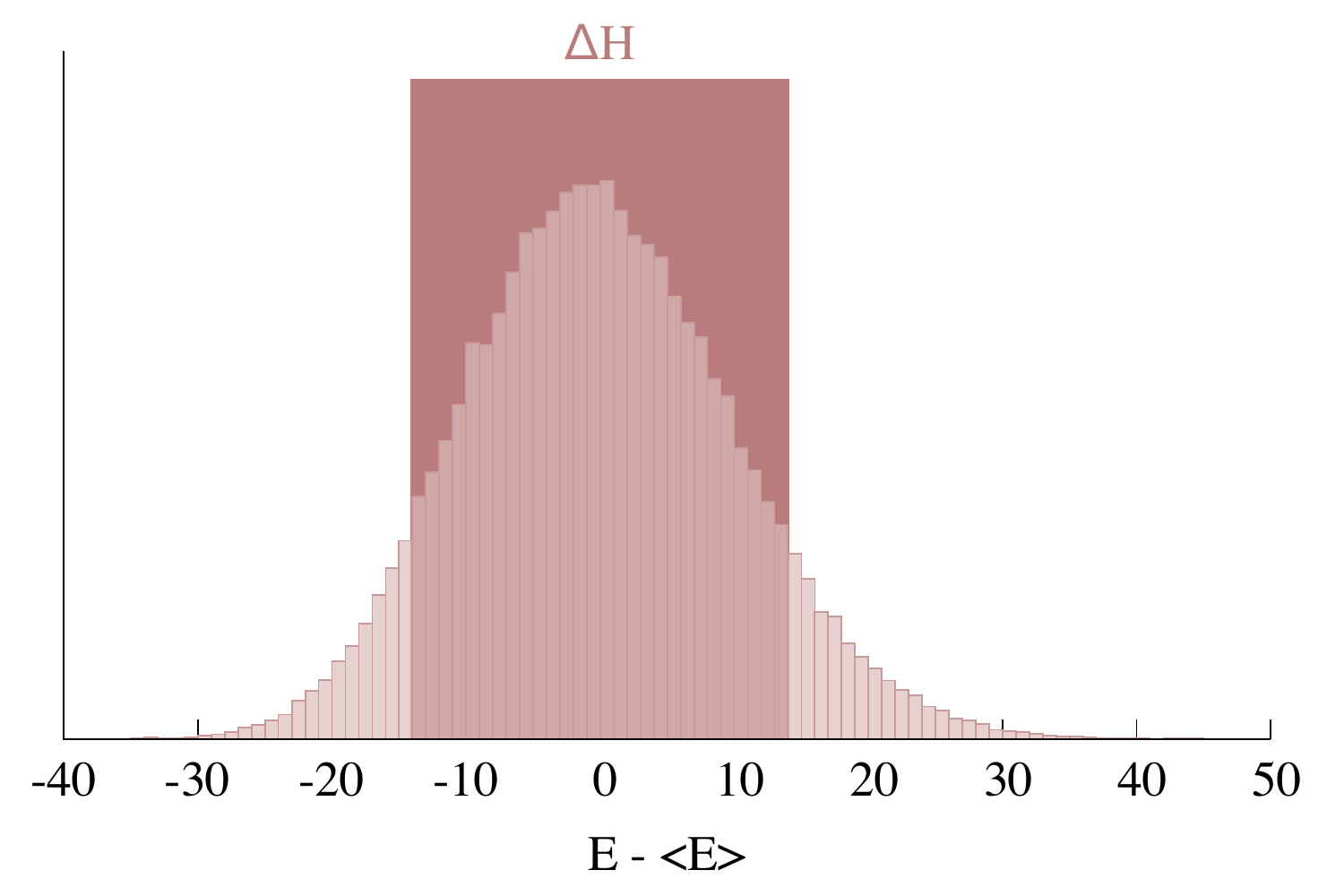} }
\subfigure[]{ \includegraphics[width=2.5in]{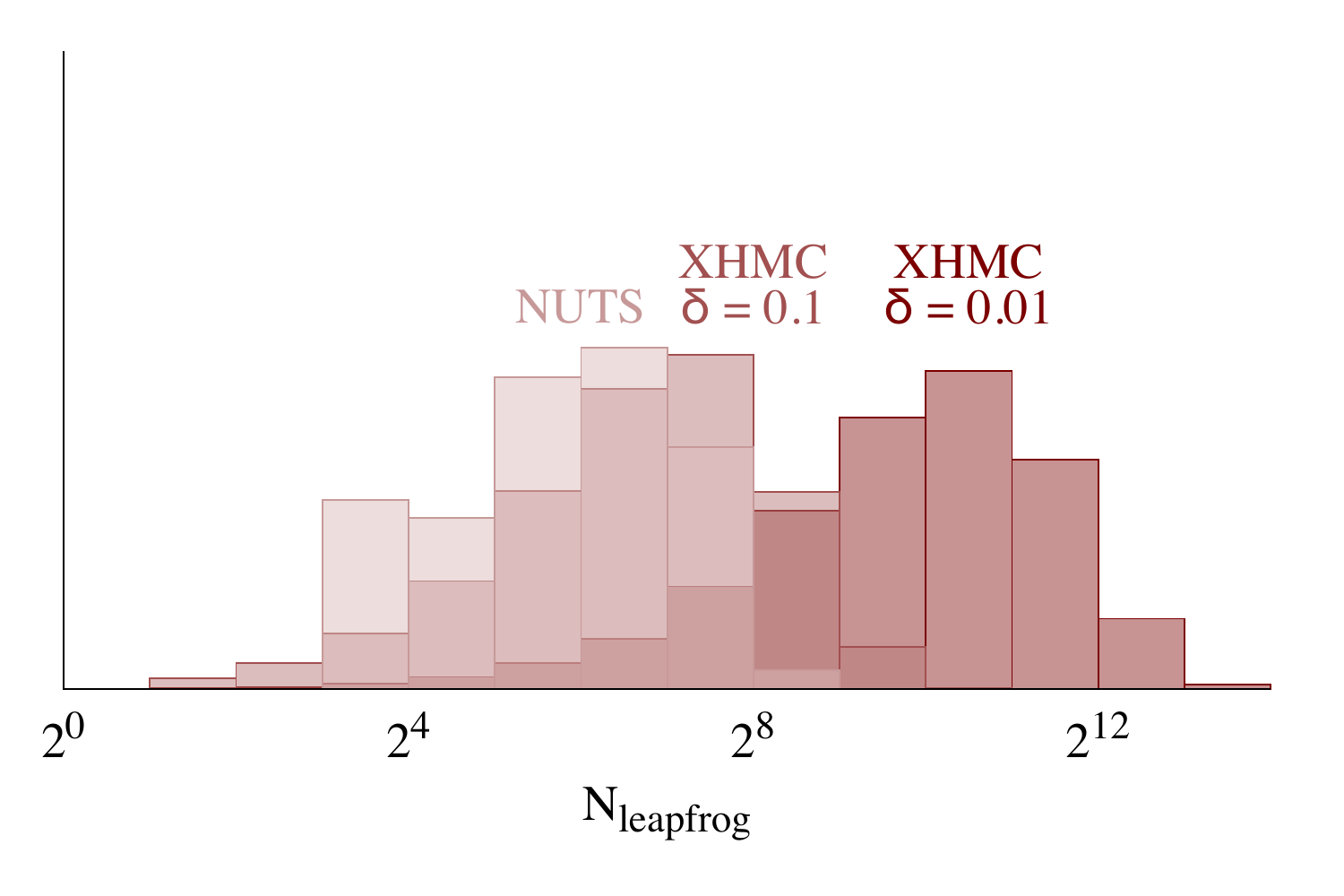} }
\subfigure[]{ \includegraphics[width=2.5in]{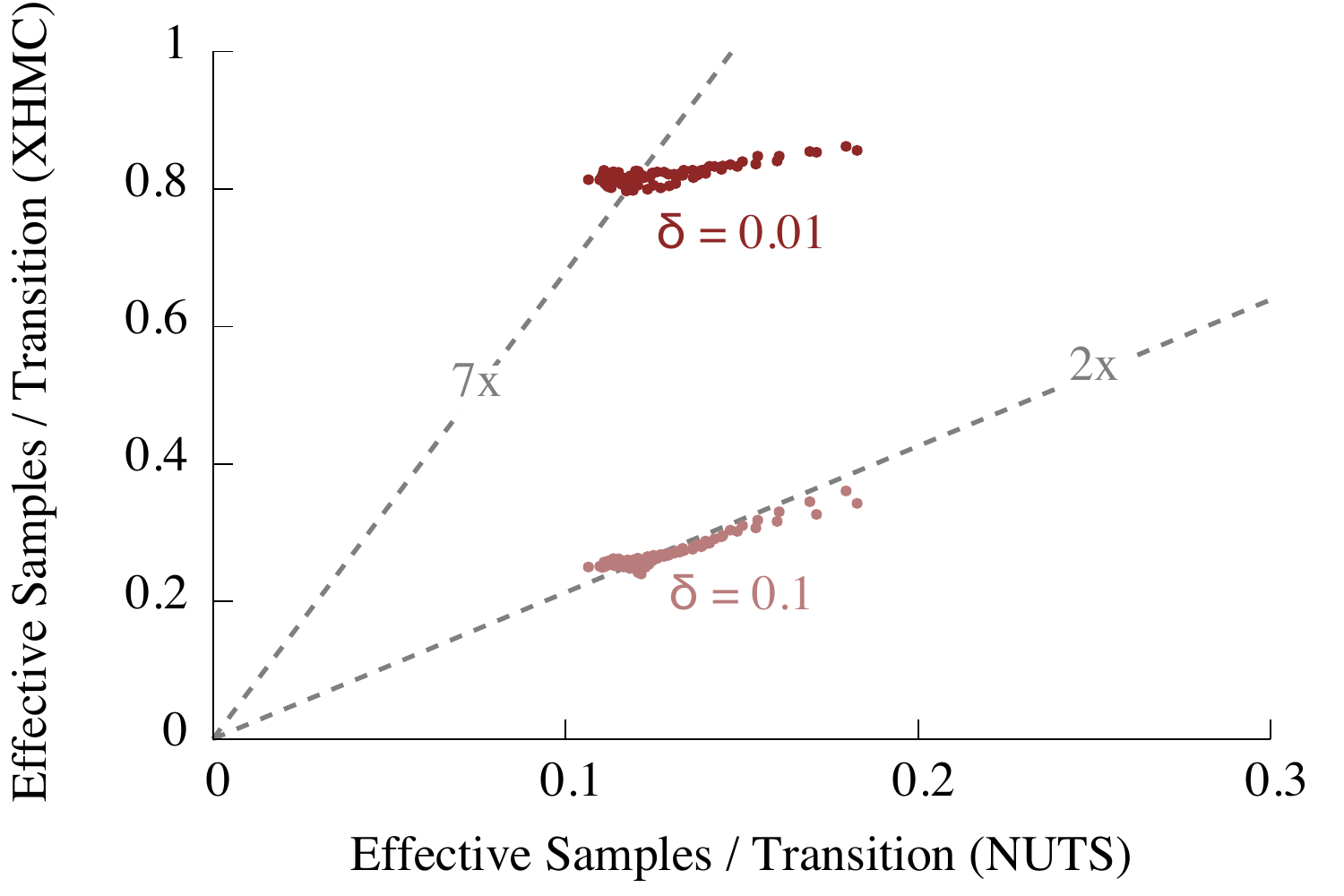} }
\subfigure[]{ \includegraphics[width=2.5in]{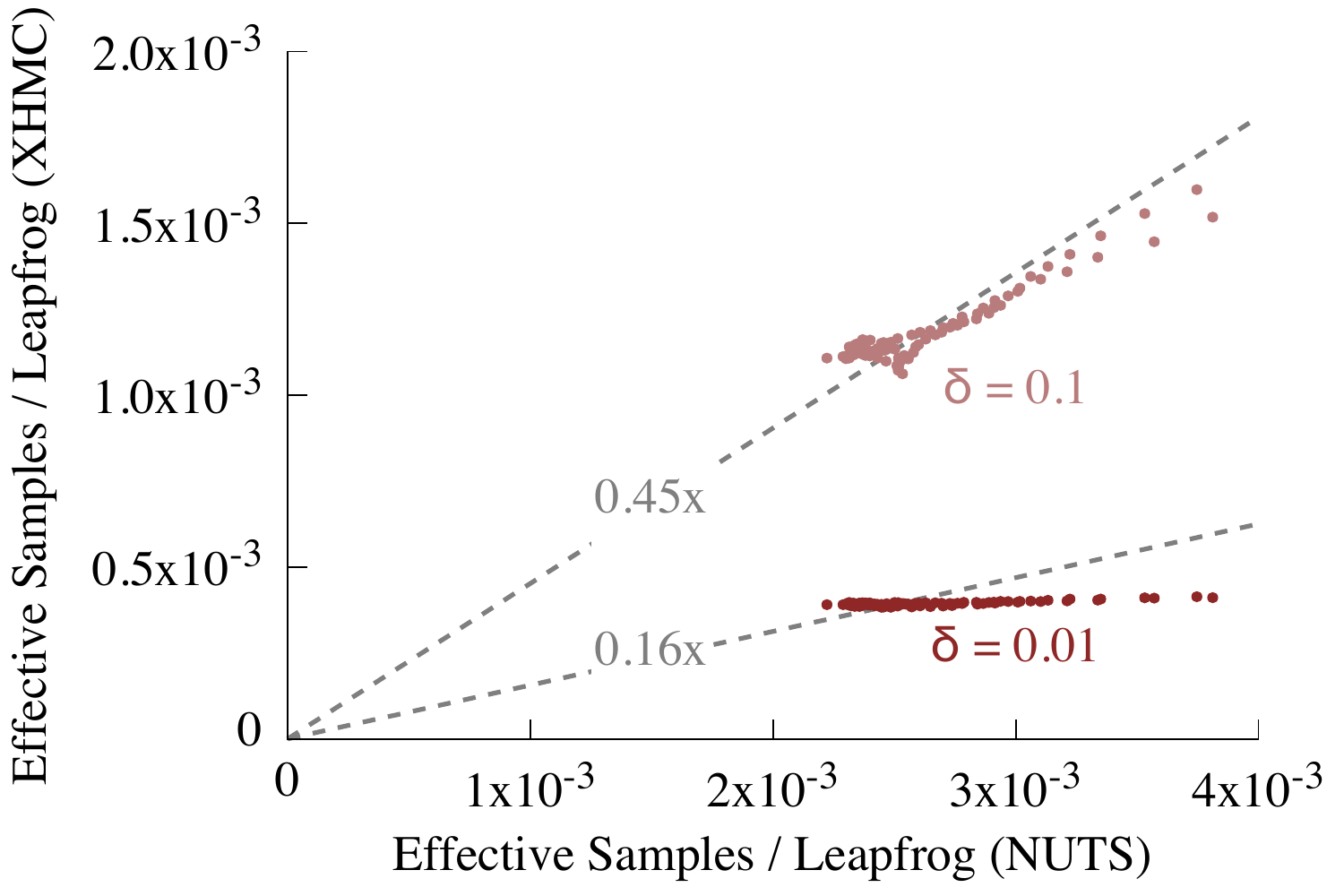} }
\caption{As in the IID case NUTS outperforms both naive tunes of XHMC
when targeting a correlated Gaussian distribution. (a) Once again the 
Euclidean-Gaussian disintegration is well-suited and (b) both XHMC tunes 
identify long integration times.  In this case the longer integration times
lead to (c) smaller autocorrelations and larger effective sample sizes,
but (d) the increase in the effective sample size is not enough to warrant
the increase computation.}
\label{fig:gauss_corr}
\end{figure}

In another strong showing, NUTS is able to identify the optimal integration
times quite well, while XHMC with the nominal tunes selects integration times
that fall into the inefficient asymptotic regime (Figure \ref{fig:gauss_corr}).
These longer integration times yield smaller autocorrelations and
larger effective sample sizes, but the increases are sublinear and hence computationally inefficient (Table \ref{tab:gauss_corr_sum}).  A 
more careful choice of the exhaustion threshold $\delta$ should lead to 
better performance, but without any guidance in selecting an optimal value 
it remains a challenging tuning problem.

\begin{table}
  \centering
  \renewcommand{\arraystretch}{1.5}
  \begin{tabular}{ccc}
    \rowcolor[gray]{0.9} 
    \textbf{XHMC} & \textbf{Increase in Total} & \textbf{Increase in Median} \\
    \rowcolor[gray]{0.9} 
    \textbf{Tune} & \textbf{Leapfrog Steps} & \textbf{Effective Sample Size} \\
    0.1 & $\approx 5$x & $\approx 2 \mathrm{x} \approx \sqrt{5} \mathrm{x}$  \\
    0.01 & $\approx 43$x & $\approx 7 \mathrm{x} \approx \sqrt{43} \mathrm{x}$
  \end{tabular}
\caption{When targeting a correlated Gaussian distribution, the nominal XHMC
tunes select long integration times that fall into the asymptotic window where the
effective sample size grows only with the square root of the number of steps
as expected.  These diminishing returns ultimately compromise the performance 
of the XHMC tunes compared to NUTS.  Larger exhaustion thresholds tuned 
to this target distribution should yield better performance, but identifying the
optimal tuning is nontrivial.}
\label{tab:gauss_corr_sum}
\end{table}

\subsubsection{Nonlinear Target}

Finally let's consider a target distribution more characteristic of applied problems: 
1-PL item response theory model for 50 students,
\begin{align*}
y_{i} 
&\sim
\mathrm{Bernoulli} \! \left( \mathrm{logistic} \left( \theta - b_{i} \right) \right)
\\
b_{i} &\sim \mathcal{N} \! \left(0, 10 \right)
\\
\theta &\sim \mathcal{N} \! \left(0, 10 \right),
\end{align*}
where the normal distributions here are specified with a mean and standard
deviation.  Because the data constrain only the sum of the $\theta$ and the
individual $b_{i}$, the likelihood is non-identified, and, although the
weakly-informative priors offer some regularization, the posterior suffers
from strong nonlinear correlations.  Because of these nonlinearities, uniform
level set exploration also requires dynamic integration times, providing a
significant challenge to the termination criteria.

\begin{figure}
\centering
\subfigure[]{ \includegraphics[width=2.5in]{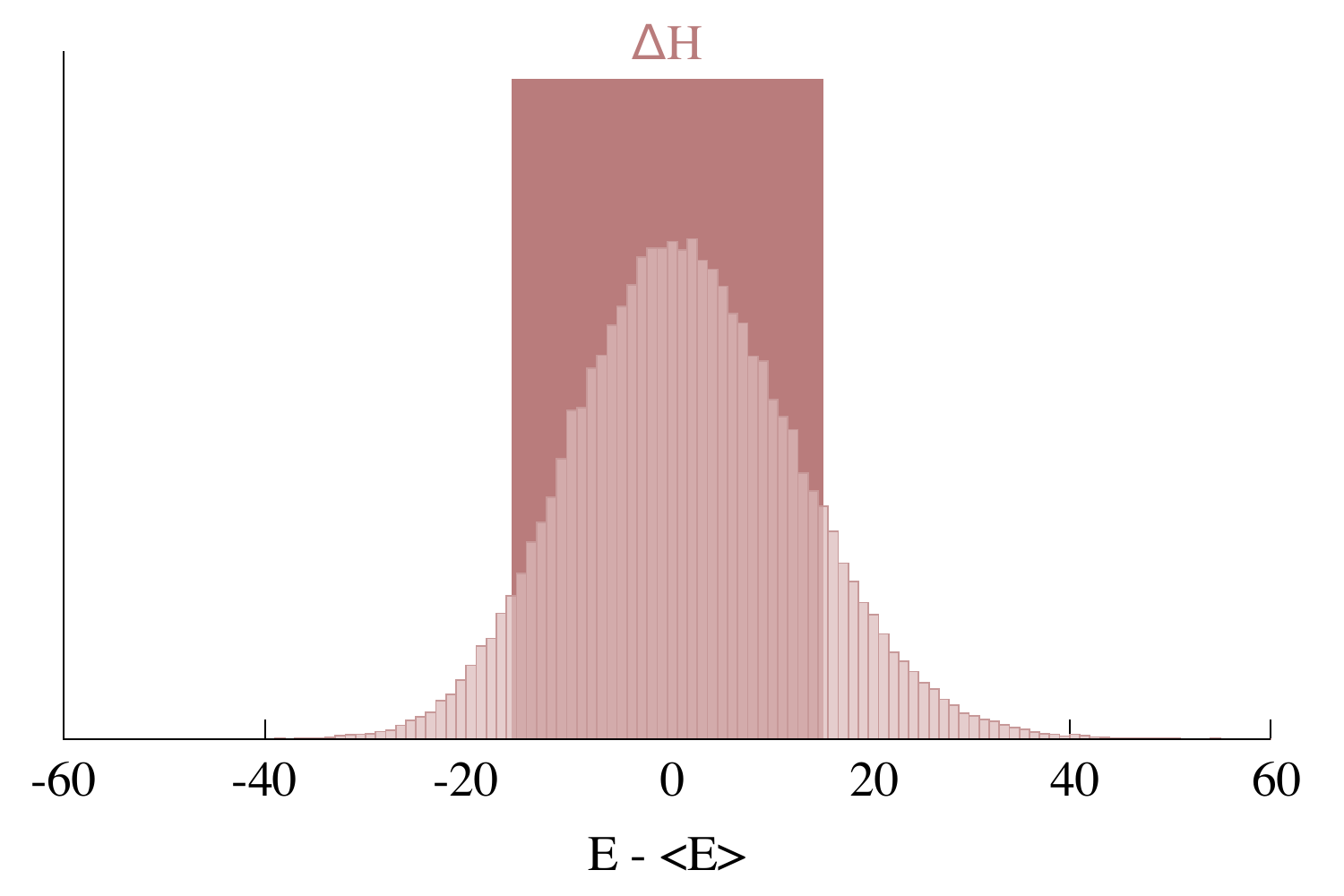} }
\subfigure[]{ \includegraphics[width=2.5in]{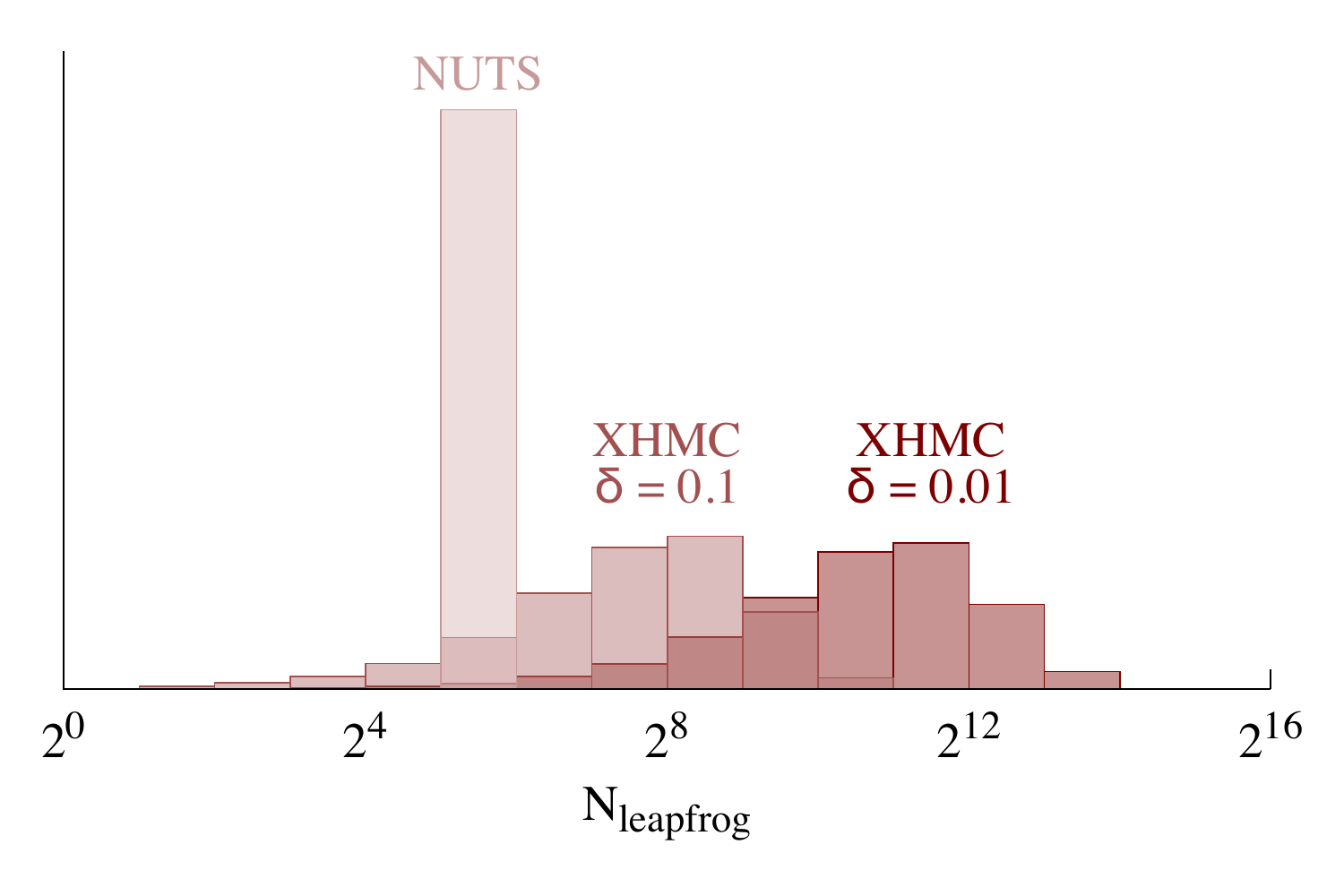} }
\subfigure[]{ \includegraphics[width=2.5in]{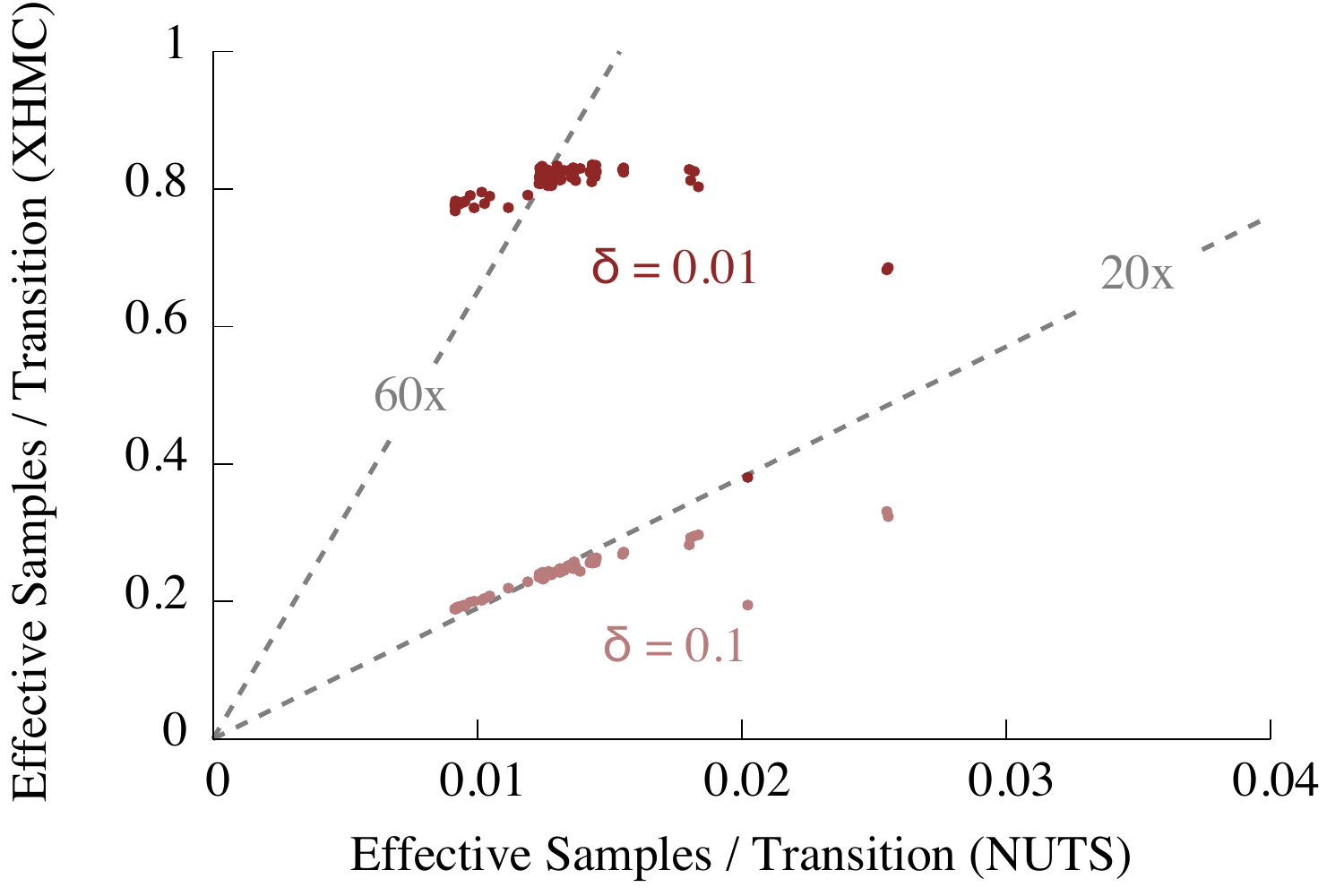} }
\subfigure[]{ \includegraphics[width=2.5in]{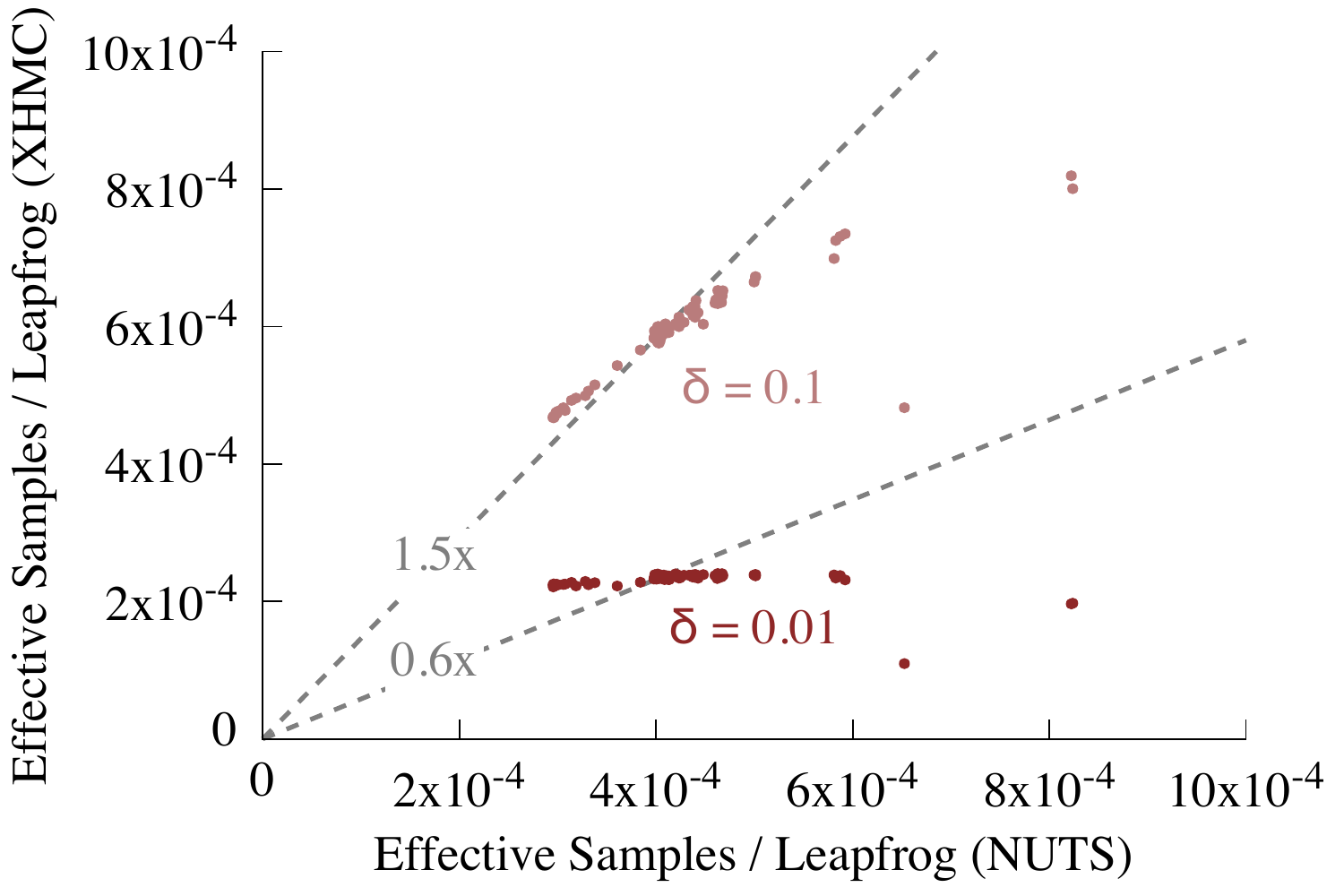} }
\caption{When targeting the highly-correlated posterior distribution of
a 1-PL item response theory model (b) XHMC integrates for much longer 
than NUTS, (c) yielding much larger effective sample sizes for each 
parameter and (d) correspondingly higher computational performance.}
\label{fig:irt}
\end{figure}

Not surprisingly, the nonlinear correlations cause the No-U-Turn Sampler to 
terminate prematurely (Figure \ref{fig:irt}b), resulting in much smaller 
effective sample sizes relative to the nominal XHMC tunes (Figure \ref{fig:irt}c) 
and correspondingly lower computationally efficiency (Figure \ref{fig:irt}d).
The superior performance of XHMC is ultimately due to the fact that the
nominal tunes identify integration times that are long without reaching the 
asymptotic regime (Table \ref{tab:irt_sum}), which is more coincidental than
deliberate.

\begin{table}
  \centering
  \renewcommand{\arraystretch}{1.5}
  \begin{tabular}{ccc}
    \rowcolor[gray]{0.9} 
    \textbf{XHMC} & \textbf{Increase in Total} & \textbf{Increase in Median} \\
    \rowcolor[gray]{0.9} 
    \textbf{Tune} & \textbf{Leapfrog Steps} & \textbf{Effective Sample Size} \\
    0.1 & $\approx 13$x & $\approx 20 \mathrm{x} > 13 \mathrm{x}$ \\
    0.01 & $\approx 110$x & $\approx 60\mathrm{x} < 110 \mathrm{x}$
  \end{tabular}
\caption{The nominal XHMC tunes not only identify longer integration times 
than NUTS when targeting the 1-PL posterior, the identified integration times
largely avoid the asymptotic regime.  In particular, $\delta = 0.1$ yields 
superlinear exploration and improved performance.  When the threshold is 
reduced to $\delta = 0.01$, however, the improvement becomes sublinear 
indicating that the increased integration times are beginning to become
asymptotic and yield only diminishing returns.}
\label{tab:irt_sum}
\end{table}

\section{Conclusions and Future Work}

Careful analysis of its rich geometric foundations demonstrates
that Hamiltonian flow efficiently explores a given target distribution, 
and admits high-performance Markov Chain Monte Carlo estimation,
when the flow is integrated long enough to avoid diffusive behavior
but not so long to waste computational resources.  This analysis not
only provides a theoretical framework for identifying optimal integration
times, it also presents new motivation for the No-U-Turn Sampler
and inspires the complementary Exhaustive Hamiltonian Monte
Carlo algorithm.  

The mixed performance of the two algorithms shows that neither
criteria is able to robustly identify optimal integration times in all cases
and suggests that better termination criteria can still be developed.  In
particular, the intriguing associations between the No-U-Turn criterion 
and Poincar\'{e} recurrence times intimates that a more explicit application
of recurrence may be critical to constructing better criteria.

One substantial benefit of exhaustive termination criterion
over the No-U-Turn criterion, however, is the stronger theoretical
foundation which makes Exhaustive Hamiltonian Monte Carlo ripe 
for rigorous formal analysis.  This includes, for example, an update of the 
step size optimality criterion of static Hamiltonian Monte 
Carlo~\citep{BetancourtEtAl:2014b} and a thorough analysis of the 
statistical ergodicity properties of the algorithm.  In particular, the
uniform exploration induced by the exhaustive termination criterion
has the potential to substantially expand the scope of target distributions
to which the implementation is geometrically ergodic.

Finally, we have not yet fully exploited the geometry of the
microcanonical disintegration.  As noted in Section \ref{sec:microcanonical}, 
for example, thorough analysis of the marginal autocorrelation on the energy 
levels can be used to identify poorly chosen cotangent disintegrations 
and, ideally, motivate optimal ones.  Additionally, the natural ergodicity 
of the Hamiltonian trajectories on their orbits suggest that we should
sample not a single point but rather average over the entire trajectory.
This averaging gives a Rao-Blackwellization of the microcanonical 
expectations with the potential to reduce the variance of the overall 
Markov Chain Monte Carlo estimators, yielding more precise estimators 
with little added computational burden.

\section{Acknowledgements}

I warmly thank Bob Carpenter, Andrew Gelman, Daniel Lee, and
Dan Simpson for helpful comments on the paper and Sam Livingstone 
and Simon Byrne for invaluable discussion of the theory and implementation.  
This work is supported under EPSRC grant EP/J016934/1 and a 2014 EPRSC 
NCSML Award for PDRA Collaboration.

\bibliography{exhaustive_hmc}
\bibliographystyle{imsart-nameyear}

\end{document}